\documentclass[12pt]{article}
\usepackage[scale=0.8]{geometry}
\usepackage[utf8]{inputenc}
\usepackage{amsfonts}
\usepackage{amsmath}
\usepackage{bbm}
\usepackage{amssymb}
\usepackage{amsthm}
\usepackage{mathabx}
\usepackage{mathrsfs}
\usepackage{natbib}
\usepackage{xcolor}
\usepackage{hyperref}
\hypersetup{
    colorlinks=true,
    linkcolor=blue,
    filecolor=magenta,      
    urlcolor=cyan,
    citecolor=blue,
}
\usepackage{graphicx}
\usepackage{subcaption}

\def\Real{\mathop{\mathbb{R}}\nolimits}

\newcommand{\be}{\boldsymbol{e}}

\newcommand{\bu}{\boldsymbol{u}}

\newcommand{\bw}{\boldsymbol{w}}
\newcommand{\bx}{\boldsymbol{x}}
\newcommand{\by}{\boldsymbol{y}}
\newcommand{\bz}{\boldsymbol{z}}
\newcommand{\bmu}{\boldsymbol{\mu}}
\newcommand{\bA}{\boldsymbol{A}}
\newcommand{\bB}{\boldsymbol{B}}
\newcommand{\bC}{\boldsymbol{C}}
\newcommand{\bD}{\boldsymbol{D}}

\newcommand{\bI}{\boldsymbol{I}}

\newcommand{\bU}{\boldsymbol{U}}
\newcommand{\bV}{\boldsymbol{V}}
\newcommand{\bW}{\boldsymbol{W}}
\newcommand{\bX}{\boldsymbol{X}}

\newcommand{\bZ}{\boldsymbol{Z}}
\newcommand{\balpha}{\boldsymbol{\alpha}}
\newcommand{\bbeta}{\boldsymbol{\beta}}

\newcommand{\bepsilon}{\boldsymbol{\epsilon}}

\newcommand{\bnu}{\boldsymbol{\nu}}

\newcommand{\btheta}{\boldsymbol{\theta}}

\newcommand{\bTheta}{\boldsymbol{\Theta}}
\newcommand{\bLambda}{\boldsymbol{\Lambda}}

\newcommand{\ha}{\hat{\balpha}}
\newcommand{\hb}{\hat{\bbeta}}

\usepackage{algorithm}
\usepackage{algpseudocode}
\newtheorem{thm}{Theorem}

\newtheorem{cor}{Corollary}

\newcommand{\indfun}[1]{\ensuremath{\mathbbm{1}_{\{#1\}}}}
\usepackage{authblk}
\title{Biconvex Clustering}
\author[1]{Saptarshi Chakraborty}
 \author[2]{Jason Xu\thanks{Correspondence to:
 \href{mailto:jason.q.xu@duke.edu}{jason.q.xu@duke.edu}}}
\affil[1]{Department of Statistics, University of California, Berkeley}
\affil[2]{Department of Statistical Science, Duke University}
\date{\vspace{-12pt}}

\begin{document}

\maketitle
\begin{abstract}
Convex clustering has recently garnered increasing interest due to its attractive theoretical and computational properties, 
but its merits become limited in the face of high-dimensional data. In such settings, pairwise affinity terms that rely on $k$-nearest neighbors become poorly specified and Euclidean measures of fit provide weaker discriminating power. %We find that recent attempts which successfully address the former difficulty still suffer from the latter, in addition to incurring high computational cost and some numerical instability. 
To surmount these issues, we propose to modify the convex clustering objective so that feature weights are optimized jointly with the centroids. The resulting problem becomes biconvex, and as such remains well-behaved statistically and algorithmically. In particular, we derive a fast algorithm with closed form updates and convergence guarantees, and establish finite-sample bounds on its prediction error. Under interpretable regularity conditions, the error bound analysis implies consistency of the proposed estimator.  Biconvex clustering performs feature selection throughout the clustering task: as the learned weights change the effective feature representation, pairwise affinities can be updated adaptively across iterations rather than precomputed within a dubious feature space. We validate the contributions on real and simulated data, showing that our method effectively addresses the challenges of dimensionality while reducing dependence on carefully tuned heuristics typical of existing approaches.
\end{abstract}
\section{Introduction and background}
\label{sec:intro}
Clustering is a cornerstone of unsupervised learning that seeks to partition unlabeled data into groups according to some measure of similarity. Classical approaches such as $k$-means clustering typically formulate the task as a non-convex optimization problem and seek a solution via a greedy algorithm. These methods are often effective in practice and have endured due to their simplicity, but also suffer from well-documented shortcomings \citep{jain2010data}. These include requiring prior knowledge or tuning of the number of clusters $k$ as input \citep{tibshirani2001estimating,hamerly2004learning}, as well as instability and sensitivity to initial conditions due to non-convexity \citep{ostrovsky2013effectiveness,pmlr-v97-xu19a}, and deteriorating performance in high dimensions \citep{witten2010framework,de2016survey,chakraborty2019strongly}.

While ongoing research continues to combat the shortcomings of algorithms such as $k$-means \citep{bachem2016fast,pmlr-v97-xu19a,zhang2020scalable}, an alternative approach is to study convex relaxations of traditionally non-convex problems \citep{tropp2006}. Convex clustering, also called sum-of-norms clustering, has garnered growing interest due to its attractive theoretical properties and viable algorithms \citep{pelckmans2005convex,hocking2011clusterpath,lindsten2011clustering}. Given data $\bX \in \Real^{n\times p}$ and denoting the $j$th row of a matrix $\bZ$ as $\bz_{j \cdot}$, the convex clustering objective is given by
\begin{equation}
    \label{convex obj}
    \min_{\bmu} \frac{1}{2} \sum_{i=1}^n \| \bx_{i\cdot} - \bmu_{i\cdot} \|^2_2 + \gamma \sum_{i<j}^n \phi_{ij} \| \bmu_{i\cdot} - \bmu_{j\cdot}\|_q .
\end{equation}
Here $\|\cdot\|_q$ denote the $\ell_q$ norm, $q \in \mathbb{N}$. Each row of the optimization variable $\bmu \in \mathbb{R}^{n \times p}$ represents a cluster center. The first term of \eqref{convex obj} is a measure of fit between the data points $\bX$ and the solution $\bmu$; the latter is a fusion term that penalizes the number of unique centers or rows of $\bmu$ by way of an $\ell_q$ norm penalty with tuning constant $\gamma>0$. The approach can be understood as a convex relaxation of hierarchical clustering. Pairwise affinities $\phi_{ij} >0$ can be chosen heuristically to accelerate computation and improve empirical performance.% detailed below.

Under this formulation, the number of clusters can be selected automatically as the number of unique rows in the solution matrix $\hat\bmu$. Indeed, the method entails a continuous solution path as a function of the parameter $\gamma$ --- a larger $\gamma$ gives the fusion penalty more relative influence, resulting in fewer unique centers or clusters \citep{chi2015splitting}. Convexity ensures a unique global minimizer, suggesting stability of any valid iterative optimization algorithm regardless of initial condition. A convex formulation is also attractive from a theory perspective; works by \cite{zhu2014convex, tan2015statistical,radchenko2017convex} provide recovery guarantees, and \cite{chi2019recovering}  establish conditions under which, the solution path recovers a tree. Computationally, \cite{chi2015splitting} provide a unified framework based on splitting methods that render the work of \cite{lindsten2011clustering} practical. % for various choices of $q$ defining the fusion penalty norm. 
The idea has been extended to many related tasks including tensors, metric versions, co-clustering, multi-view and histogram-valued data. \citep{wu2016general,chi2018provable,park2019convex,wang2019integrative}.

Despite the recent success of convex clustering, two notable and related shortcomings include its limitations in high dimensions and its strong reliance on careful specification of affinities $\phi_{ij}$. Typically, users follow a recommendation that combines $k$-nearest neighbors ($k$-NN) and Gaussian kernels \citep{chi2015splitting}: that is,
\begin{equation}\label{eq:affinity}
\phi_{ij} = \indfun{i,j}^k \exp\bigg\{-\frac{\|\bx_i-\bx_j\|^2}{\tau}\bigg\}
\end{equation}
where the indicator $\indfun{i,j}^k$ is equal to $1$  if $\bx_j$ is among the $k$-NNs of $\bx_i$ with respect to $\|\cdot\|$ 
and $0$ otherwise, and the constant $\tau$ is a bandwidth parameter. 
The success of such heuristics for choosing $\phi_{ij}$ varies depending on the application, and improper specification may lead to pathological behavior including splits in the clustering path or abrupt merging to the overall mean \citep{hocking2011clusterpath,chi2015splitting}, and %Further, this Gaussian ``blurring" 
becomes ineffective in high dimensions where pairwise distances appearing in the kernel as well as $k$-NN evaluations become less informative \citep{aggarwal2001surprising}. Though a convex formulation ensures a unique global optimum, the instability it seeks to address is conserved in a sense as the method becomes fragile with respect to good choice of $\phi_{ij}$.

Toward high dimensional problems, a sparse variant of convex clustering has been developed by \cite{wang2018sparse}, and proposes to include an additional group lasso penalty on the \textit{columns} of the centroid matrix as follows:
\begin{equation}\label{wang}
    \min_{\bmu} \frac{1}{2}\sum_{i=1}^n\|\bx_{i \cdot}-\bmu_{i \cdot}\|_2^2+\gamma_1 \sum_{i<j}\phi_{ij}\|\bmu_{i \cdot}-\bmu_{j \cdot}\|_q+\gamma_2 \sum_{l=1}^p u_l \|\bmu_{\cdot l}\|_2.
\end{equation}
Applying such penalties to the columns of $\bmu$ preserves convexity, but also introduces unwanted shrinkage toward the origin, and comes with higher computational costs and additional tuning parameters. Methodologically, this approach as well as extensions to convex clustering mentioned above borrow heavily from the splitting framework introduced by \cite{chi2015splitting}, and similarly rely on careful choice of $\phi_{ij}$. 
The penalty formulations in convex clustering and sparse convex clustering are largely inspired by techniques that have proven successful in regression \citep{tibshirani2005sparsity,yuan2006model} but do not fully leverage some of the good intuition established in the existing clustering literature. 
In this article, we advocate maneuvering to a biconvex objective, which will enable us to bridge the benefits of convex clustering to ideas that have successfully improved classical clustering schemes. Specifically, we introduce an additional optimization variable $\bw$ to enable feature weighing and selection.  Such \textit{feature weights} \citep{desarbo1984synthesized,modha2003feature}, have been used to improve many classical clustering schemes \citep{de2016survey}: for instance, the weighted $k$-means ($W$-$k$-means) method \citep{huang2005automated} seeks to solve the following minimization problem: 
\[\min_{\bmu, \bw}\bigg\{\sum_{i=1}^n\min_{1\leq j \leq k}\sum_{l=1}^p w_l^2 \, d(x_{il},\mu_{jl})\bigg\}, \hspace{5pt} \text{subject to } \sum_{l=1}^p w_l=1, \]  
where $\bw=[w_1,\dots, w_p]^\top$ is the vector of non-negative feature weights modifying the usual $k$-means objective, and  $d(\cdot,\cdot)$ is a similarity measure typically chosen as the Euclidean norm. Variations using entropy regularization
\citep{jing2007entropy,chakraborty2020entropy} and non-linear distance measures \citep{de2012minkowski} have also been explored. A framework for sparse clustering by \cite{witten2010framework} also employs such feature weights in seeking to maximize the between cluster sum-of-squares
\begin{equation}
\sum_{l=1}^p w_l \Bigg(\frac{1}{n}\sum_{i=1}^n\sum_{i^\prime=1}^n d(x_{il}, x_{i^\prime l})- 
\sum_{j=1}^k \frac{1}{|C_j|}\sum_{i,i^\prime \in C_j}d(x_{il} , x_{i^\prime l})\Bigg)
\label{eq: witten}
\end{equation}
with respect to clusters $C_j$, subject to $\ell_1$ and $\ell_2$ constraints on weights  $\|\bw\|_2 \le 1$, $\|\bw\|_1 \le s .$  
We remark that feature weights should not be confused with the weight terms appearing in sparse regression techniques such as the adaptive lasso \citep{zou2006adaptive} --- in particular, they are disjoint from the optimization variable and are penalized separately. Our approach will leverage this fact to avoid %allow for feature weighing and selection without 
excess shrinkage of the centroids toward the global mean.

These advantages come at the cost of sacrificing convexity, yet biconvex formulations inherit many of the same appealing properties such as stability and robustness compared to general non-convex problems \citep{gorski2007biconvex}. Indeed, optimization routines for solving biconvex objectives come with attractive convergence guarantees \citep{tseng2001convergence} and provide efficient solutions to biconvex statistical learning tasks such as non-negative matrix factorization \citep{lee2001}. In contrast to the variable splitting approaches used by \cite{wang2018sparse,chi2015splitting}, we show that our objective function can be optimized directly via block-coordinate descent, reflecting the simpler alternating update form that typifies many now canonical clustering algorithms. While convergence is straightforward due to biconvexity, the theoretical analysis of the statistical properties of our estimator is more challenging. Similar results in the literature \citep{tan2015statistical,wang2018sparse,chi2018provable} rely on a classical Hanson-Wright  argument \citep{hanson1971bound}, but the dependence between the learned feature weights and the noise term (assumed in the model) renders this approach ineffective in our case. Instead we appeal to uniform versions of such inequalities \citep{10.1214/20-EJP422}, and newly derive interpretable conditions under which the method is consistent. In particular, under the standard assumption of merging only $\mathcal{O}(n)$ many centroids, %(for the $k$-NN scheme, we give incentive to merge no more that $kn$ centroids), 
the proposed method is shown to be $\sqrt{n}$-consistent.

%Moreover, the theoretical analysis undertaken in this paper is relatively more challenging compared to the analogous theoretical contributions in the literature \citep{tan2015statistical,wang2018sparse,chi2018provable}. In particular the dependence of the found feature weights with the added noise term (assumed in the model) renders the classical Hanson-Wright \citep{hanson1971bound} argument ineffective. 

In addition to improving clustering performance and providing interpretable estimates of feature relevance, biconvex clustering also ameliorates sensitivity to ad-hoc specifications of the affinities $\phi_{ij}$.
As a sparse set of weights is maintained throughout the clustering task, pairwise distances and nearest neighbors can be considered within a lower-dimensional, learned feature space that \textit{adaptively} informs $\phi_{ij}$. 
Before further detailing the proposed framework, we begin with a simple toy example to motivate our contributions.

\begin{figure}[htbp!]
        \begin{subfigure}[t]{0.23\textwidth}
    \centering
        \includegraphics[height=.9\textwidth,width=.9\textwidth]{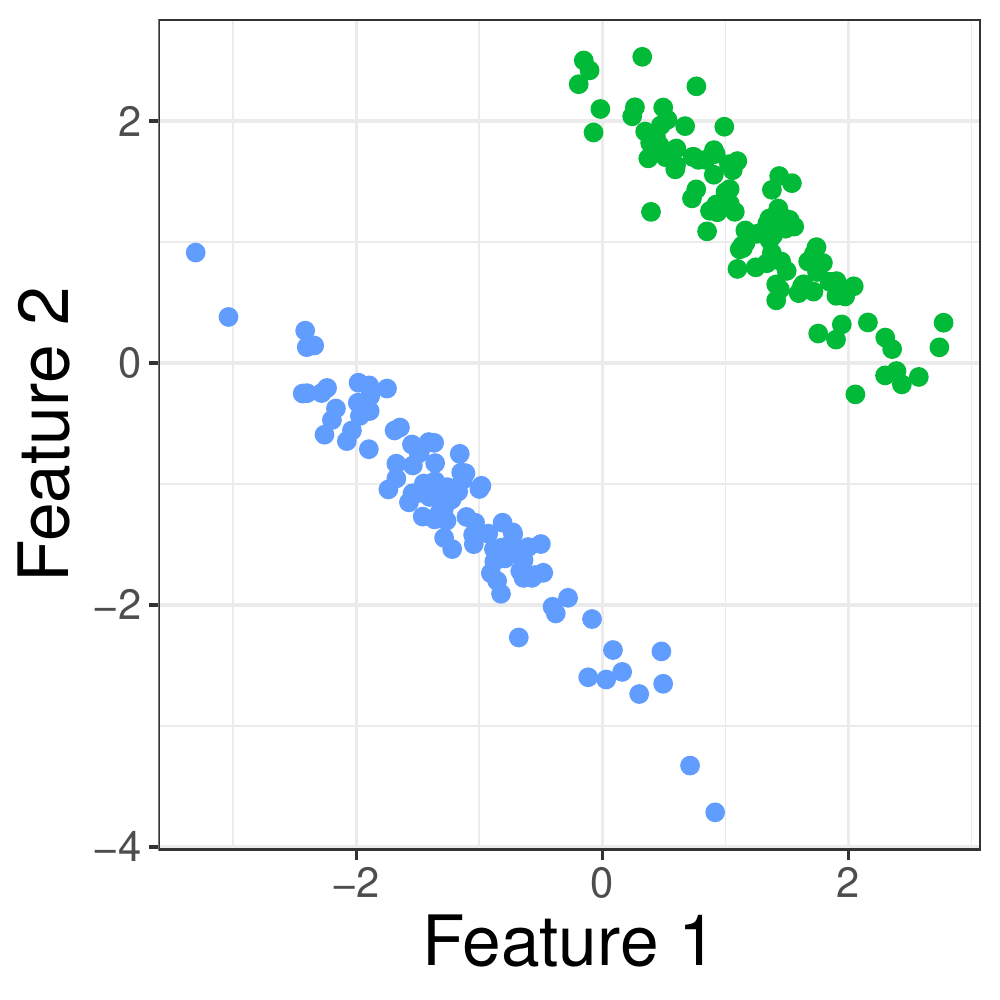}
        \caption{Ground Truth}
        \label{fig: gt}
        \end{subfigure}
        ~
        \begin{subfigure}[t]{0.23\textwidth}
    \centering
        \includegraphics[height=.9\textwidth,width=.9\textwidth]{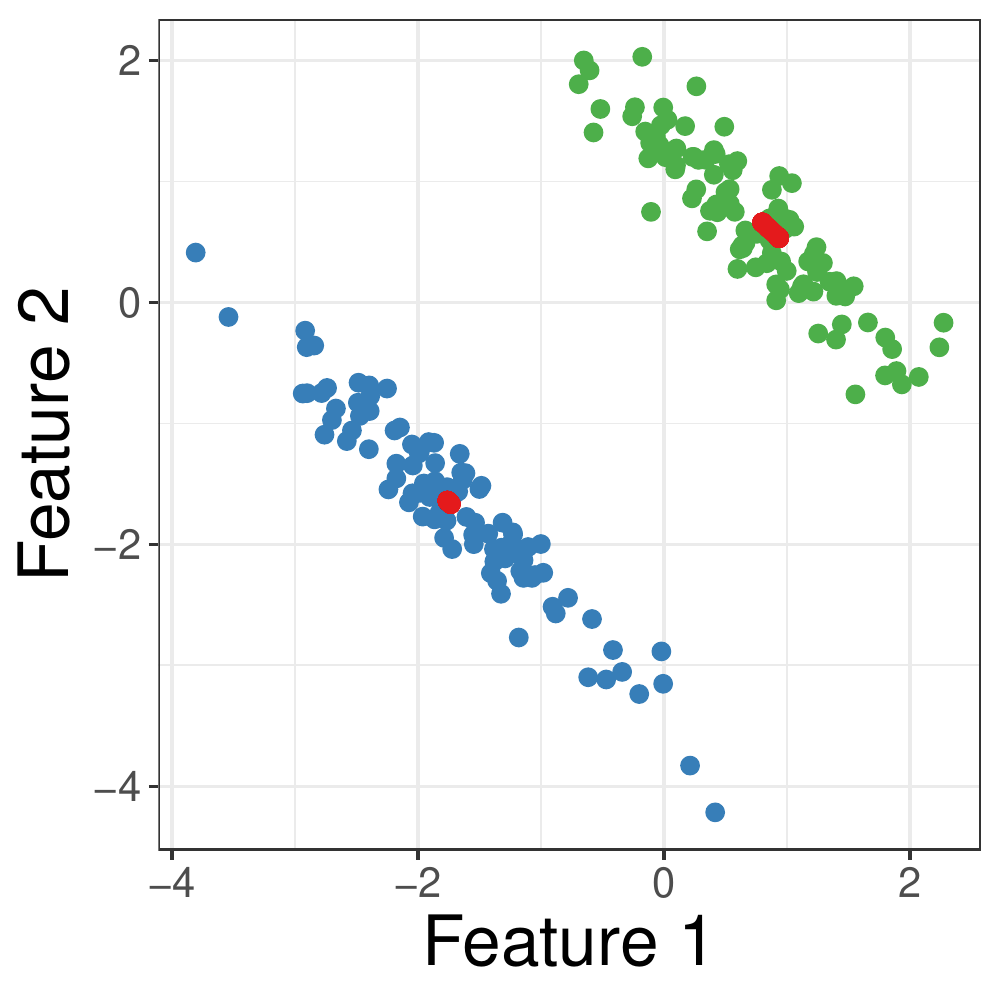}
        \caption{Biconvex}
        \end{subfigure}
    ~
    \centering
    \begin{subfigure}[t]{0.23\textwidth}
    \centering
        \includegraphics[height=.9\textwidth,width=.9\textwidth]{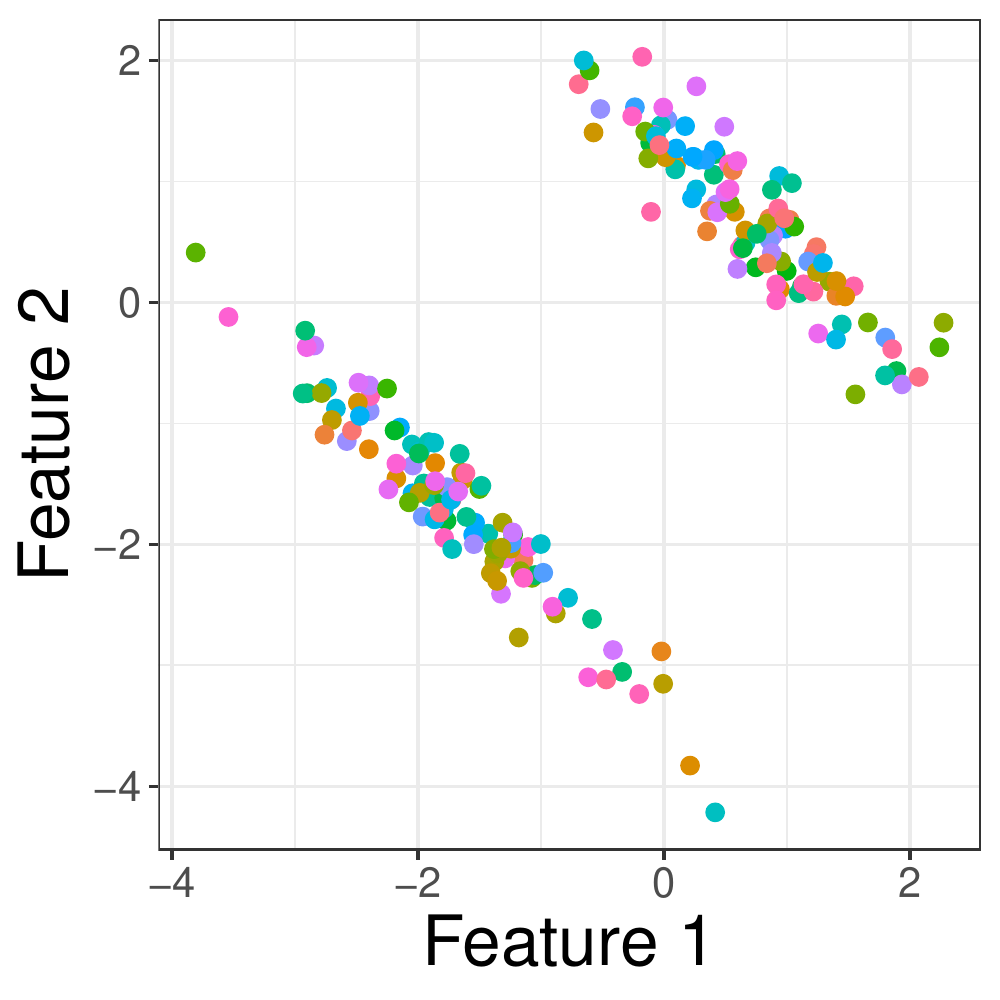}
        \caption{Convex Clustering}
    \end{subfigure}
    ~
    \begin{subfigure}[t]{0.23\textwidth}
    \centering
        \includegraphics[height=.9\textwidth,width=.9\textwidth]{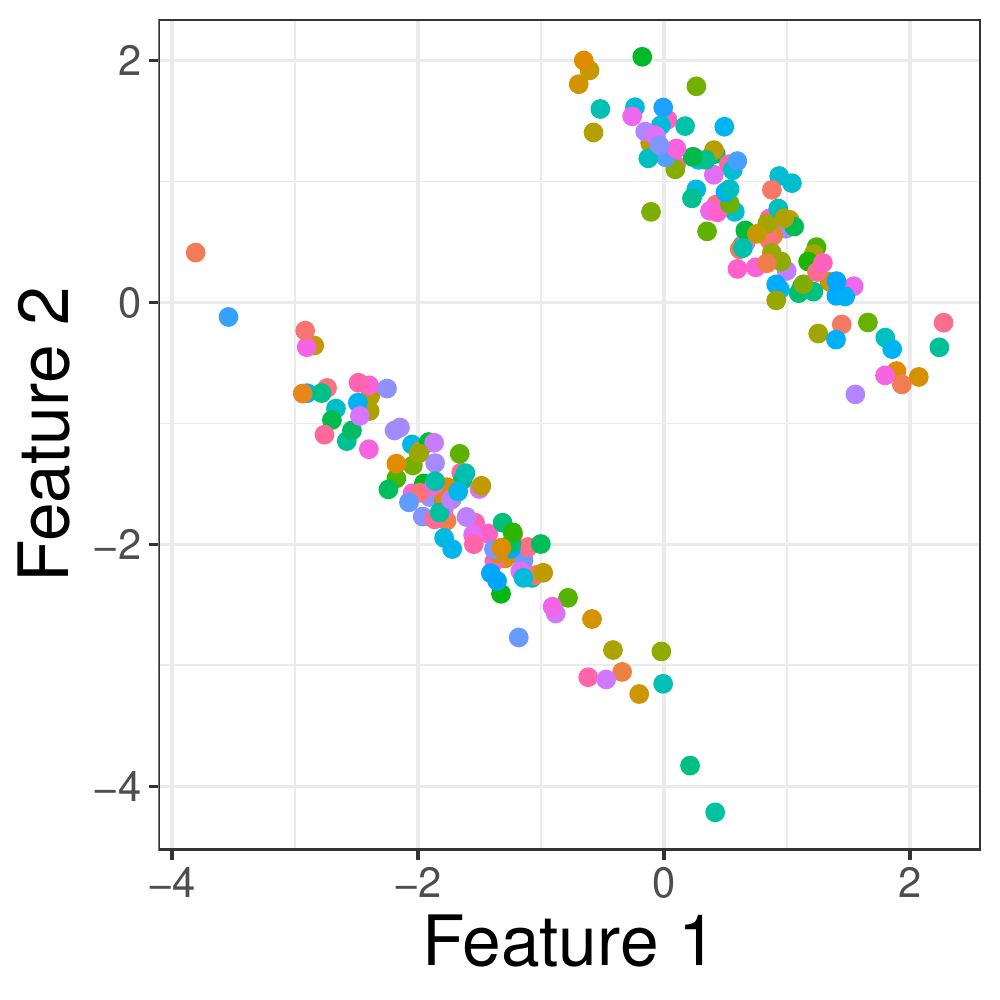}
        \caption{Sparse Convex}
        \label{motiv:spcvx}
    \end{subfigure}
    ~
        \begin{subfigure}[t]{0.23\textwidth}
    \centering
        \includegraphics[height=.9\textwidth,width=.9\textwidth]{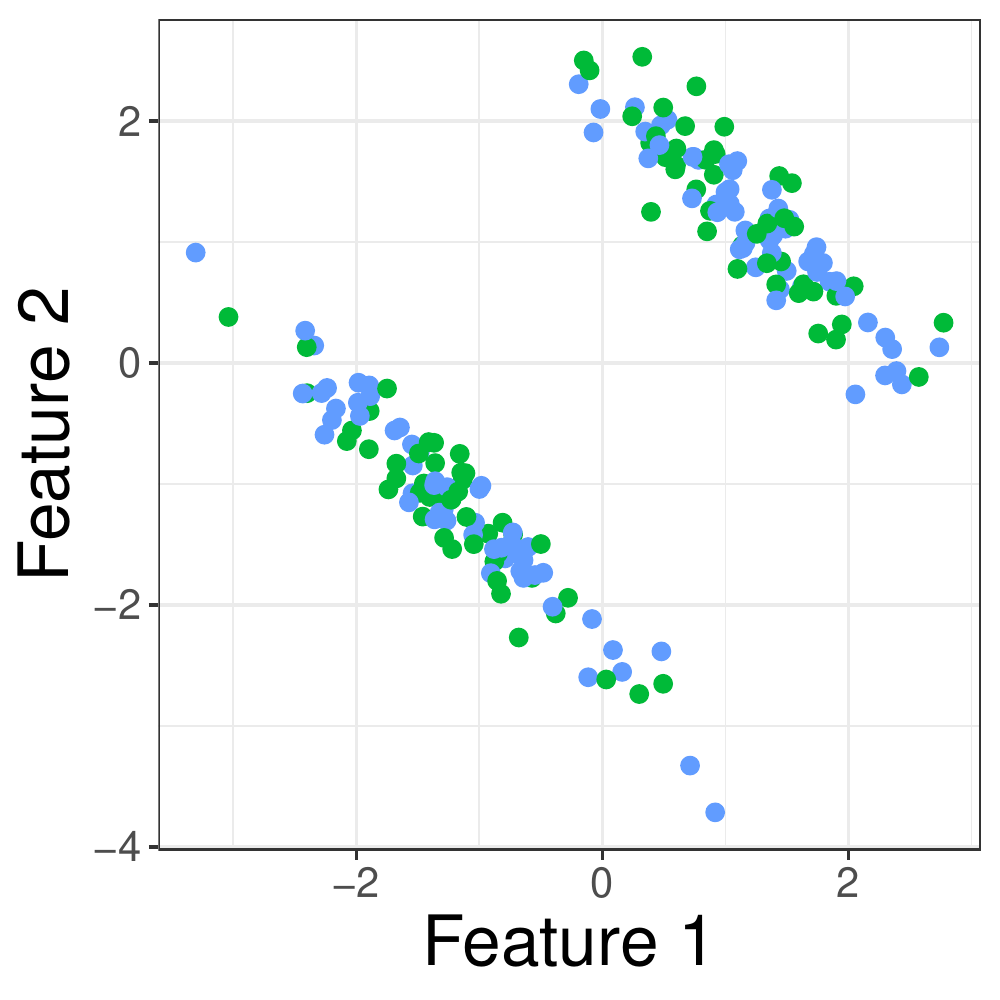}
        \caption{Sparse $k$-means}
    \end{subfigure}
        ~
    \begin{subfigure}[t]{0.23\textwidth}
    \centering
        \includegraphics[height=.9\textwidth,width=.9\textwidth]{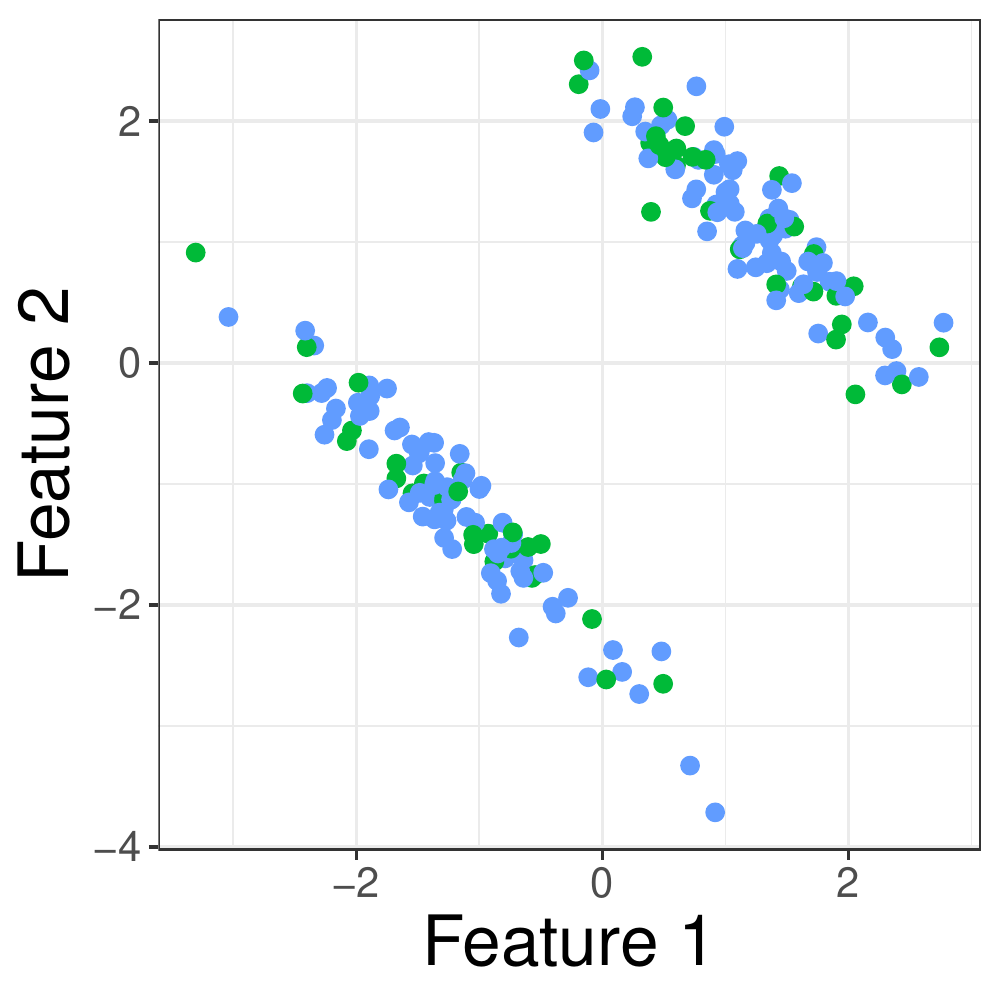}
        \caption{Sparse Hierarchical}
    \end{subfigure}
    ~
     \begin{subfigure}[t]{0.23\textwidth}
    \centering
        \includegraphics[height=.9\textwidth,width=.9\textwidth]{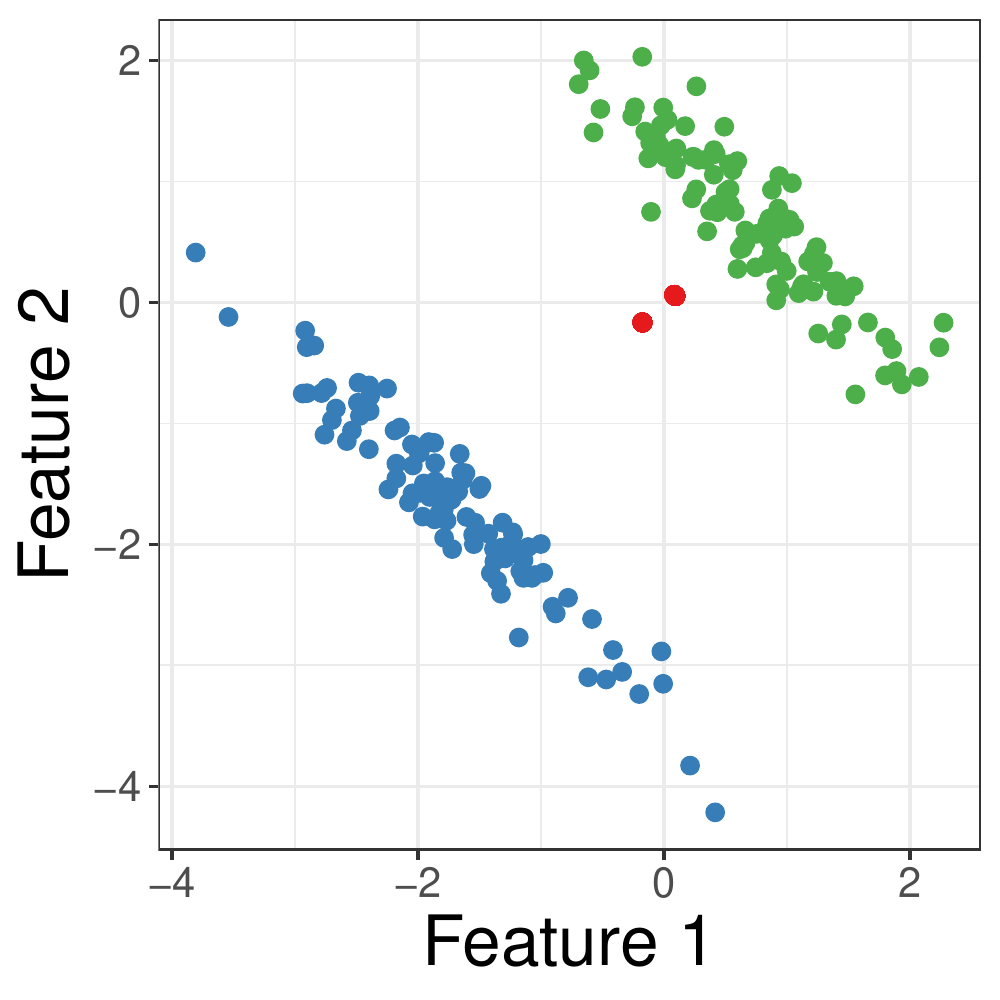}
        \caption{Sparse Convex, with learned affinities}
        \label{scvx2}
    \end{subfigure}
    ~
    \begin{subfigure}[t]{0.23\textwidth}
    \centering
        \includegraphics[width=.9\textwidth,height=.9\textwidth]{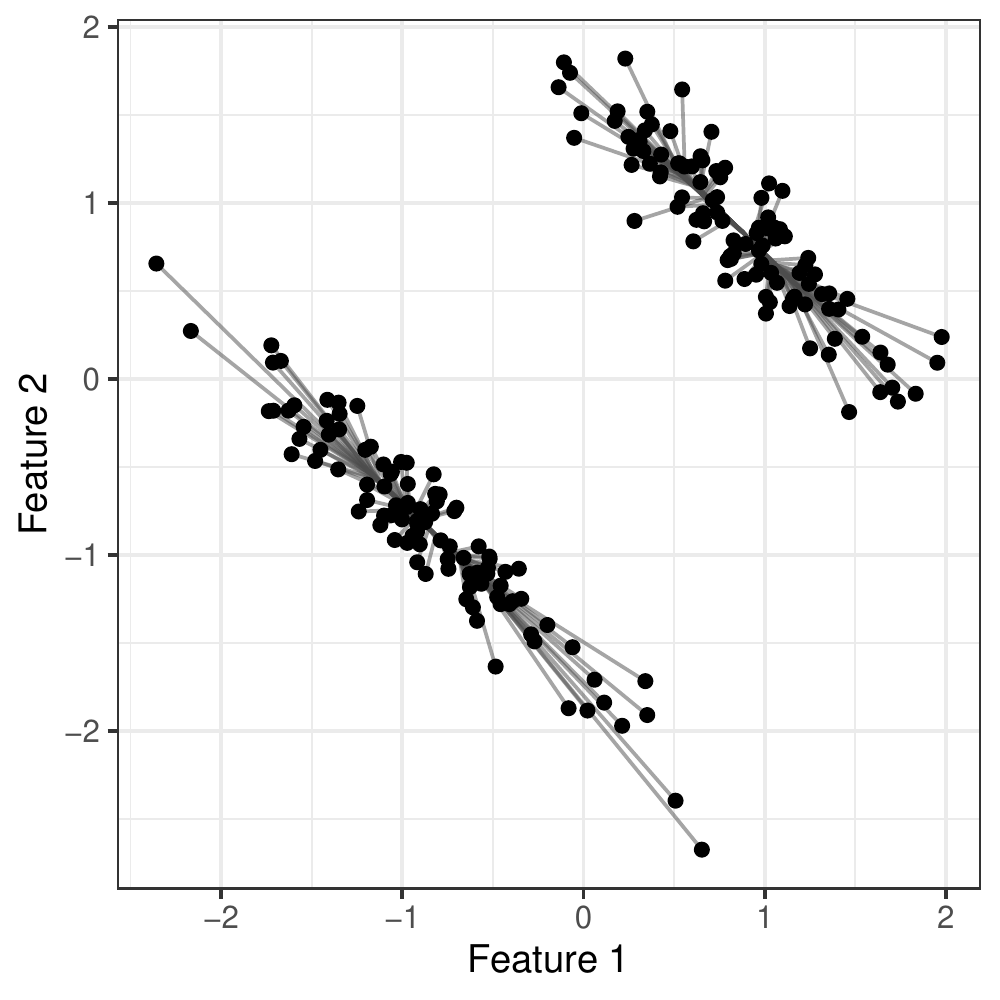}
        \caption{Solution paths of Biconvex Clustering}
        \label{my dendro}
    \end{subfigure}
    \caption{Performance of different peer algorithms on a toy dataset, showing the efficacy of biconvex clustering in a low signal-to-noise-ratio setting.}
    \label{motivation1}
\end{figure}

\paragraph{Motivating Example}
Consider a toy dataset simulated from two ground truth clusters each with $100$ associated points. The data are designed so that they can be discriminated along only the first two features, but we add twelve uninformative features  from a standard normal distribution that serve only to decrease the signal-to-noise ratio. We study the results of various clustering algorithms by visualizing their solutions projected onto the two relevant dimensions. The performances of different peer algorithms are shown in Figure~\ref{motivation1}. 

 \begin{figure}
 \centering
         \begin{subfigure}[t]{0.44\textwidth}
         \centering
        \includegraphics[height=0.5\textwidth,width=0.7\textwidth]{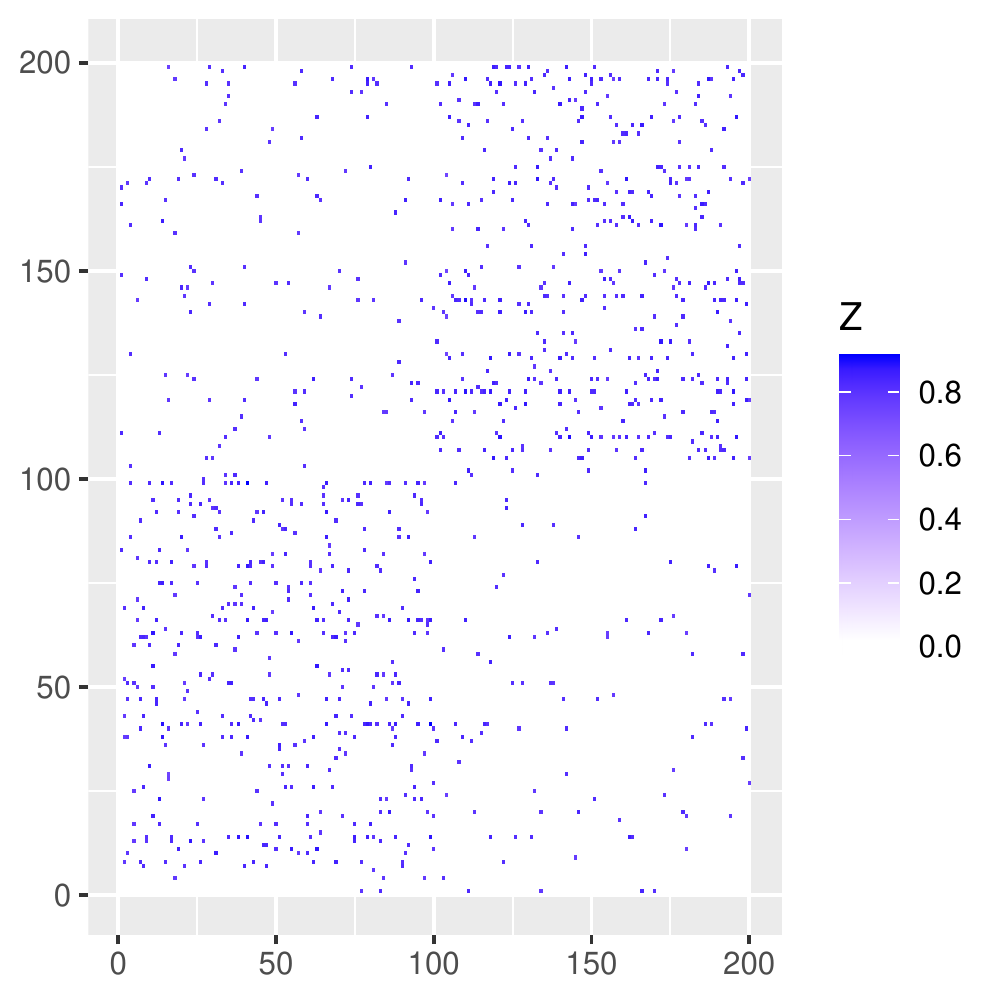}
            %\caption{Distance Matrix}
            \caption{Default Affinities (as in equation \eqref{eq:affinity})}

        \label{fig:g1_dist}
    \end{subfigure} 
    ~
        \begin{subfigure}[t]{0.44\textwidth}
         \centering
        \includegraphics[height=0.5\textwidth,width=0.7\textwidth]{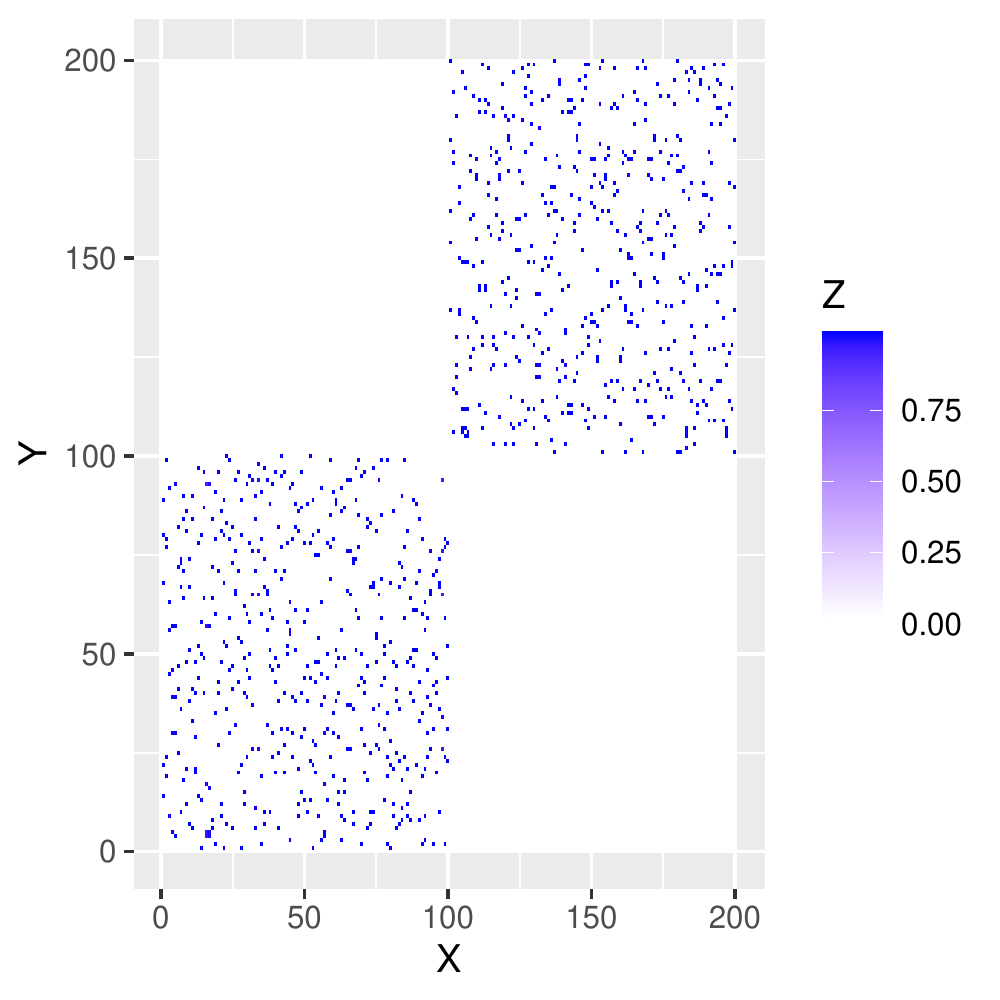}
    \caption{Learned by Biconvex clustering}
        \label{modified phi}
    \end{subfigure}
        \caption{Heatmap of affinities $\phi_{ij}$ show that default affinities can be misleading, while those learned by biconvex clustering reveal the cluster structure of the toy dataset.}
        \label{fig: eg2}
\end{figure}

Our method successfully recovers the ground truth (Figure \ref{fig: gt}), while prior work that seeks a sparse solution,  solves a convex relaxation, or both fail due to the large number of noise variables. 
We examine the performance of sparse convex clustering   \citep{wang2018sparse} in particular to further emphasize the effect of updating affinities by way of the learned feature weight vector.  
To understand why sparse convex clustering fails to recover the true cluster structure here despite attempting to perform feature selection as evident in Figure \ref{motiv:spcvx}, we examine a heatmap of the affinities $\phi_{ij}$ computed under Euclidean distance as defined in \eqref{eq:affinity}. Figure \ref{fig:g1_dist} shows that because of the large relative influence of noise features, pairwise distances computed in the original Euclidean space provide only limited information, reflected in the noisy pattern of $\phi_{ij}$ in the left panel. In contrast, updating $\phi_{ij}$ according to \eqref{eq:affinity} where distances are defined under the \textit{learned} feature space induced by $\bw$ reveals clear structure (Figure \ref{fig: eg2}). To further highlight this point, we may re-run sparse convex clustering, provided with the affinities learned by our algorithm as inputs. The method is now able to recover the ground truth, displayed in Figure \ref{scvx2}. Note that despite achieving a perfect clustering on this relatively simple synthetic data, the estimated centers are noticeably biased toward the origin. This behavior arises from applying shrinkage penalties directly to the centroids, and can negatively affect clustering accuracy in more realistic and challenging data settings. 

The rest of the paper is organized as follows. Section \ref{sec:mainmethods} introduces the biconvex clustering formulation, and establishes some intuition in relation to prior work. We next derive a coordinate descent algorithm with simple, closed-form updates toward solving the resulting optimization problem with respect to the centroids and feature weights. Notably, as the feature representation adapts under the learned feature weights, users have the option to update pairwise affinities $\phi_{ij}$ across iterations rather than rely on their \textit{a priori} initialization throughout the clustering task. Next, the properties of the proposed method are studied closely in Section \ref{sec:properties}, where we derive finite sample prediction bounds that imply consistency of the estimates. We also establish convergence guarantees and discuss the computational complexity and practical considerations of the algorithm.  These results are validated empirically in Section \ref{simulation}, where performance is thoroughly assessed by standard clustering metrics as well as in terms of feature selection over several simulation studies. We then apply biconvex clustering to analyze several case studies in Section \ref{real analysis}, including a high-dimensional movement data corpus and a DNA microarray study of human leukemia, followed by discussion.

\section{Biconvex Clustering and adaptive feature selection}\label{sec:mainmethods}
In this section, we present biconvex clustering and discuss its formulation. We then derive a block coordinate descent algorithm for efficiently minimizing its objective function. 
\subsection{Problem formulation and intuition}
\label{problem formualtion}
Let $\bX\in \mathbb{R}^{n \times p}$, where the row $\bx_{i \cdot }$ contains feature values of the $i^\text{th}$ data point.
Consider the objective function
\begin{equation}\label{eq:objective1} f(\bmu,\bw)=  \sum_{i=1}^n \| \bx_{i \cdot} - \bmu_{i \cdot } \|_{\bw}^2 + \gamma \sum_{i\neq j}^n \phi_{ij} \| \bmu_{i \cdot} - \bmu_{j \cdot}\|_2 \qquad \text{subject to} \quad w_l\geq 0, \sum_{l=1}^p w_l = 1,
\end{equation}

\noindent where the optimization variables $(\bw$, $\bmu)$ denote the vector of feature weights and an $n \times p$ matrix whose rows contain the centroids, respectively.  For a vector $\by \in \Real^p$, we define the norm induced by $\bw$ as \,$\| \by \|_{\bw}^2~:=~\sum_{l=1}^p (w_l^2 + \lambda w_l) y_l^2\,$. We see that the first component $\, \sum_{i=1}^n \| \bx_{i \cdot} - \bmu_{i \cdot} \|_{\bw}^2\,$ of \eqref{eq:objective1} assesses the fit between the centroids $\bmu$ and the data $\bX$ measured in a way that is weighed by $\bw$ \citep{huang2005automated}. 
Note that if we fix all $w_l \propto 1$ throughout, $\| \by \|_{\bw}$ becomes a scalar multiple of the Euclidean norm $\| \by \|_2$, and thus minimizing \eqref{eq:objective1} is equivalent to convex clustering as a special case. 
Like convex clustering, the procedure begins with $n$ centroids---one at each data point---which merge toward each other due to the term $\,\sum_{i<j}^n \| \bmu_{i \cdot} - \bmu_{j \cdot}\|_2 \,$. It is straightforward to show that the solution path is unique and depends continuously on $\gamma$ \citep{chi2015splitting}. 

This objective is convex in either of the optimization variables as the other is held fixed: that is, it is \textit{biconvex} in $(\bmu, \bw)$. To establish some intuition, we focus on the variable $\bmu$ and fix all $w_l= 1/p$ now for exposition.
Recall for sufficient $\gamma$, the fusion penalty encourages  centroids to merge so that only a subset of the rows of $\bmu$ are unique; denote this resulting set of centers as $\bC \in \Real^{c \times d}$. As a result of merging, the measure of fit term
\begin{equation}\label{eq:merge} \sum_{i=1}^n \| \bx_{i \cdot} - \bmu_{i \cdot} \|_2^2 \quad \text{ gradually tends to } \quad  \sum_{i=1}^n \min_j \| \bx_{i \cdot} - \bC_{j \cdot} \|_2^2.
\end{equation} The familiar quantity on the right hand side appears in classic formulations such as $k$-means, and is equivalent to the within-cluster variance \citep{lloyd1982least}. 
Now removing the restriction $w_l =1/p$, reveals how our formulation  improves upon existing continuous clustering methods. The weights $\bw$ enable us to learn which features are helpful for distinguishing these clusters: fixing $\bmu$ and expanding the measure of fit yields $$ \sum_{i=1}^n \sum_{l=1}^p (w_l^2  + \lambda w_l) (x_{il} - \mu_{il})^2.$$  
Under the simplex constraint $\sum_{l=1}^p w_l = 1$, the $\lambda w_l$ term promotes sparsity, while without the $w_l^2$ term, the linear cost in $w$ under the sum constraint would assign all weight to  the most important feature and result in a degenerate solution \citep{witten2010framework}. Thus the objective in $\bw$ entails feature weighing as well as selection. 

With $\bmu$ fixed, the terms $\sum_{i=1}^n (x_{il} - \mu_{il})^2 := u_l$ that had played the role of the objective when optimizing $\bmu$ now act as ``weights" themselves to inform the optimizatiton of $\bw$.  To again relate to the familiar within-cluster variance, we can rewrite 
\begin{equation}\label{eq:ul} 
\sum_{i=1}^n \min_j \| \bx_{i \cdot} - \bC_{j \cdot} \|_2^2 = \sum_{l=1}^p  \widetilde{u}_l, \qquad \text{where we define} \qquad \widetilde{u}_l = \sum_{i=1}^n \sum_{j=1}^c \indfun{i,j}(x_{i,l} - C_{j,l})^2;
\end{equation}
here $\indfun{i,j}$ is an indicator function of whether data point $i$ is closest (assigned) to center $j$. As the $\bmu_{i\cdot}$ terms coalesce via the fusion penalty and approach a set of unique centers $\bC$, our definition of $u_l$ coincides with this familiar interpretation when $w_l$ are equal. 
We see that $u_l$ can be interpreted as the contribution to within-cluster variance along dimension $l$ across clusters,  related to the dispersion measures used by \citet{friedman2004clustering}. 

In other words, a large $u_l$ indicates that feature $l$ is less useful toward discriminating clusters, and promotes setting $w_l$ to zero. Further note that upon making the change of variables $z_l = w_l/u_l$, the unit simplex constraint $\sum_{l=1}^p w_l = 1$ on $\bw$ can now be reinterpreted as a \textit{weighted simplex} in $\bz$. Intuitively, this means that distances to activating constraints are driven by the learned cluster information rather than static. This adaptive behavior expands upon prior work on sparse clustering \citep{witten2010framework, wang2018sparse} and imports intuition behind the celebrated adaptive lasso for regression \citep{zou2006adaptive}. Where these related ideas in regression entail two-stage estimation procedures, our framework allows for $\bw, \bmu$ to be learned \textit{jointly}.

Before proceeding with the optimization method, we make some remarks regarding fusion penalties of the form $\sum_{i<j}^n \| \bmu_{i \cdot} - \bmu_{j \cdot}\|_q$. Historically, the namesake \textit{fusion} is borrowed from an unpublished technical report \citep{land1997variable} which explores variable fusion under penalties on a vector $\bnu$ of the form \begin{equation}\label{eq:fusionexponent}
\|\nu_j - \nu_{j-1}\|_q^\alpha.
\end{equation} 
A prominent example is the fused lasso or total variation penalty $\sum_{j=2}^p \| \nu_j - \nu_{j-1}\|_1$ \citep{tibshirani2005sparsity} for seeking local constancy when $\bnu$ is a vector of ordered regression coefficients by promoting sparsity in $\bnu$ as well as its successive differences. 
 \citet{tibshirani2005sparsity} mention they do not consider other values of $\alpha$ in \eqref{eq:fusionexponent} because  piecewise constant coefficient solutions under $\alpha=1$ are desirable and easily interpretable in their context. 
 Other natural choices such as $\alpha=2$ have appeared much less frequently in the literature in contexts such as precision estimation \citep{hebiri2011smooth,price2015ridge,lam2016fused}. When exact sparsity is not the end goal, such alternate choices may produce the desired behavior yet yield more elegant solutions. We argue that this is the case in our present context. The following sections show that squaring the norms appearing in the fusion penalty will confer substantial computational advantages. 
 
 This slight ``ridge fusion" modification entails the resulting objective
\begin{equation}\label{eq:compactobjective} f(\bmu,\bw)=  \sum_{i=1}^n \| \bx_{i \cdot} - \bmu_{i \cdot} \|_{\bw}^2 + \gamma \sum_{i\neq j}^n \phi_{ij} \| \bmu_{i \cdot} - \bmu_{j \cdot}\|^2_2 \qquad \text{subject to} \quad w_l \geq 0, \, \sum_{l=1}^p w_l = 1 ,
\end{equation}
which now encourages centroids to merge toward  each other by way of penalizing the \textit{quadratic} variation between row pairs of $\bmu$ rather than the total variation distance. Doing so no longer results in exact merging--- subsets of rows now agglomerate closely rather than coalescing to a single point. Though the resulting cluster assignment under a na\"ive interpretation would be trivial---each point is assigned to its own centroid---we demonstrate how to obtain a nontrivial clustering from the solution to \eqref{eq:compactobjective} analogous to that under exact merging in Section \ref{complexity}. In particular, assigning clusters as so shares the same worst-case complexity as simply reading off the unique rows of the solution under \eqref{eq:objective1}.

\subsection{Optimization}\label{optimization}
We now derive closed form coordinate descent updates for the subproblems in $\bmu$ and $\bw$. We begin by rewriting objective \eqref{eq:compactobjective}, expanding and decomposing the first term into two components as written below (still subject to the simplex constraint on $\bw$):
\begin{equation}
    \label{objj}
    f(\bmu,\mathbf{w})=\sum_{i=1}^n\sum_{l=1}^pw_l^2 (x_{il}-\mu_{il})^2+\gamma \sum_{i,j=1;i\neq j}^n\sum_{l=1}^p \phi_{ij} (\mu_{il}-\mu_{jl})^2+\lambda \sum_{l=1}^p w_l \sum_{i=1}^n (x_{il}-\mu_{il})^2.
\end{equation}

The block coordinate descent updates for minimizing the objective \eqref{objj} are given by the solutions to the following two subproblems:
\begin{itemize}
    \item {\it Problem} $P_1$: Fix $\bw=\bw_0$, minimize $f(\bmu,\bw_0)$ w.r.t. $\bmu$.
    \item {\it Problem} $P_2$: Fix $\bmu=\bmu_0$, minimize $f(\bmu_0,\bw)$ w.r.t. $\bw$, subject to $\sum_{l=1}^p w_l = 1; w_l \geq 0$.
\end{itemize}
Let us solve {\it Problem} $P_1$. First, observe that if $w_l=0$, one can simply choose $\mu_{il}=0$ for all $i \in \{1,\dots,n\}$ in order to decrease the value of the objective function. Thus take $w_l>0$; differentiating the objective function with respect to $\mu_{il}$, we obtain  
\begin{equation}
\label{eq:p1sol}
\mu_{il}=\frac{\gamma \sum_{i \neq j}\phi_{ij} \mu_{jl}+\gamma \sum_{i \neq j}\phi_{ji} \mu_{jl}+w_l^2 x_{il}+\lambda w_l x_{il}}{\gamma \sum_{i \neq j}\phi_{ij} +\gamma \sum_{i \neq j}\phi_{ji} +w_l^2 +\lambda w_l}.
\end{equation}
Fixing $\bmu$, the minimization of $\bw$ is summarized below, with more details of its derivation in Appendix A.
\begin{thm}
The solution to {\it Problem} $P_2$ is given by
\begin{equation}
\label{eq:p2sol}
w_l^*=%\bigg(
\frac{1}{2}S\bigg(\frac{\alpha^*}{\sum_{i=1}^n(x_{il}-\mu_{il})^2},\lambda\bigg)%\bigg)^\frac{1}{\beta-1},$$
\end{equation}
where for any $y \ge 0$, and $x \in \Real$, $S(x,y)$ denotes the soft thresholding function 
%defined 
%\[S(x,y)=\begin{cases}
%x-y & \text{ if } x \ge y\\
%x+y & \text{ if } x \le -y\\
%0 & \text{ Otherwise.}
%\end{cases}\] 
and  $\alpha^*$ satisfies the equation $\sum_{l=1}^p\frac{1}{2}S\bigg(\frac{\alpha}{\sum_{i=1}^n(x_{il}-\mu_{il})^2},\lambda\bigg)=1$.%\bigg)^\frac{1}{\beta-1}=1$
\end{thm}
\noindent In particular, solutions to each of the subproblems are parameter-separated; that is, all computations can be carried out component-wise, and the univariate problems can be executed in parallel. 

\begin{algorithm}
\caption{Biconvex Clustering algorithm (BCC)}
 \label{algo}
 \begin{algorithmic}
 \State \textbf{Input}: $\mathbf{X}\in \mathbb{R}^{n \times p}$, $\lambda>0$, $\gamma>0$
 \qquad  \textbf{Output}: $\bmu$\\
 \textbf{Step 0.}
 Initialize $\bmu=\bX$,  $w_l = 1/p, \, l = 1, \ldots, p$, and \\
 (optional) %$\phi_{ij}$,  
$\phi_{ij}=\exp\{-\|\bx_i-\bx_j\|_2^2\} \indfun{ j \in  k\text{-NN of }i \text{ under }\| \cdot \|_2 } \quad i=1,\ldots,n \, ; l = 1 ,\ldots p$. 
\Repeat
 \State \textbf{Step 1.}
 Update $\bmu$ %(coordinate-separable, can be parallelized) 
 by $\, \, \displaystyle \mu_{il} \leftarrow \frac{\gamma \sum_{i \neq j}\phi_{ij} \mu_{jl}+\gamma \sum_{i \neq j}\phi_{ji} \mu_{jl}+w_l^2 x_{il}+\lambda w_l x_{il}}{\gamma \sum_{i \neq j}\phi_{ij} +\gamma \sum_{i \neq j}\phi_{ji} +w_l^2 +\lambda w_l}.$
 
 \State \textbf{Step 2.}
 Update $\alpha^\ast$ so that  $\displaystyle \sum_{l=1}^pS\big(\frac{\alpha}{\sum_{i=1}^n(x_{il}-\mu_{il})^2},\lambda\big)=2$ (i.e. via bisection). \\
 \State \textbf{Step 3.}
 Update $\bw$ % (coordinate-separable, can be parallelized) 
 by
 $ \displaystyle \, \, w_l\leftarrow \frac{1}{2}S\bigg(\frac{\alpha^\ast}{\sum_{i=1}^n(x_{il}-\mu_{il})^2},\lambda\bigg).$%\bigg).^\frac{1}{2-1}.$$
 \vspace{8pt}
 \State \textbf{Step 4.} (Optional) Update $\displaystyle \phi_{ij}=\exp\{-\|\bx_i-\bx_j\|_{\bw}^2/p\} \indfun{ j \in  k\text{-NN of }i \text{ under }\| \cdot \|_{\bw} }$ %\quad i=1,\ldots,n \, ; l = 1 ,\ldots p. \]
 
\Until{ convergence criterion based on objective \eqref{objj} is reached}
%Iterate Steps 1, 2, 3 and 4 until convergence.
\end{algorithmic}
\end{algorithm}

Before analyzing the properties and complexity of the proposed method, we pause to discuss notable differences from prior work. Recall the sparse convex clustering   objective \eqref{wang} addresses high-dimensionality through penalizing the columns of the centroid matrix $\bmu$ directly. This approach is intuitive and nicely preserves convexity in the overall objective. However, the formulation can introduce significant shrinkage to the origin, an effect that may lead to bias as well as spurious selection. Computationally, the solution method in sparse convex clustering modifies the same two approaches outlined in \cite{chi2015splitting}, namely sparse variants of the alternating minimization (AMA) and alternating directions (ADMM) methods. Computation becomes more complicated, however:  one step within the alternating directions method requires an additional iterative algorithm of fitting $p$ group lasso regressions in the simplest case when $q=2$. Perhaps more troubling is that not only do the two implementations S-AMA and S-ADMM differ in speed, but may lead to quite different clustering solutions \citep{wang2018sparse}. The stability properties that convex formulations originally aimed to provide thus do not carry over. In contrast, our biconvex formulation can be solved via alternating between simple solutions to each subproblem given by \eqref{eq:p1sol} and \eqref{eq:p2sol}. A return to form reflecting the structure of classic methods such as Lloyd's algorithm, the proposed method marks a departure from the majority of existing work on convex clustering and its variants, which rely on variable splitting that entail additional dual variables and step-size selection.
This affords us a transparent and efficient block coordinate descent method, summarized in  Algorithm \ref{algo}. We establish convergence guarantees, finite sample bounds that imply consistency, and desirable computational properties in the following section.

\section{Convergence and statistical properties}\label{sec:properties}
This section establishes the theoretical properties of the (global) optimal solutions of the proposed objective function. We also analyze computational complexity and discuss convergence properties of our method.

\subsection{Finite sample bounds and prediction consistency}\label{finite sample bounds}
We begin our statistical analysis of biconvex clustering by providing finite sample error bounds on the prediction error. In particular, these bounds provide sufficient conditions for consistency of the centroid and weight estimates. Unlike prior work by \cite{tan2015statistical,wang2018sparse,chi2018provable}, our theoretical analysis must account for an additional feature weight term, which renders the widely used Hanson-Wright argument ineffective. We resort to more recent uniform versions of Hanson-Wright inequalities \citep{10.1214/20-EJP422} to surmount this difficulty. \par
Subject to the simplex constraint on $\bw$, recall the biconvex clustering objective
  \begin{equation}
  \label{opt1}
  %\min_{\bmu \in \mathbb{R}^{n \times p}}
  \min_{\bmu, \bw}
  \bigg\{\sum_{i=1}^n \|\mathbf{x}_{i \cdot}-\bmu_{i \cdot}\|_{\bw}^2+\gamma \sum_{(i,j) \in \mathcal{E}} \|\bmu_{i \cdot}-\bmu_{j \cdot}\|^2_2\bigg\}.
  \end{equation} 
%  where, $\| \by \|_{\bw}^2 : = \sum_{j=1}^p (w_j^\beta + \lambda |w_j|) y_j^2$.\par
Here $\mathcal{E} \subseteq \{(i,j): i,j \in \{1,\dots,n\}\} $ is an index set.  Throughout this section, we will assume that $\phi_{ij}$ are kept fixed and for clarity of exposition will proceed after vectorizing the problem setup. To this end let $\bx=vec(\mathbf{X})$ and $\bu=vec(\bmu)$,  where $vec(\cdot)$ denotes the function that flattens a matrix by appending its columns together. Now $\bx,\bu \in \mathbb{R}^{np}$, with $\bx_{(i-1)p+j}=X_{ij}$ and $\bu_{(i-1)p+j}=\mu_{ij}$. Similarly, let $\bW=diag(w_1^2 + \lambda w_1,\dots, w_p^2 + \lambda w_p)\otimes \bI_n$, where $\otimes$ denotes the Kronecker product. Let $\mathcal{W}=\{\bW=diag(w_1^2 + \lambda w_1,\dots, w_p^2 + \lambda w_p)\otimes \bI_n: w_1,\dots,w_p \ge 0 \text{ and } \bw^\top \mathbf{1}_p=1  \}$. To deal with the penalty term, let $\bD \in \mathbb{R}^{[p n(n-1)]\times np}$ be such that $\bD_{\mathcal{C}(i,j)}\bu=\bmu_{i \cdot}-\bmu_{j \cdot}$, where $\mathcal{C}(i,j)$ is an index set: then the optimization problem  (\ref{opt1}) can now be rewritten as
 \begin{equation}
 \label{opt2}
 \min_{\bu \in \mathbb{R}^{np}, \bW \in \mathcal{W}} \bigg\{ (\bx - \bu)^\top \bW (\bx - \bu) + \gamma \sum_{ (i,j) \in \mathcal{E}} \|\bD_{\mathcal{C}(i,j)} \bu \|^2_2 \bigg\}.
 \end{equation}
 We will assume the model $\bx=\bu+\bepsilon$, where $\bepsilon \in \mathbb{R}^{np}$ is a vector of independent noise variables and $\mathbb{E}(\bepsilon)=\mathbf{0}$.  This model is quite standard in analyzing the large sample behaviors of convex clustering methods \citep{tan2015statistical,wang2018sparse}. For all practical purposes, one may assume that the error terms are bounded almost surely, i.e. for some $M \ge 0$, $|\epsilon_i| \le M$, for all $i=1,\dots,np$. For notational simplicity, we write, $\vvvert \by \vvvert_{\bA}^2 = \by^\top \bA \by$, for any positive semi-definite matrix $\bA$. The goal of this analysis is to find probabilistic bounds on $\frac{1}{2np} \vvvert\hat{\bu}-\bu\vvvert_{\widehat{\bW}}^2$, where, $\hat{\bu}$ and $\widehat{\bW}$ are the solutions to the minimization problem \eqref{opt2}. Because the additional term $\widehat{\bW}$ depends on $\bepsilon$, the classical Hanson-Wright argument used in \citep{tan2015statistical,wang2018sparse} is not enough. One needs to appeal to uniform Hanson-Wright inequalities \citep{10.1214/20-EJP422}, which additionally assume only that errors are bounded. The following theorem provides a finite sample bound on the prediction error. For notational simplicity, let  $\gamma'=\frac{\gamma}{np}$.
 \begin{thm}\label{main thm}
 	Suppose $\bx=\bu+\bepsilon$, where,  $\bepsilon \in \mathbb{R}^{np}$ is a vector of independent bounded random variables, with mean $0$ and $|\epsilon_i| \le M$, for all $i=1,\dots,np$. Suppose that $\hat{\bu}$ and $\hat{W}$ are obtained form minimizing \eqref{opt2}, then if $\gamma' \ge  \frac{(1+\lambda)M}{\sqrt{n^3p}}$
 	\begin{align}
 	\frac{1}{2np} \vvvert\hat{\bu}-\bu\vvvert_{\widehat{\bW}}^2\leq & M^2 (1+\lambda) \left[\frac{1}{\sqrt{n}} + c \sqrt{\frac{\log(1/\delta)}{np}} + c \frac{\log(1/\delta)}{np}\right]+\frac{\gamma' |\mathcal{E}|}{4} + \gamma' \sum_{(i,j) \in \mathcal{E}}\|\bD_{\mathcal{C}(i,j)} \bu\|_2 \nonumber \\
 	& +\gamma' \sum_{(i,j) \in \mathcal{E}}\|\bD_{\mathcal{C}(i,j)} \bu\|^2    \label{oracle}
 	\end{align}
 	holds with probability at least  $1-\delta$.
 \end{thm} 
The complete proof is provided in Appendix B. 
We pause to examine and discuss the result: Theorem~\ref{main thm} shows that the average prediction error is bounded by oracle terms 
 $$\frac{\gamma^\prime |\mathcal{E}|}{4} + \gamma' \sum_{(i,j) \in \mathcal{E}}\|\bD_{\mathcal{C}(i,j)} \bu\|_2+\gamma' \sum_{(i,j) \in \mathcal{E}}\|\bD_{\mathcal{C}(i,j)} \bu\|^2.$$ 
 Note the first summand in the right hand side of \eqref{oracle} tends to $0$ as $n,p \to \infty$. Therefore, the estimator is consistent whenever $\frac{\gamma^\prime|\mathcal{E}|}{4} + \gamma^\prime \sum_{(i,j) \in \mathcal{E}}\|\bD_{\mathcal{C}(i,j)} \bu\|_2+\gamma' \sum_{(i,j) \in \mathcal{E}}\|\bD_{\mathcal{C}(i,j)} \bu\|^2 = o(1)$. Regarding when these oracle terms are $o(1)$: suppose $p>n$ and there exists a fixed number of informative features. In this setting, $\|\bD_{\mathcal{C}(i,j)} \bu\|_2=\mathcal{O}(1)$, and so $\sum_{(i,j) \in \mathcal{E}}\|\bD_{\mathcal{C}(i,j)} \bu\|_2$ as well as $\sum_{(i,j) \in \mathcal{E}}\|\bD_{\mathcal{C}(i,j)} \bu\|^2_2$ are $\mathcal{O}(|\mathcal{E}|)$. This in turn requires that $\gamma^\prime \mathcal{O}(|\mathcal{E}|) = o(1)$, %, i.e. $\gamma^\prime = o(1/|\mathcal{E}|)$. 
 which is satisfied for any fixd $\lambda$ if we take $\gamma^\prime = \mathcal{O}(n^{-3/2}p^{-1/2})$. Thus, for any given $\lambda$, we see that a sufficient condition for consistency is to have 
 \begin{equation}\label{condition}
     \sqrt{\frac{1}{n^3p}}|\mathcal{E}|=o(1).
 \end{equation} %for the condition to hold, which we may expect in sufficiently high-dimensional settings. % a reasonable assumption that applies in high-dimensional settings. 
 Noting that  $|\mathcal{E}| \le n^2$, for condition \eqref{condition} to hold, it is sufficient to have, $\sqrt{\frac{n}{p}} =o(1) \iff n=o(p)$, a reasonable assumption for high-dimensional settings. Next, note that in practice we take $|\mathcal{E}| = \mathcal{O}(n)$---for instance if we take $k$ nearest neighbors, $|\mathcal{E}| \le kn$. In this case, condition \eqref{condition} holds whenever $\frac{1}{\sqrt{np}}=o(1)$, which applies for not only high-dimensional data but also classical low-dimensional datasets. We thus arrive at the following two corollaries, whose complete proofs are available in the Appendix. We write $\xi_n \xrightarrow{\mathbb{P}}\xi$ if $\xi_n$ converges in probability to $\xi$, and write $\xi_n = \mathcal{O}_p(a_n)$ if the family $\{\frac{\xi_n}{a_n}\}_{n \in \mathbb{N}}$ is tight.
 \begin{cor}
Suppose $\|\bD_{\mathcal{C}(i,j)}\bu\| \le C$, for all $1 \le i,j \le n$, for some constant $C$ and $\gamma^\prime = \mathcal{O}(n^{-3/2}p^{-1/2})$. If $n=o(p)$, then $\frac{1}{2np} \vvvert\hat{\bu}-\bu\vvvert_{\widehat{\bW}}^2 \xrightarrow{\mathbb{P}}0$, as, $n,p \to \infty$.  
 \end{cor}
 \begin{cor}\label{cor2}
Suppose $\|\bD_{\mathcal{C}(i,j)}\bu\| \le C$,  for all $1 \le i,j \le n$,  for some constant $C$ and $\gamma^\prime = \mathcal{O}(n^{-3/2}p^{-1/2})$. If $|\mathcal{E}|=\mathcal{O}(n)$, then $\frac{1}{2np} \vvvert\hat{\bu}-\bu\vvvert_{\widehat{\bW}}^2 =\mathcal{O}_p\left(n^{-1/2}\right)$.    
 \end{cor}
In particular, Corollary \ref{cor2} reveals that $\hat{\bu}$ is $\sqrt{n}$-consistent for $\bu$ in the $\vvvert\cdot\vvvert_{\widehat{\bW}}$ norm.
\subsection{Convergence and complexity analysis}
Here we examine the algorithmic aspects of biconvex clustering. The following convergence result is proven straightforwardly in the Appendix. %For the sake of simplicity, through out this section, we will assume that $\phi_{ij}$'s are kept fixed and not updated through out the algorithm.  \jxadd{We should be clear about which parts of the theory are assuming that the $\phi_{ij}$ are fixed}
\begin{thm}
Algorithm \ref{algo} converges to a coordinate-wise minimum of \eqref{eq:compactobjective} in finite steps.
\end{thm}
% \begin{proof}
% Let $g_t$ be the value of the objective function \eqref{eq:compactobjective} at iteration $t$. From the update steps of Algorithm \ref{algo}, it is clear that $g_{t+1} \le g_t$, for all $t \in \mathbb{N}$. Thus the sequence $\{g_t\}_{t=1}^\infty$ is forms a decreasing sequence, and moreover $g_t \ge 0, \, \forall t\in \mathbb{N}$. Thus $\{g_t\}_{t=1}^\infty$ converges by the monotone convergence theorem. Thus, for all $\epsilon>0$, there exists $T \in \mathbb{N}$, such that $g_T -g_{T+1} \le \epsilon$, so that an absolute or relative convergence criterion based on \eqref{eq:compactobjective} is satisfied in finite iterations . %for all $t \ge T$. The 
% Because the objective is biconvex, applying Theorem 5.1 of \cite{tseng2001convergence} immediately implies that the limit point is a coordinate-wise minimum of \eqref{eq:compactobjective}.
%\end{proof}

\paragraph{Complexity}
\label{complexity}
 Evaluating $\mu_{il}$ in step 1 takes $\mathcal{O}(k)$ steps where $k$ is the number of nonzero summands, i.e. number of nonzero $\phi_{ij}$ coefficients. Since there are $n \times p$ many computations, it takes $\mathcal{O}(npk)$ time to complete the centroid updates, which can be executed in parallel. Computing $\sum_{i=1}^n(x_{il}-\mu_{il})^2$ requires $\mathcal{O}(n)$ time for each $l=1,\dots,p$, contributing $\mathcal{O}(np)$ total, while 
solving for $\alpha^\ast$ via bisection is $\mathcal{O}(p)$, as is the soft-threshold update of $\bw$. Therefore, the  per-iteration complexity of the Algorithm \ref{algo} is $\mathcal{O}(npk+np+p)=\mathcal{O}(npk).$ Note for vanilla convex clustering, the computational complexity is $\mathcal{O}(n^2p)$ for ADMM and $\mathcal{O}(npk)$ for AMA, where $k$ is also quadratic in $n$ in the worst case. Thus we are able to either match or improve upon its complexity despite handling the additional task of variable weighing and selection simultaneously.

\paragraph{Cluster assignments from $\bmu$} 
Convex clustering promotes centroids to merge at convergence, but one must ``read off" the unique rows of a matrix $\bV$ containing all difference pairs appearing in the penalty term to make cluster assignments. Doing so calls breadth-first search (BFS) to identify connected components of the graph induced by $\bV$ \citep{chi2015splitting}, which has linear complexity in the cardinality of the edge set plus the vertex set $\mathcal{O}(| \mathcal{E}| + |\mathcal{V}|) = \mathcal{O}(n^2)$.
Recall that squaring the penalty term in our approach does not lead to exact coalescence of centroids; in place of BFS, we instead determine cluster assignments from $\bmu$  by a dynamic tree cut \citep{langfelder2008defining} on the resulting dendrogram. The bottleneck computation involves creating a distance matrix, which also has $\mathcal{O}(n^2)$ complexity but is trivially parallelizable. 
Even though our algorithm does not rely on exact fusion of centroids, the assignment step in our algorithm does not entail higher computational cost. The efficacy of this approach is demonstrated in Section \ref{simulation}.

\paragraph{Nearest neighbor affinities}
When users apply the optional Step 4, Algorithm \ref{algo}  allows for affinities $\phi_{ij}$ to adapt with the learned feature space. Existing approaches to convex clustering have been criticized for their strong dependence on choice of $\phi$, which remain fixed throughout the algorithm and strongly influence performance. %The graph induced by $\phi$ not only affects computational complexity, but significantly determines the shape of the solution path. 
Under dense choice of $\phi$, clusters may not merge until the trivial solution at the origin, and convex clustering may perform worse than standard $k$-means \citep{tan2015statistical}. Further computational considerations arise, for instance in choosing step-sizes within splitting methods based on the eigenvalues of the Laplacian associated with the edge set $\mathcal{E}$ induced by $\phi$. Our algorithm can take advantage of the same recommended heuristics for initializing $\phi$, but allow for these \textit{a priori} choices to correct throughout the task %in light of learned cluster information 
by recomputing all distances under the induced norm $\|\cdot\|_{\bw}$ as the estimate for $\bw$ improves.  %$\lVert \cdot \rVert_{\bw}$.
%This is described in the optional Step 4 of Algorithm \ref{algo}. 
The advantages of adaptivity, which occurs in Step 4 of Algorithm \ref{algo}, are seen empirically in the following section. 

%Both the cluster assignment and adaptivity of affinities described here are empirically investigated in the following section. \par

\section{Simulation studies}\label{simulation}
In this section, we assess the clustering and feature selection accuracy of our method, and compare to peer algorithms. %we demonstrate how to assign cluster labels from the solution to \eqref{eq:compactobjective} in practice. 
We begin by highlighting the reduced reliance on sensitivity to heuristic specification of $\phi_{ij}$,  revisiting the simple motivating example from Section \ref{sec:intro} and examine the dendrograms produced under various settings.  
\begin{figure*}[h]
    \begin{subfigure}[t]{0.234\textwidth}
    \centering
        \includegraphics[height=\textwidth,width=\textwidth]{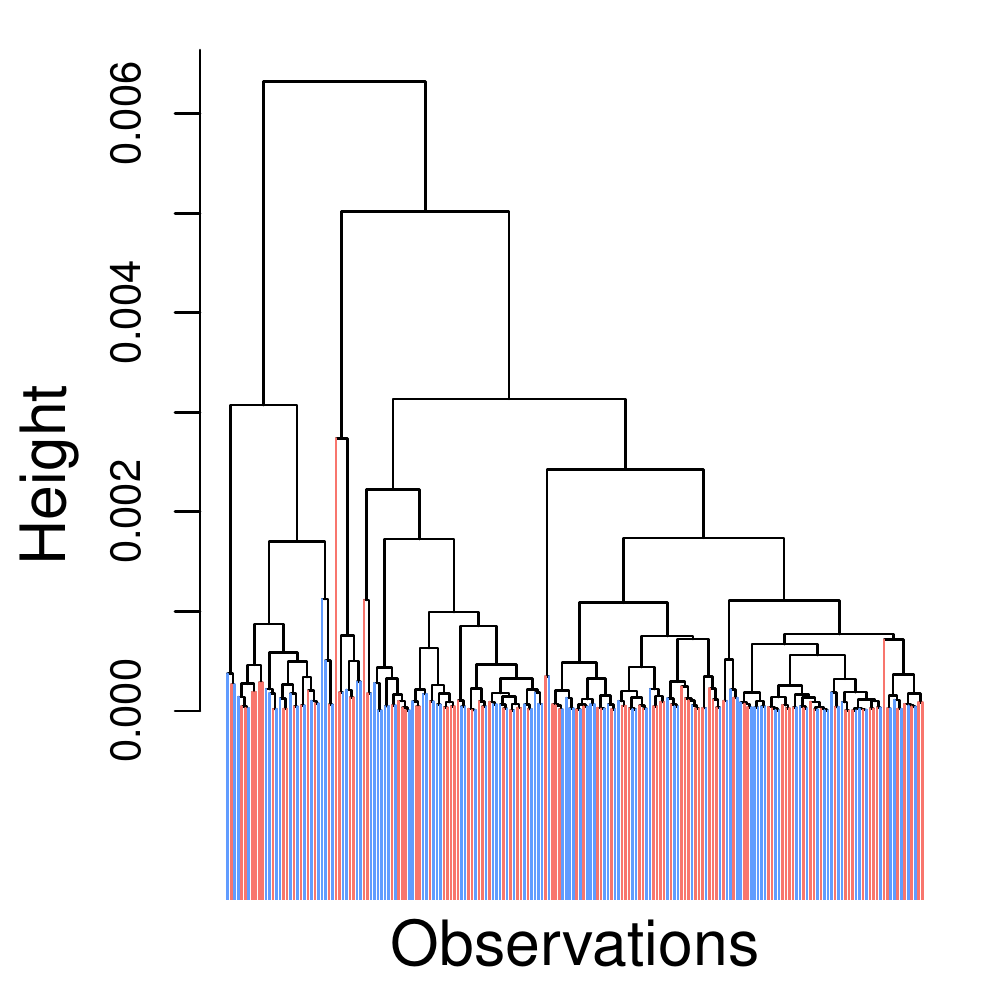}
        \caption{ Sparse Hierarchical Clustering}
       % \label{scvx modified}
    \end{subfigure}
    ~
    \begin{subfigure}[t]{0.234\textwidth}
    \centering
        \includegraphics[width=\textwidth,height=\textwidth]{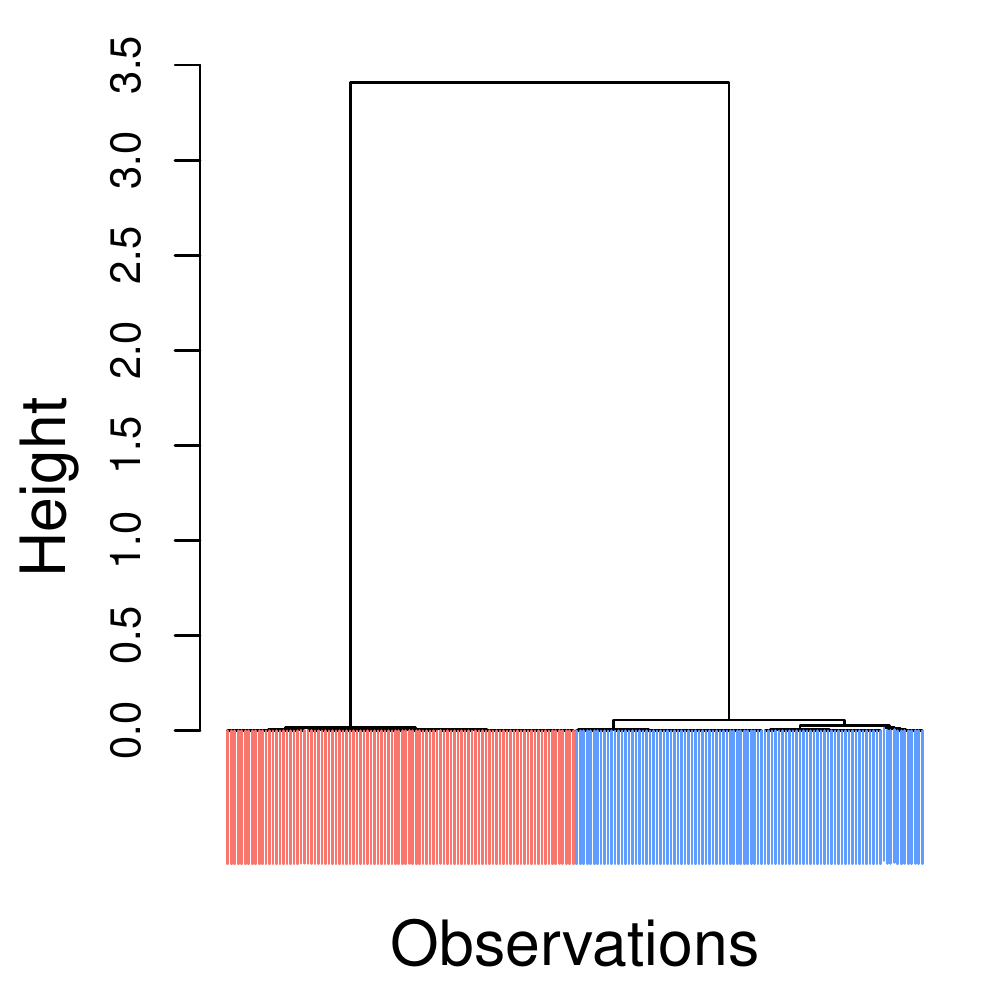}
        \caption{Biconvex Clustering}
        %\label{my dendro}
    \end{subfigure}
    ~  
    \begin{subfigure}[t]{0.234\textwidth}
    \centering
        \includegraphics[height=\textwidth,width=\textwidth]{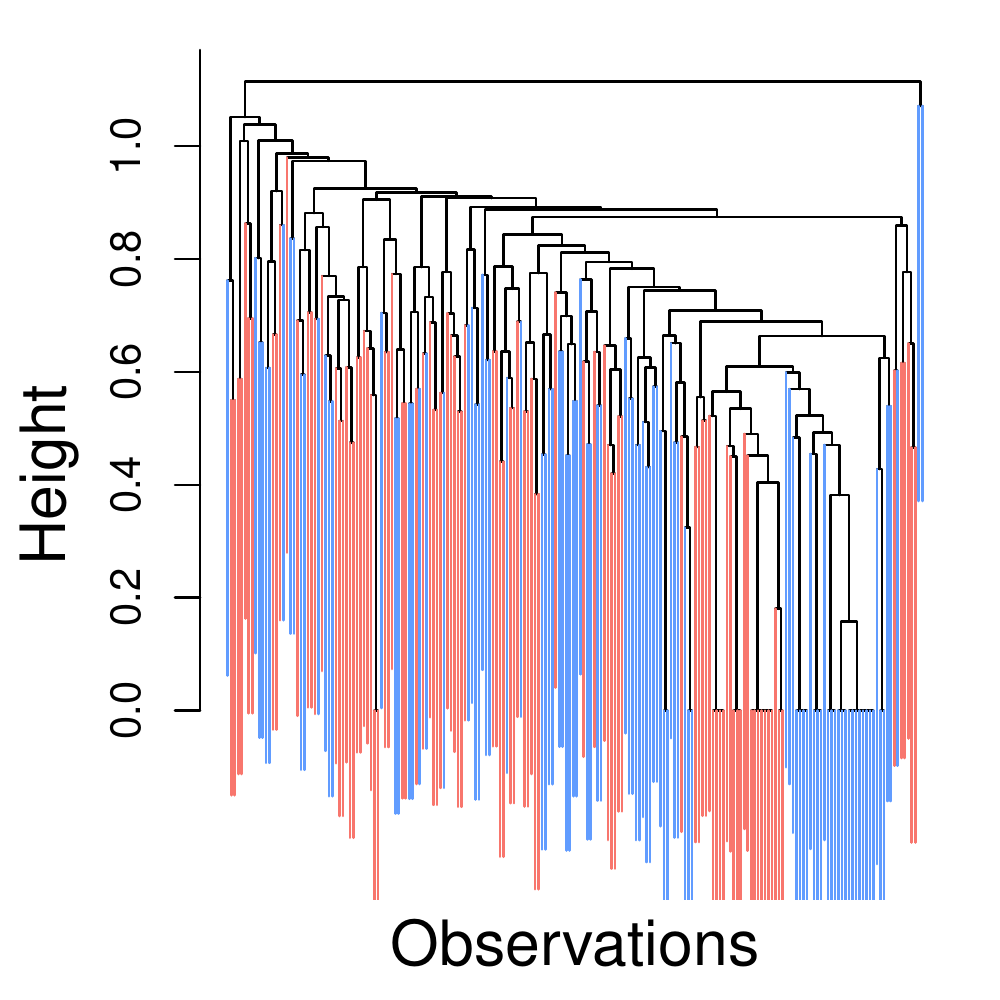}
        \caption{Sparse Convex Clustering}
      %  \label{scvx dendro}
    \end{subfigure}
    ~
     \begin{subfigure}[t]{0.234\textwidth}
    \centering
        \includegraphics[height=\textwidth,width=\textwidth]{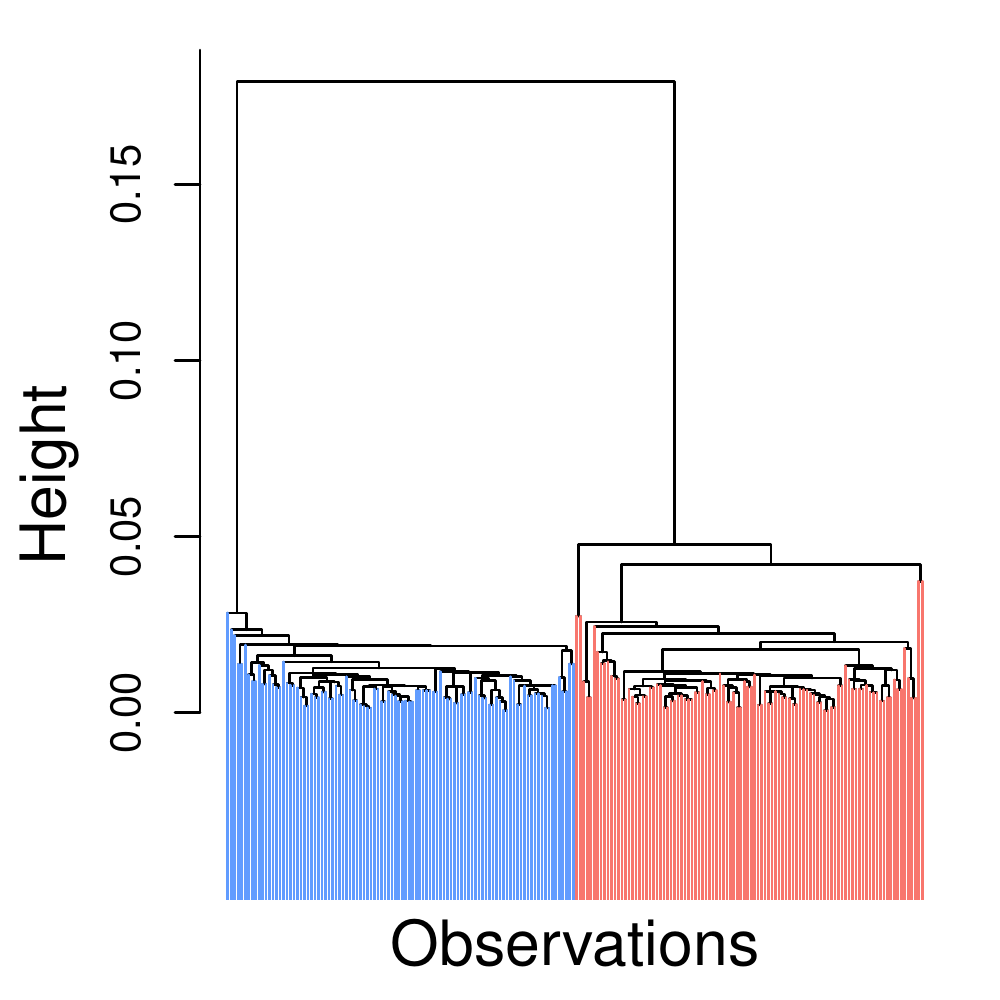}
        \caption{Sparse Convex with learned affinities}
        \label{scvx modified dendro}
    \end{subfigure}
    \caption{Dendrogams for motivating example (Sec. \ref{sec:intro}) show that identifying the two true clusters via a tree cut is straightforward for biconvex clustering.}
    \label{fig:fig3}
\end{figure*}
Figure \ref{fig:fig3} displays the resulting dendrograms obtained under peer algorithms from the motivating example, showing that the height of the branches corresponding to  correct separation into the two true clusters is markedly larger under our approach than the competitors. This clear separation suggests that any reasonable choice of dynamic tree cut provides a stable way to convert the solution $\hat\bmu$ to accurate cluster assignments. We do not observe this to be the case when examining dendrograms of competing sparse clustering methods. When initializing the competing sparse convex clustering approach using our learned affinities $\phi_{ij}$ induced by $\hat\bw$ (discussed further below), the dendrogram structure it attains becomes largely rectified, though still visibly more sensitive to choice of tree cut than our method. Under both settings for choice of $\phi_{ij}$, sparse convex clustering did not lead to exact merging despite the $\ell_1$ penalty on columns (and hence giving incentive to eliminate the noisy features); thus its solutions are also visualized as dendrograms rather than applying breadth-first search to identify unique rows of the centroid matrix. 
\begin{figure}[b]
    \centering
    \begin{subfigure}[t]{0.21\textwidth}
        \includegraphics[height=\textwidth,width=\textwidth]{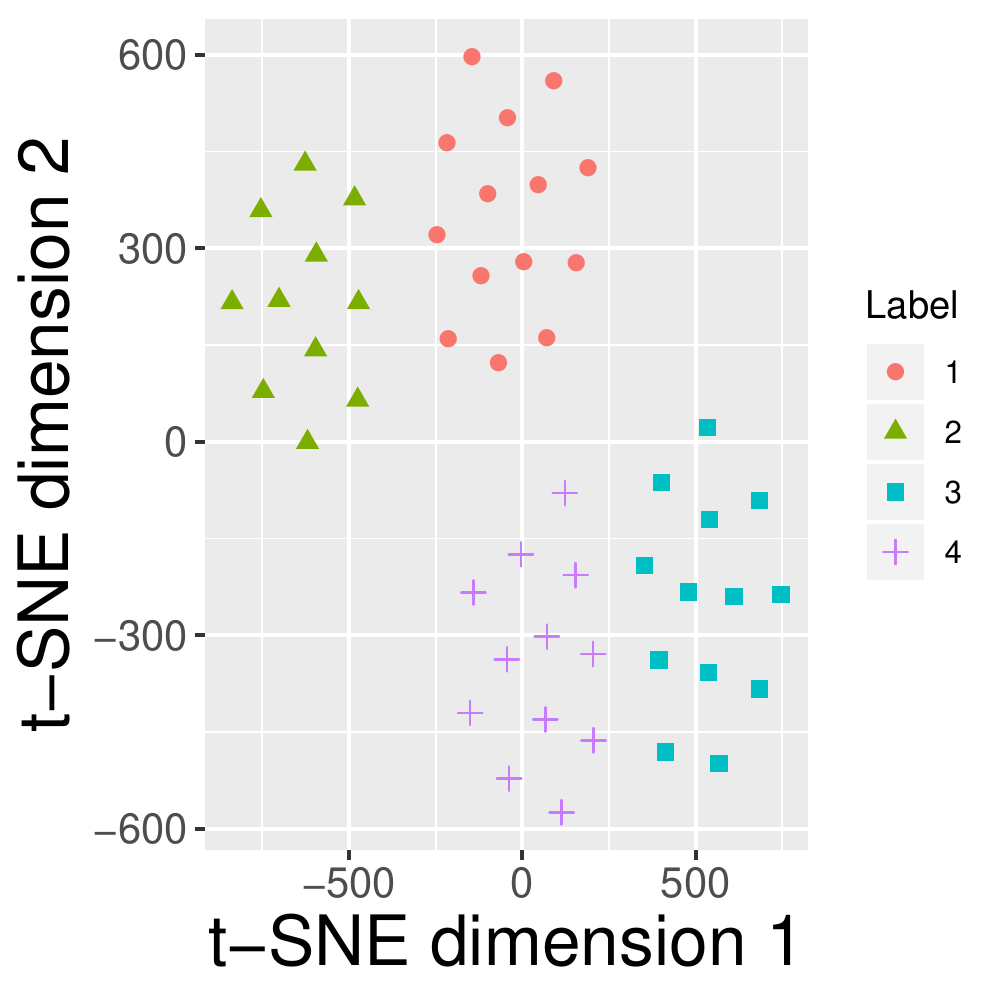}
        \caption{t-SNE plot}
        \label{fig:com:tsne}
    \end{subfigure}
    ~
    \begin{subfigure}[t]{0.24\textwidth}
        \includegraphics[height=\textwidth,width=\textwidth]{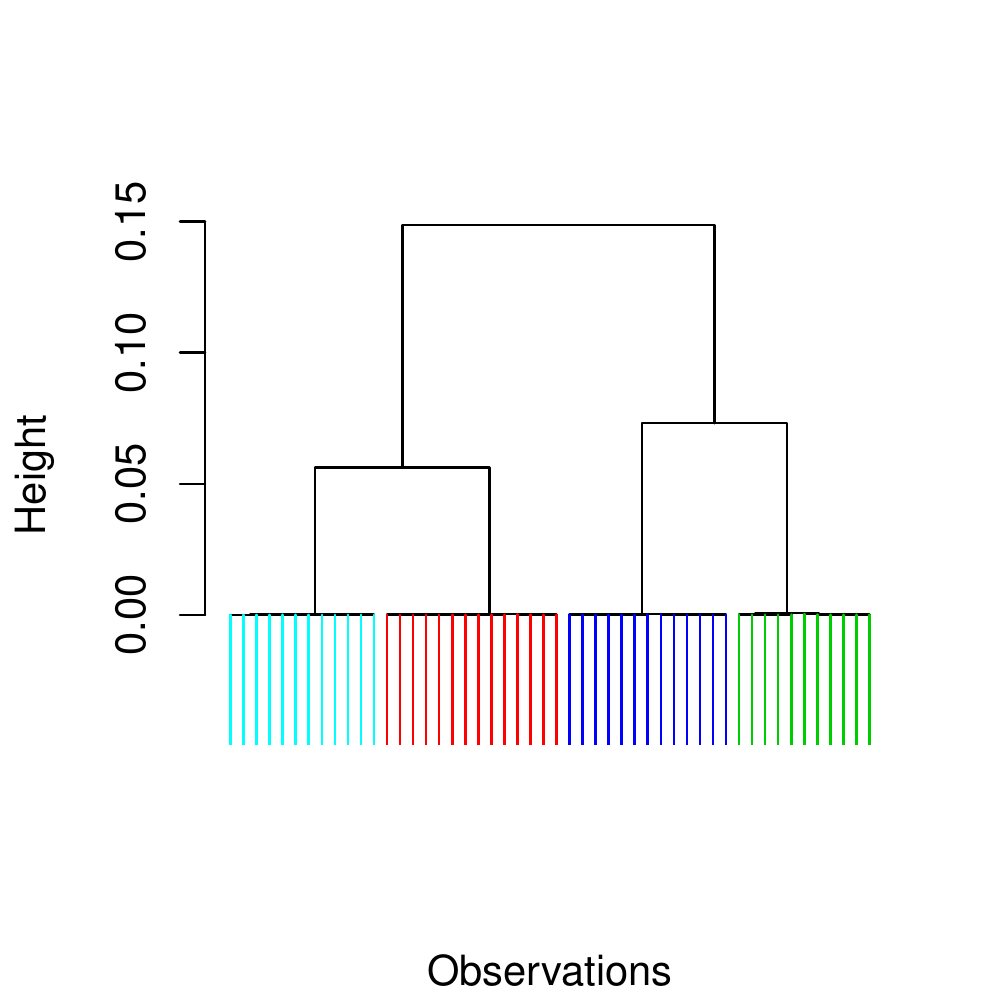}
        \caption{$k$-NN updates}
    \end{subfigure}
        ~
    \begin{subfigure}[t]{0.24\textwidth}
        \includegraphics[height=\textwidth,width=\textwidth]{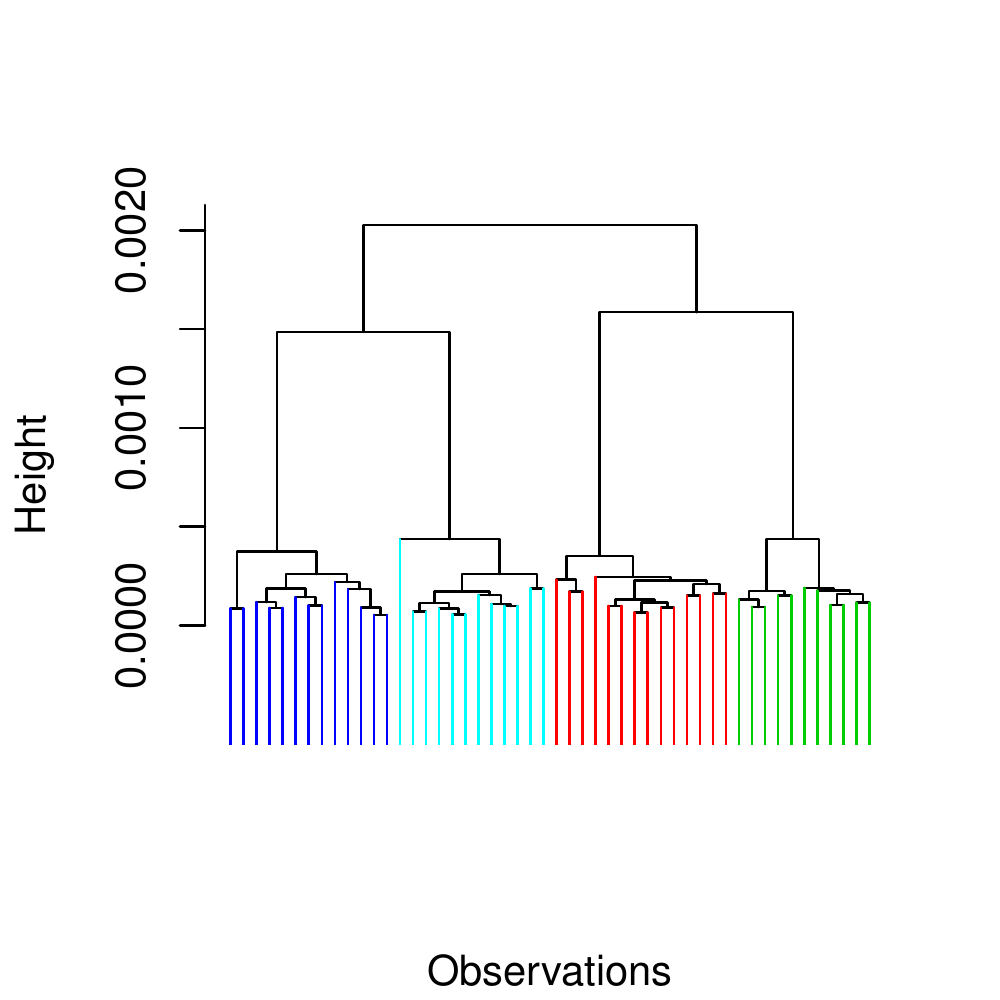}
        \caption{Dense updates}
    \end{subfigure}
     ~
    \begin{subfigure}[t]{0.24\textwidth}
        \includegraphics[height=\textwidth,width=\textwidth]{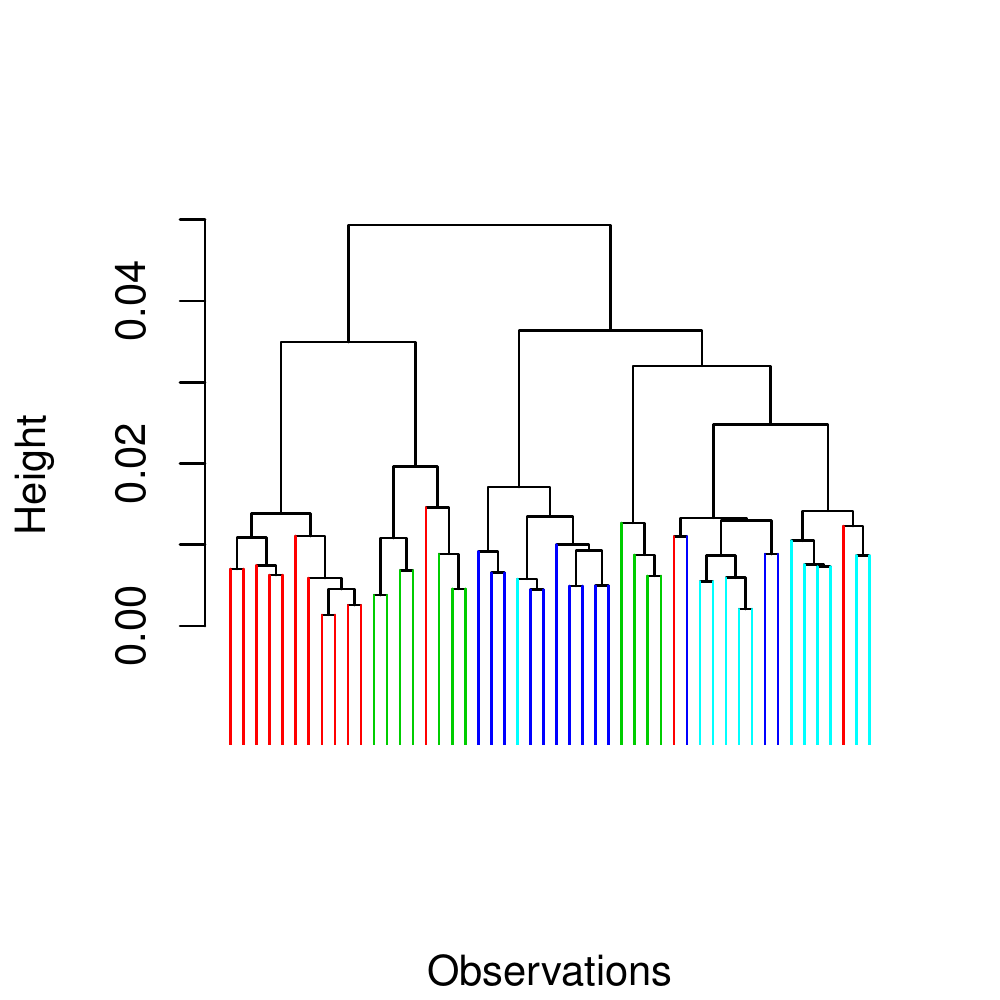}
        \caption{No  updates}
    \end{subfigure}
    \caption{Dendrograms produced by biconvex clustering under various update rules; best result is obtained when affinities $\phi_{ij}$ are updated using $k$-NN (with $k=5$). }
    \label{fig:complexity}
\end{figure}

Next, we take a closer look at the effect of various choices of $\phi_{ij}$ on performance in a controlled setting. We consider a  simulated dataset in a fairly low signal-to-noise regime, with ambient dimension $p=200$ but only $5$ features relevant to clustering, and $n=50$ observations. %Amongst the 200 features, only the first 5 of them contain the cluster structure of the dataset. The rest of the 195 variables are simulated from a $\mathcal{N}(0,1)$ distribution. 
The data are visualized using t-SNE in the first panel of Figure \ref{fig:com:tsne}. % which shows that the signal to noise ratio is fairly low. 
Panel (b) shows that when updating affinities $\phi_{ij}$ with the choice of $k=5$ nearest neighbors, biconvex clustering produces an estimate whose dendrogram reveals the true structure clearly. In contrast to convex clustering and its existing variants, this remains true even without sparsifying the affinities via $k$-nearest neighbors: that is, setting $k=n$ so that the neighbor graph is dense, panel (c) reveals that the relative height of the dendrogram that corresponds to a perfect recovery of the four true clusters is again very pronounced.  Finally, biconvex clustering without any affinity updates fails to perfectly recover the true partition (Figure \ref{fig:complexity}d). This study highlights the robustness of our method to heuristic choices in $\phi_{ij}$--- the algorithm produces dendrograms that are able to reveal the ground truth whether or not we sparsify the choice of affinities. At the same time, the example demonstrates the power of adapting within the learned feature space, without which it is not possible to cuts the dendrogram and obtain the true clustering. % Form Figure \ref{fig:complexity}, it is easily observed that the the BCC algorithm with affinity updates gives a perfect clustering of the dataset, even in a low signal to noise ratio scenario. On the other hand the initial choice of the affinity parameters misguides the BCC algorithm since the nearest neighbours concept fades away in high-dimensions. We also will show that our algorithm succeeds even without sparsifying $\phi$, i.e. choosing $K=n$ in the k-nearest neighbor procedure \jxadd{This is shown in dendrogram-- rewrite section}.
We next consider several simulation studies that  more closely examine empirical performance in terms of feature selection accuracy and clustering accuracy.
%\subsection{Feature selection}

\paragraph{Performance with increasing feature dimension}\label{inc_irr}
This simulation study aims to assess the performance of biconvex clustering and competitors as the number of non-informative features increases. The first two features  are drawn from four bivariate Gaussian distributions with means $(\pm 1, \pm 1)$ and standard deviation $0.25$ with equal probability; these corners of the cube with side length $2$ can be thought of as the true centers along the first two dimensions. The remaining $d$ features are uninformative, drawn from a standard normal distribution; we will examine performance as $d$ increases. 

\begin{figure}[!htbp]
    \centering 
    \includegraphics[height=0.35\textwidth,width=0.55\textwidth]{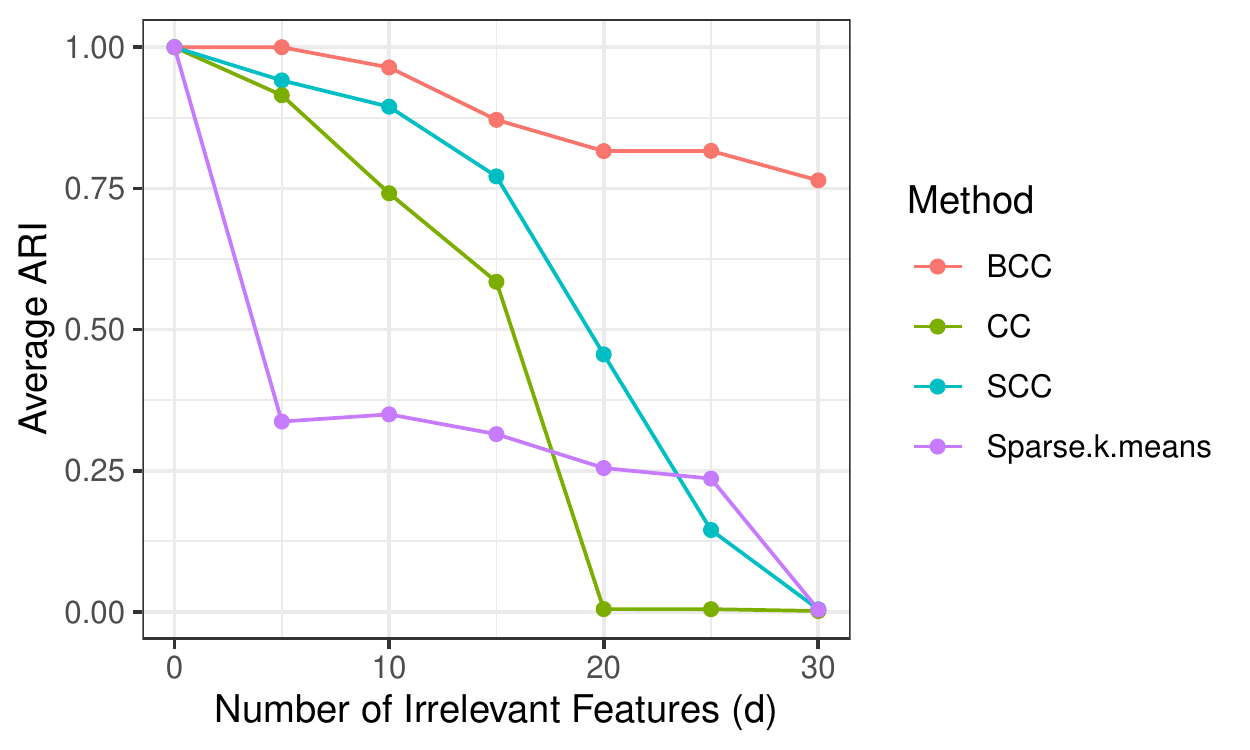}
    \caption{Performance in a low signal-to-noise ratio setting (Section~\ref{inc_irr}).  %showing the efficacy of BCC algorithm even in a low signal-to-noise-ratio situation. 
    Here SCC, BCC, CC denote Sparse Convex Clustering, Biconvex Clustering and Convex Clustering.}
    \label{fig:inc_fe}
\end{figure}

% To be more precise, each row of the data matrix is simulated as follows:
% \begin{align*}
%     (X_{i1},X_{i2})^\top  \sim & \frac{1}{4} \mathcal{N}((-1,-1)^\top,0.2^2I_2) +\frac{1}{4} \mathcal{N}((-1,1)^\top,0.2^2I_2) + \frac{1}{4} \mathcal{N}((1,-1)^\top,0.2^2I_2) \\
%     &+ \frac{1}{4} \mathcal{N}((1,1)^\top,0.2^2I_2)\\
%     X_{il}  \sim & \mathcal{N}(0,1), \quad \text{if } l \in \{3,\dots,d+2\}.
% \end{align*}
%Here $I_2$ denotes the $2 \times 2$ identity matrix. Thus, the data contains four clusters and only the first two features of the data matrix are informative. The rest $d$ many contain no information about the cluster structure of the data. 
We report the  performance of each algorithm over 20 restarts as the number of noise variables $d$ varies from $0$ to $30$ in Figure~\ref{fig:inc_fe}. On average, biconvex clustering performs better than its competitors, while sparse convex clustering performs nearly as well, with the performance gap widening as the number of noise variables increases. The proposed method is the only approach remaining moderately successful in the more challenging settings: without sparsity, convex clustering performs as poorly as random assignment, while the other existing approaches are only marginally better.  % Moreover, the average performance of the other competing algorithms deteriorate as the number of noise variables increases, performing almost as poorly as random assignment at $d=30$. %This is because as $d$ increases, the signal-to-noise-ratio drops down rapidly, making the SCC and Sparse $k$-means harder to detect the right features. The convex clustering (CC) fails due to the fact that irrelevant features misguide the solution paths of the algorithm. 
%It should be noted that biconvex clustering remains effective even in regimes as the signal-to-noise-ratio continues to drop, yielding  superior performance as $d$ is further increased (not shown) in this simulation setting. 

\paragraph{Feature Selection}
\label{feature selection}
The following simulation examines feature selection performance of our method compared to the widely used sparse $k$-means clustering method of \citet{witten2010framework}. Our simulation study begins with $n=1000$, $p=100$ and $k=5$. The first five features of each true center $\btheta_1, \ldots \btheta_k$ are sampled uniformly in $(0,1)$, and the remaining $\theta_{j,l}=0$.
% The matrix $\Theta_{k \times p}$ whose rows contain the true cluster centroids is generated as follows:%\vspace{-0.4cm}
% \begin{enumerate}
% %\item Select $5$ relevant features $l_1,\dots,l_5$ at random.%\vspace{-0.1cm}
% \item Simulate $\theta_{j,l} \sim Unif(0,1)$ for all $j=1,\dots,k$ and $l=1,\dots,5$.%\vspace{-0.1cm}
% \item Set $\theta_{j,l}=0$ for all $l \not\in \{1,\dots,5\}$ and all $j$.%\vspace{-0.3cm}
% \end{enumerate}
After defining $\Theta$, we generate data $\bX$ so that only the first $5$ features are informative %toward distinguishing clusters, simulated 
as follows:
\begin{equation*}
x_{il}\sim \frac{1}{k}\sum_{j=1}^k\mathcal{N}(\theta_{j,l},0.015) \text{ if }   l \in \{1,\dots,5\}; \qquad 
x_{il}\sim \mathcal{N}(0,1) \text{ if }   l \not\in \{1,\dots,5\}.
\end{equation*}
We standardize all features and repeat the simulation to generate $30$ simulated datasets.  We compare results under biconvex clustering with tuning parameters $\lambda=0.2$ and $\gamma=100$, and sparse $k$-means with parameter $s^\ast$ selected using the gap statistic \citep{tibshirani2001estimating} over a fine grid of 100 values of $s \in [1.01,10]$, using the \texttt{R} package \texttt{sparcl}. 

\begin{figure}[htbp]
    \centering
            \begin{subfigure}[t]{0.23\textwidth}
    \centering
        \includegraphics[height=\textwidth,width=\textwidth]{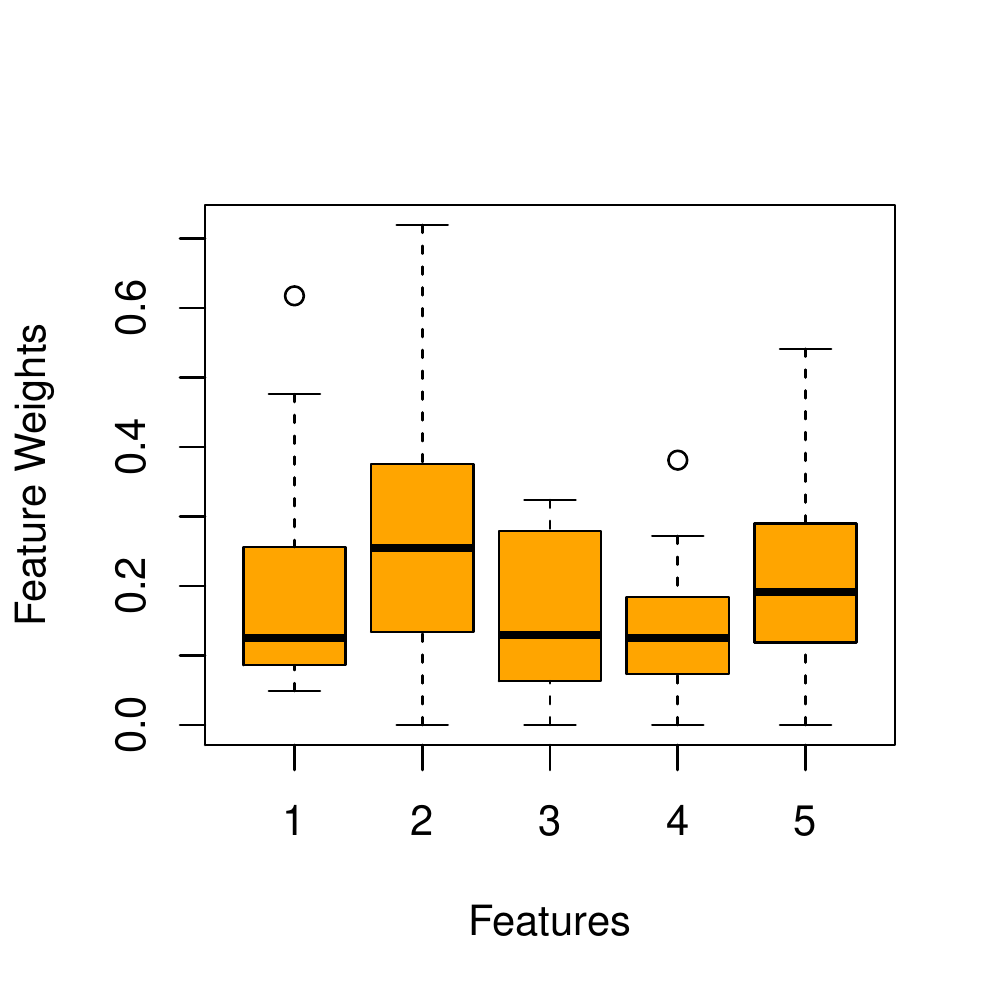}
        \caption{Biconvex}
        \label{sim1_a}
%        \label{libras_gt}
    \end{subfigure}
    ~
    \begin{subfigure}[t]{0.23\textwidth}
    \centering
        \includegraphics[height=\textwidth,width=\textwidth]{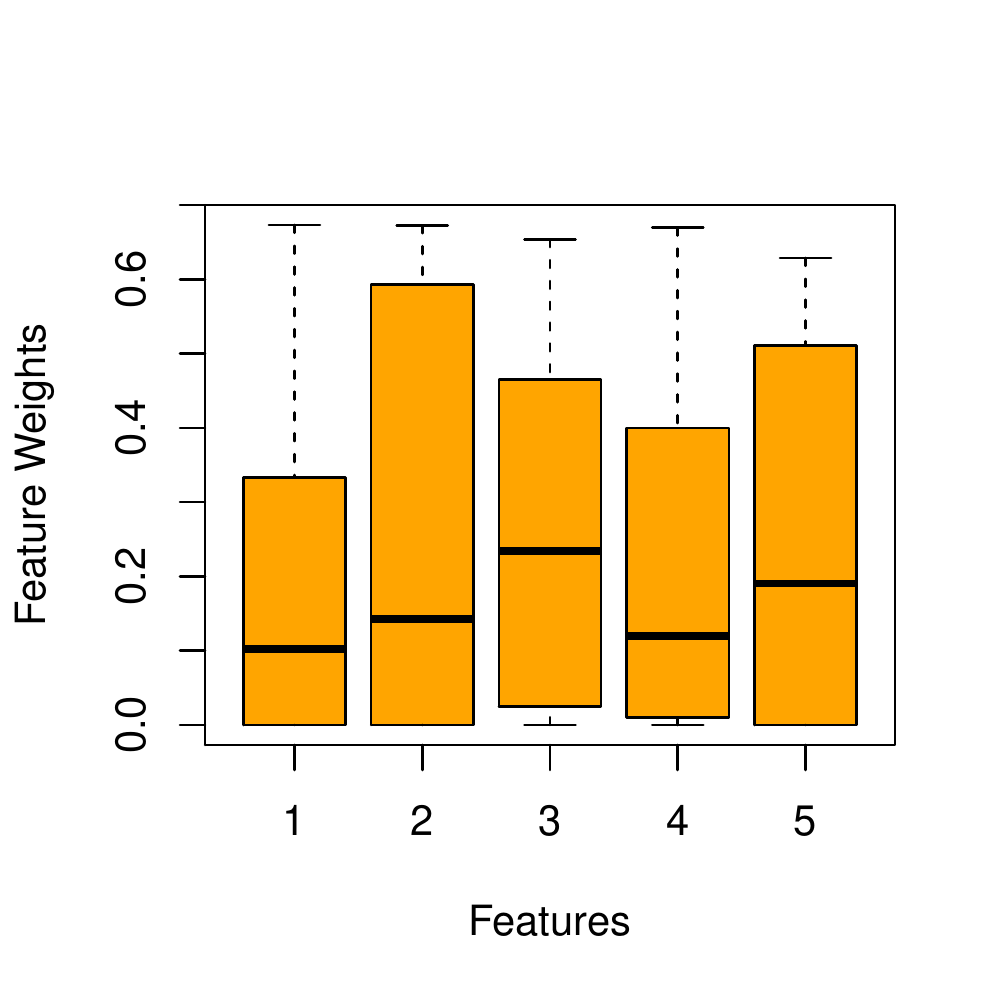}
        \caption{Sparse $k$-means}
\label{sim1_b}
    \end{subfigure}
            ~
        \begin{subfigure}[t]{0.23\textwidth}
    \centering
        \includegraphics[height=\textwidth,width=\textwidth]{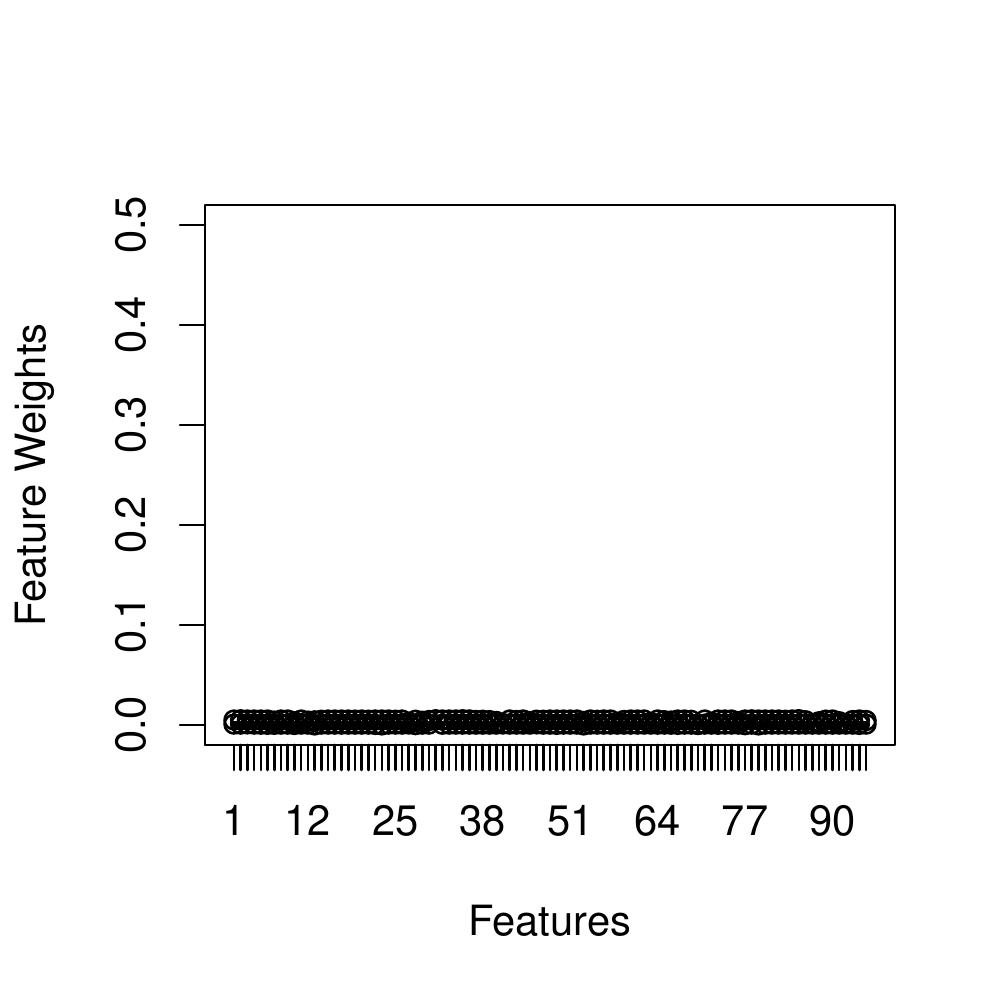}
        \caption{Biconvex}
       \label{sim1_c}
       
    \end{subfigure}
    ~
        \begin{subfigure}[t]{0.23\textwidth}
    \centering
        \includegraphics[height=\textwidth,width=\textwidth]{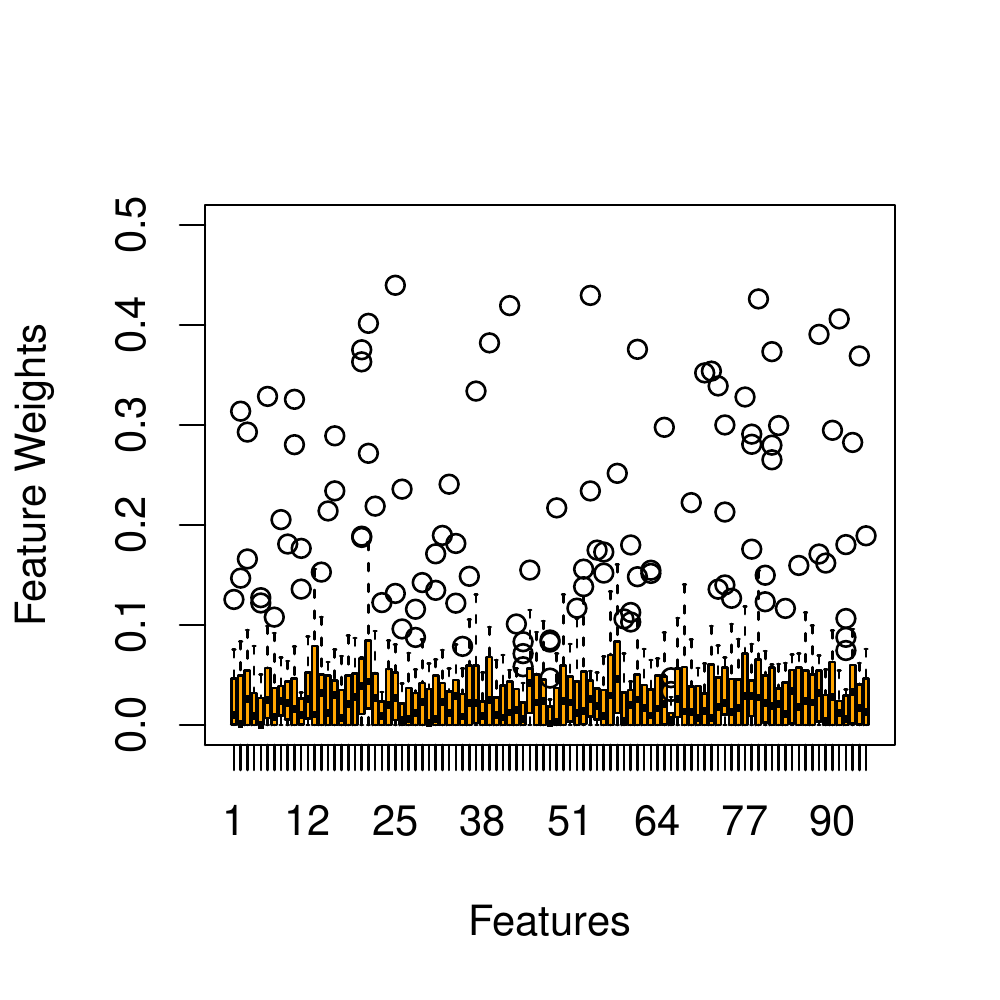}
        \caption{Sparse $k$-means}
       \label{sim1_d}
    \end{subfigure}
    \caption{Figures \ref{sim1_a} and \ref{sim1_b} show the boxplots of the feature weights for the 5 relevant features , while \ref{sim1_c} and \ref{sim1_d} show the same for the 95 irrelevant features. Biconvex clustering consistently zeroes out the irrelevant features while sparse $k$-means fails to do so.}
\label{Simulation 1}
    \end{figure}
Boxplots of the resulting feature weights obtained under both algorithms, separated by relevant and irrelevant features, are displayed in Figure \ref{Simulation 1}. From panels  \ref{sim1_a} and \ref{sim1_c}, we see that biconvex clustering consistently assigns weight to all $5$ relevant features, while correctly zeroing out the remaining irrelevant features. On the other hand, while sparse $k$-means assigns a large chunk of weight to the relevant features (Figure \ref{sim1_b}), there is much more variance across datasets. More to the point, Figure \ref{sim1_d} shows that sparse $k$-means fails to consistently zero out irrelevant dimensions, often assigning them weight values on par with the relevant features. Finally, though our formulation loses convexity, we observe that in practice the method is quite insensitive to initial guess; this stability is showcased via extended simulations in Appendix C.2. % of the Supplement, demonstrating that solutions remain stable to random initializations of $\bw$. 

\paragraph{Clustering accuracy}\label{sec:acc}
%Having seen that our method is more stable even when the number of unimportant features is increased in the previous simulation study, we now shift attention to examine the quality of clusterings produced by the various peer algorithms more closely. 
%Non-convex objective functions such as those based on $k$-means are known to become increasingly susceptible to poor local minima when the number of clusters grows \citep{lloyd1982least,pmlr-v97-xu19a}. Here we show that our biconvex formulation enjoys robustness to this phenomenon compared to competing nonconvex approaches. On the other hand, though we can no longer guarantee that solutions are necessarily global minimizers, our method delivers more accurate solutions than convex clustering and its sparse counterpart. Our empirical studies indicates that this tradeoff is well worth it, and suggests that when the convex relaxation may be far from the original formulation for which it serves as proxy, its global solution may not translate to desirable clustering performance.

\begin{figure}[t]
    \centering
            \begin{subfigure}[t]{0.31\textwidth}
    \centering
        \includegraphics[height=.6\textwidth,width=.6\textwidth]{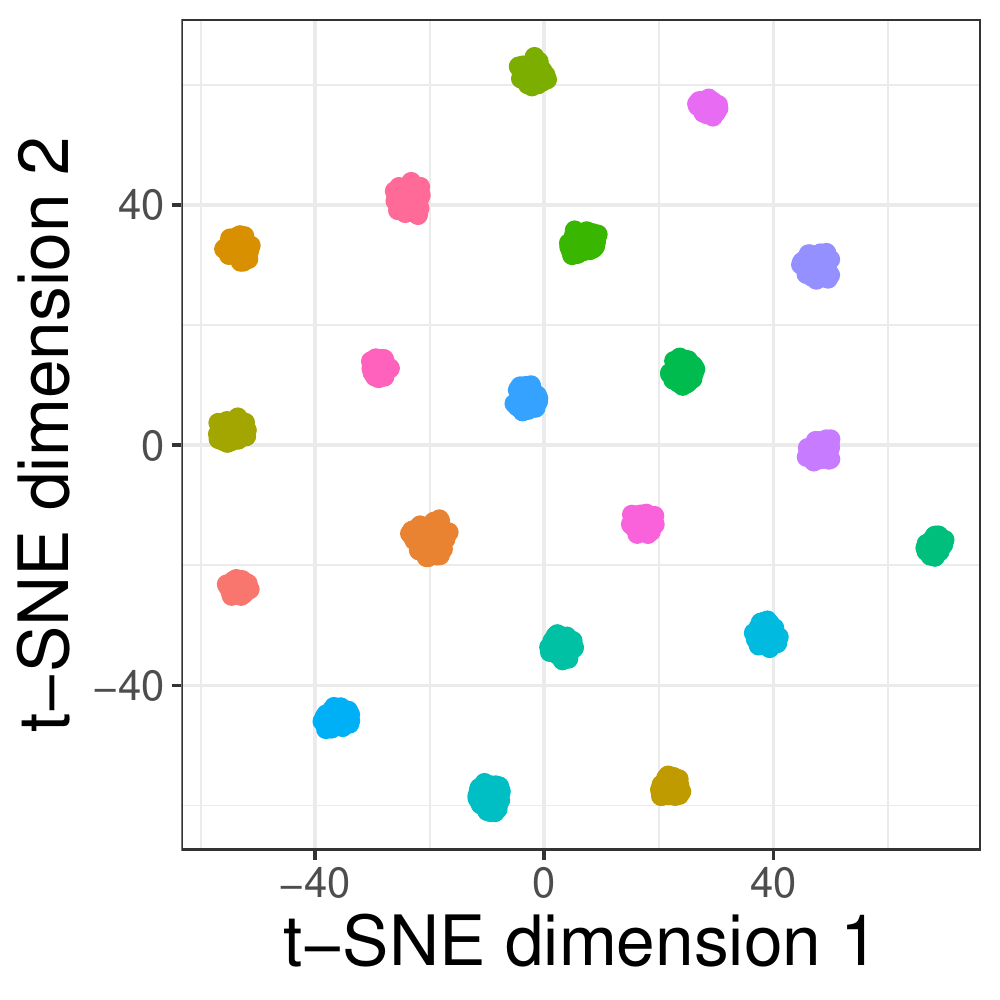}
        \caption{Biconvex}
        \label{sim2_a}
%        \label{libras_gt}
    \end{subfigure}
    ~
%     \begin{subfigure}[t]{0.23\textwidth}
%     \centering
%         \includegraphics[height=\textwidth,width=\textwidth]{images/eff_bcc.pdf}
%         \caption{BCC}
% \label{sim2_b}
%     \end{subfigure}
%     ~
        \begin{subfigure}[t]{0.31\textwidth}
    \centering
        \includegraphics[height=.6\textwidth,width=.6\textwidth]{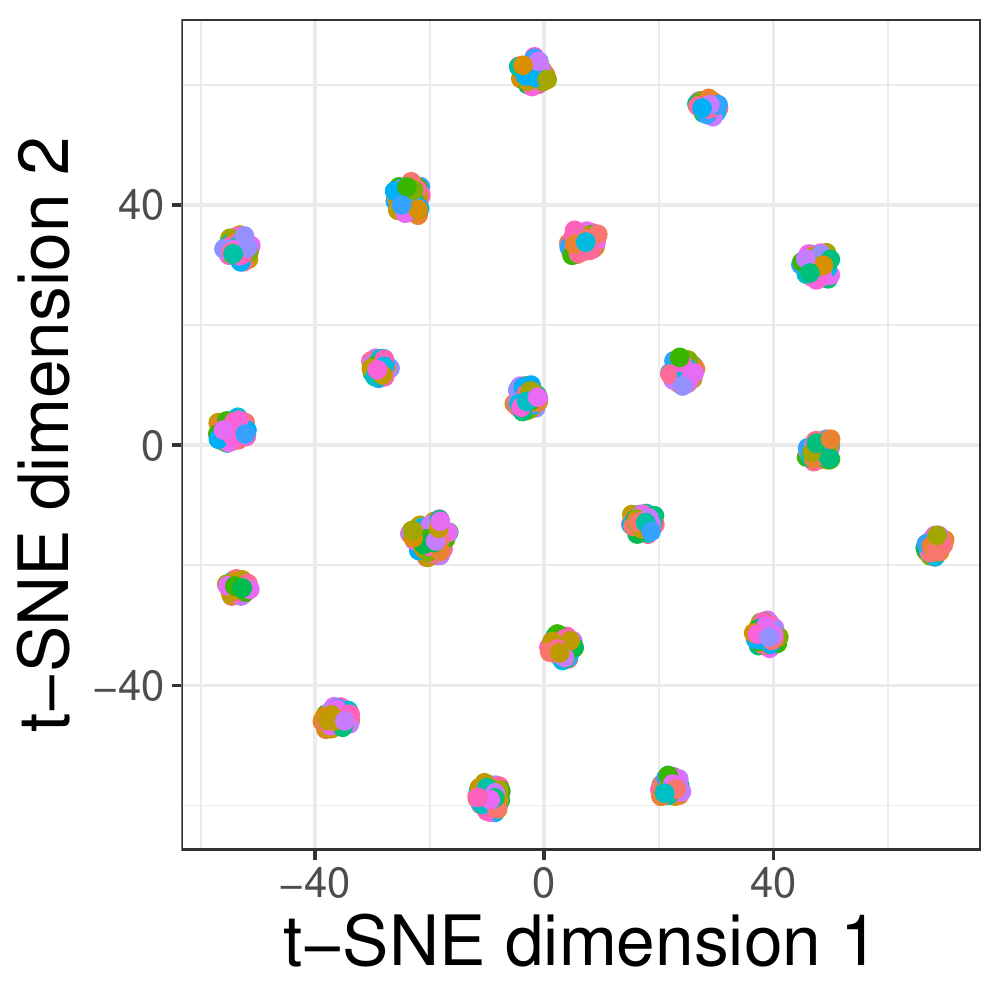}
        \caption{Sparse Convex (SCC)}
    \end{subfigure}
            ~
        \begin{subfigure}[t]{0.31\textwidth}
    \centering
        \includegraphics[height=.6\textwidth,width=.6\textwidth]{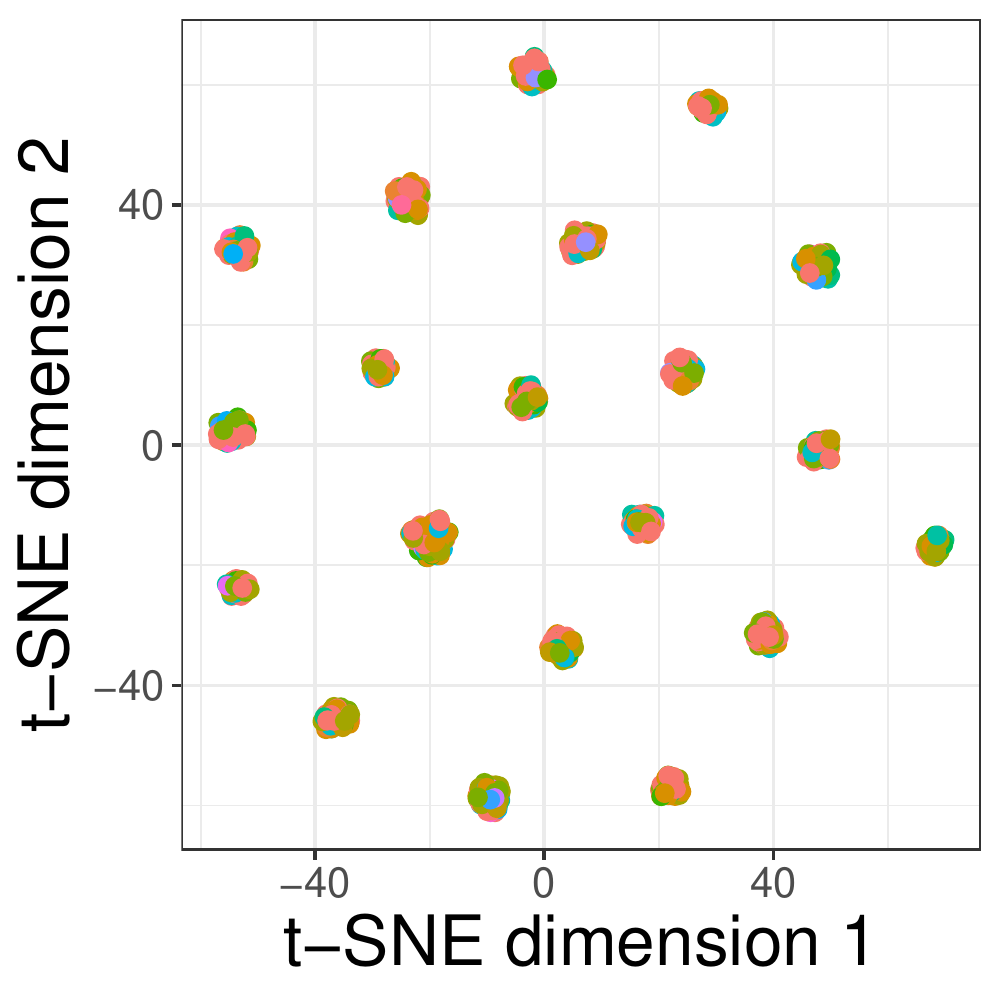}
        \caption{Sparse Hierarchical}
       \label{sim2_c}
    \end{subfigure}
    ~ 
    \begin{subfigure}[t]{0.31\textwidth}
    \centering
        \includegraphics[height=.6\textwidth,width=.6\textwidth]{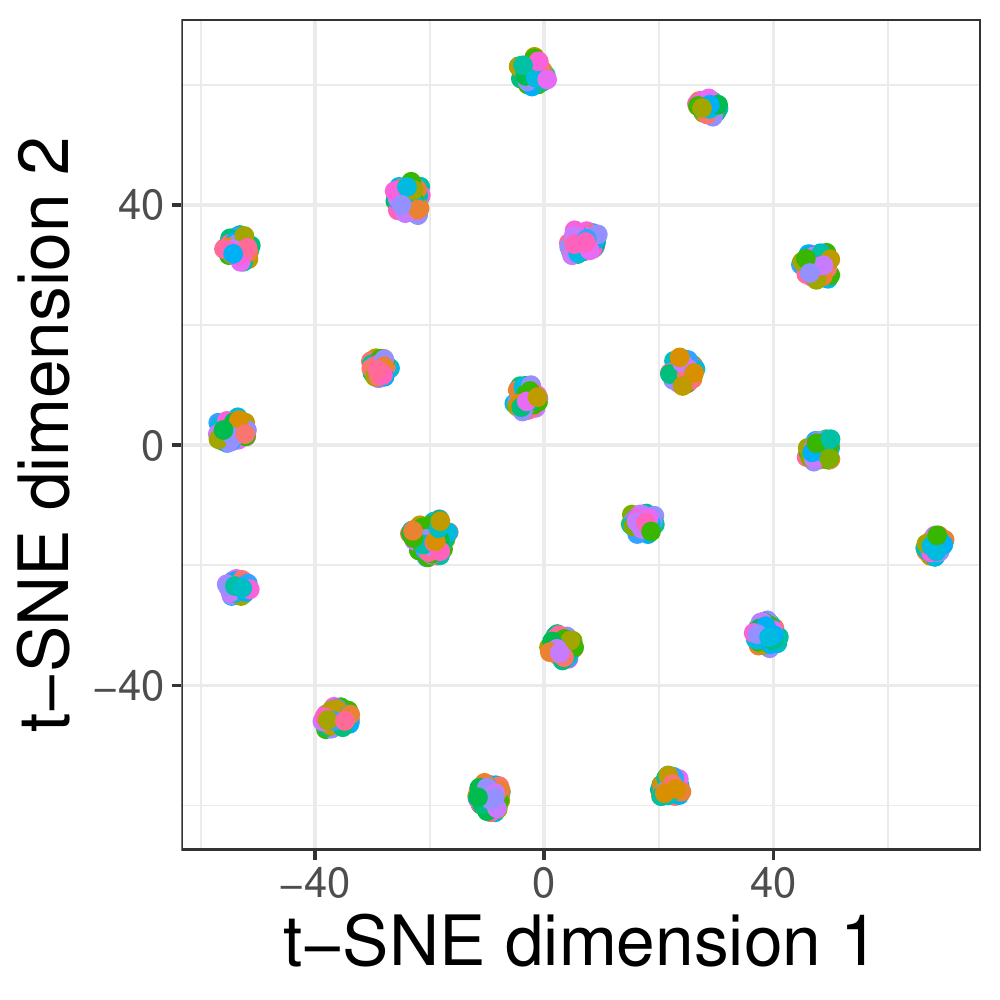}
        \caption{Sparse $k$-means}
       \label{sim2_c}
    \end{subfigure}
    ~
        \begin{subfigure}[t]{0.31\textwidth}
    \centering
        \includegraphics[height=.6\textwidth,width=.6\textwidth]{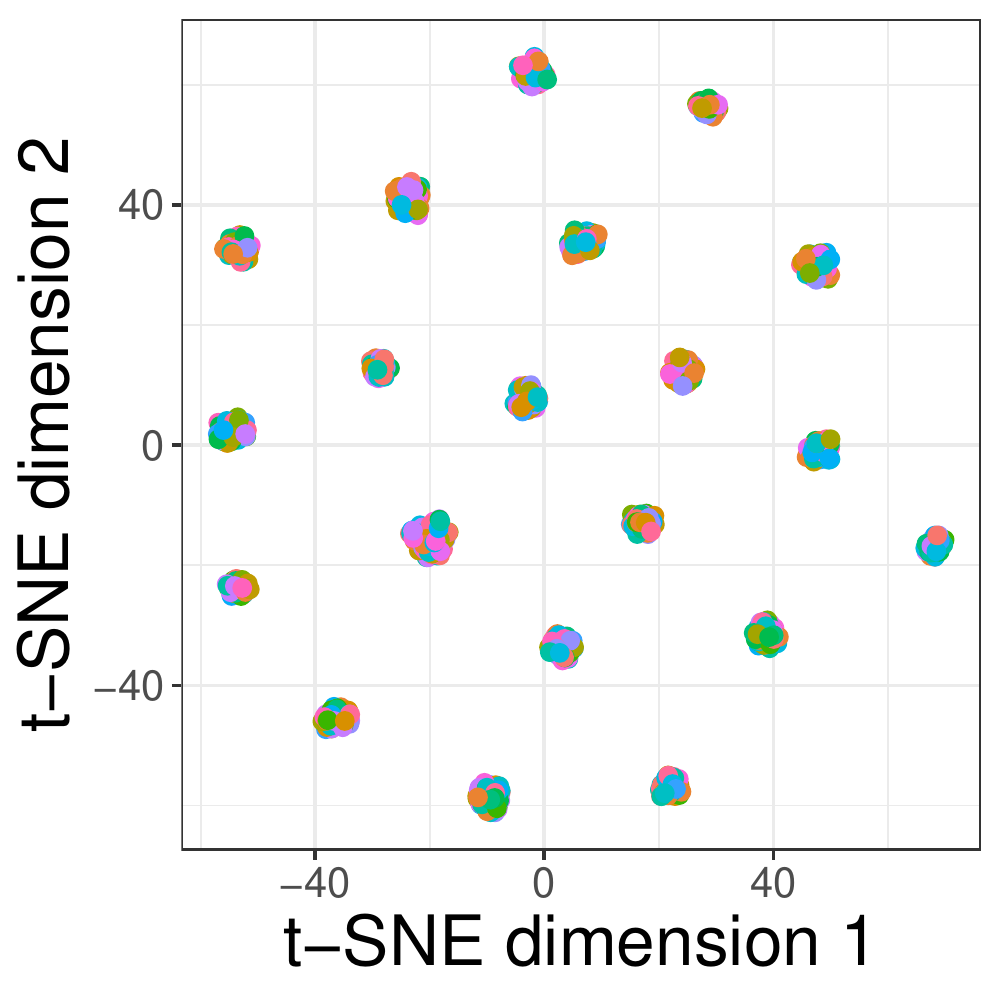}
        \caption{Convex Clustering}
       \label{sim2_d}
    \end{subfigure}
    ~
            \begin{subfigure}[t]{0.31\textwidth}
    \centering
        \includegraphics[height=.6\textwidth,width=.6\textwidth]{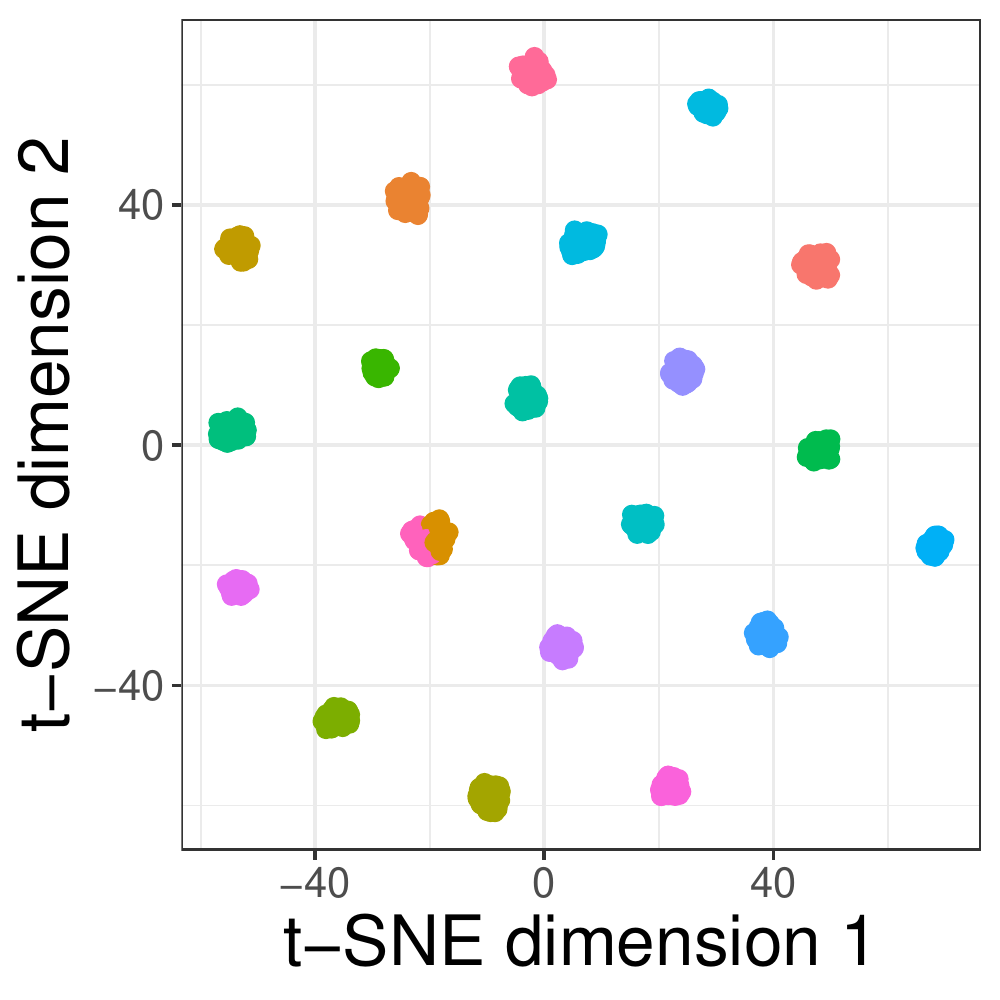}
        \caption{SCC, learned affinities}
       \label{sim2_e}
    \end{subfigure}
    ~
    \caption{$t$-SNE plots showing the performance of the peer algorithms when the number of clusters is high in a sparse clustering scenario.}
    \label{sim2}
\end{figure}

Under the same setup as above but increasing $k=20$, %A $t$-Stochastic Neighbourhood Embedding ($t$-SNE) plot, which is constructed using the five relevant features, is shown in Figure \ref{sim2_a} along with the ground truth partitioning of the dataset. 
the %performance (best performance out of 20 independent runs for sparse $k$-means and sparse hierarchical clustering)  
best of $20$ restarts for each algorithm are shown using $t$-Stochastic Neighbourhood Embedding ($t$-SNE) \citep{maaten2008visualizing} in Figure \ref{sim2}.  Convex clustering, sparse convex clustering, and the sparse variants of $k$-means and hierarchical clustering are all implemented using their respective \texttt{R} packages. Sparse $k$-means tends to deteriorate when it  is trapped by a poor local optimum of the objective function, which becomes highly non-convex when $k$ is large. Convex clustering unsurprisingly fails because it is not designed for high-dimensional scenarios with many uninformative features. \par

The poor performance of sparse convex clustering \citep{wang2018sparse} may arise largely due to the strong dependence on choice of the affinity parameters, since the signal-to-noise-ratio is quite low and the selection of nearest neighbours is significantly influenced by the irrelevant dimensions. To assess this possibility, we try sparse convex clustering again under the \textit{learned} affinities resulting from the estimate $\hat\bw$ obtained by biconvex clustering. Indeed, panel \ref{sim2_e} shows that the performance of sparse convex clustering is rescued when provided with $\hat\phi_{ij}$ defined with respect to the weighted feature space induced by $\hat\bw$.
%In case of sparse convex clustering \citep{wang2018sparse}, we used the tuning parameters prescribed by the authors. However we observed that the algorithm always gives singleton clusters. This might be due to the choice of the affinity parameters, since in this scenario, the signal to noise ratio is quite low and the nearest neighbours are highly influenced by the irrelevant dimensions. However, when we use the affinity parameters generated by the BCC as an input for sparse convex clustering algorithm, it performs significantly well. The resulting clustering is shown in Figure \ref{sim2_e}. %The average NMI for the sparse convex clustering using this modification turns out to be 0.93. \par
Finally, the solution under our biconvex clustering method is identical to the ground truth partition, displayed in  \ref{sim2_a}. Not shown in the plot, biconvex clustering also assigns positive feature weights only to the true discriminative features. The resulting dendrograms as well as extended simulations as dimension continues to increase are displayed in the Appendix. % which provides further insight validating this claim. 
Due to differences in implementation, we omit a detailed runtime comparison, but note that even a naive implementation of biconvex clustering runs significantly faster than both convex clustering and sparse convex clustering.% Finally, our algorithm remains stable as the feature dimension increases while these existing methods falter; for length considerations, a detailed simulation study showcasing this is included in the Supplement.

%Our empirical studies indicates that this tradeoff is well worth it, and suggests that when the convex relaxation may be far from the original formulation for which it serves as proxy, its global solution may not translate to desirable clustering performance.

%\jxadd{Can we be more specific about these results, ie reported as best out of 10 trials, etc. It may be confusing to readers when we plot just one image but mention the ``average NMI" of SCC. Of course we will also have plenty of time to report either average or best performance over a number of trials in the line plots to be added}.

\section{Case studies}
\label{real analysis}
%We turn our attention to several case studies on real data. We evaluate performance of the algorithms via the Adjusted Rand Index (ARI) %\citep{hubert1985comparing} 
%between the output partition and the ground truth. An ARI value of 1 indicates perfect clustering whereas a value of 0 indicates complete mismatch between the ground truth and the partition obtained by the algorithm. 

\paragraph{Movement data}
We begin by analyzing the Libras movement data corpus, which consists of 15 classes each containing 24 observations. Each class represents a hand movement type, and observations are described via 90 features representing the coordinates of movement.  
The data are publicly available \citep{Dua:2019} and a subset of the full dataset was chosen to  showcase the sparse convex clustering algorithm \cite{wang2018sparse}, enabling us to consider a conservative comparison to prior work.
Following  \cite{wang2018sparse}, we normalize all features and consider 6 movement types for evaluation purposes, chosen to avoid excess overlap among the highly correlated classes. Each of the peer algorithms is tuned and run using their recommended settings, as implemented in their respective \texttt{R} packages. We run all sparse $k$-means variants from $20$ initial guesses and report the best performing trial. The hyperparameters of convex and sparse convex clustering are selected using the stability selection procedure outlined by \citet{fang2012selection}.

We apply biconvex clustering with $\lambda=0.2$ and $\gamma=100$ and use the \texttt{dynamicTreeCut} package in \texttt{R} to assign cluster labels to the resulting centroid estimates \citep{langfelder2008defining}. 
We then run the average linkage algorithm on the dataset, again using \texttt{dynamicTreeCut} on the data, as well as sparse convex clustering algorithm and sparse $k$-means clustering. The Adjusted Rand Index (ARI) values obtained by each algorithm are displayed in Table \ref{libras tab}, with further t-SNE visualizations in the Appendix. We see a clear difference  in performance between the proposed method and existing competitors. Interestingly, we find that biconvex clustering does not zero out any features in this case, though proper feature weighing nonetheless helps it discover the true cluster structure in the dataset. 

\begin{table}[t]
	\centering
	\caption{Feature selection and clustering performance, Libras movement data.}
	\label{libras tab}
	\begin{tabular}{c c c c}
		\hline
		%\multicolumn{2}{l}
		Algorithm & \# Nonzero Weights &  \# Clusters & ARI \\
		\hline
		Sparse $k$-means & 83 & 6 & 0.46\\
		Sparse Hierarchical Clustering & 14 & 6 & 0.11\\
		Average Linkage & 90 & 4 & 0.36\\
		Convex Clustering & 90 & 8 & 0.61\\
	    Sparse Convex Clustering (S-AMA) & 63 & 3 & 0.31\\
		Sparse Convex Clustering (S-ADMM) & 13 & 3 & 0.31\\
		Biconvex Clustering & 90 & 5 & \textbf{0.79}\\
		\hline
	\end{tabular}
\end{table}

\paragraph{Leukemia classification}
We next revisit a classic DNA microarray dataset from a study of leukemia on human subjects. The data are collected and described by \cite{golub1999molecular} and after a standard preprocessing, consist of $3571$ gene expression levels collected over $72$ samples (comprised of 62 bone marrow samples and 10 from the peripheral blood). Out of the 72 samples, 47 correspond to acute lymphoblastic leukemia (ALL) and 25 to acute myeloblastic leukemia (AML). The observations are centered and scaled before use. 

To distinguish between these two classes on the basis of gene expression, we apply biconvex clustering with hyperparameters $\lambda=0.01$ and $\gamma=0.1$. At convergence, our method selects $456$ relevant genes among the original space of $3571$ features, and is able to correctly classify all but one one point. Prior work has suggested that this point is likely a potential outlier \citep{chakraborty2020hierarchical}. 
In this case study, we emphasize the success of gene selection in addition to recovering the ground truth partition almost perfectly. Among $50$ genes that were deemed potentially relevant to distinguishing between leukemias in the original study by \cite{golub1999molecular}, $46$ are also selected by our algorithm. A later study by \citet{chow2001identifying} revisits the data with a focus on identifying the discriminative genes,  employing heuristic criteria based on a mix of three relevance measures: a naive Bayes score, a median vote, and mean aggregate relevance. These measures are combined in a supervised learning framework based on Support Vector Machines (SVMs) that requires true labels as inputs. They reported that the top $50$ genes under each of the relevance measures had little overlap; we find all genes in the intersection of the three criteria are selected by our method. In particular, the genes deemed most significant in this prior study--- Adipsin (M84526) and Cystatin C (M27891)--- were assigned the $3$rd and $7$th largest feature weights among the $456$ selected by biconvex clustering. While findings from neither of these previous studies should be taken as ground truth, it is reassuring that their results are consistent with our feature selection results. Though it is perhaps unsurprising that our method may outperform those applied to these data at that time--- these studies predated the sparse clustering framework of \citet{witten2010framework}---our simultaneous clustering and feature weighing method offers new insights in a reexamination of the data. A detailed list of the top $10$ genes identified by our method is included in Table \ref{tab:leukemia}, and  visualization of solutions via t-SNE is provided in Figure A3 of the Appendix.
\begin{table}[htbp!]
{\footnotesize
    \centering
    \begin{tabular}{c c c}
    \hline
        Gene ID & Gene Annotation & Feature Weight \\
        \hline
         M27891\_at & CST3 Cystatin C (Amyloid angiopathy and cerebral hemorrhage) & 0.012265694\\
         M28826\_at & CD1B antigen (Thymocyte antigen) & 0.050758810\\
         M84526\_at & DF D component of complement (Adipsin) & 0.014803023\\
         U05259\_rna1\_at &  MB-1 gene & 0.010192411 \\
         U46499\_at &  Gluthanione S-Transferase, Microsoma & 0.013318948\\
         X04145\_at &  CD3G antigen, gamma polypeptide (TiT3 complex) & 0.010848832 \\
         X14975\_at & GB DEF5CD1 R2 gene for MHC-related antigen & 0.011330957\\
         X87241\_at & HFat protein & 0.012668398\\
         X76223\_s\_at &  GB DEF5MAL gene exon 4& 0.015623870\\
         M31523\_at &  TCF3 Transcription factor 3 &0.013837447  \\
        % & (E2A immunoglobulin enhancer binding factors E12/E47) & \\
         \hline
    \end{tabular}
    \caption{Top 10 genes identified by biconvex clustering on leukemia dataset, along with their gene IDs, annotations and feature weights.}
    \label{tab:leukemia}
    }%
\end{table}

\section{Discussion}\label{discussion}
The proposed methodology seeks to address three chief challenges in clustering: high-dimensionality via feature weighing and selection, sensitivity to initialization via a biconvex formulation, and selection of cluster centers via a penalty formulation that encourages centroids to agglomerate along the solution path. Biconvex clustering can be seen as occupying a middle ground between the convex clustering objective and the original non-convex objective arising from hierarchical clustering. We build upon recent algorithmic developments for convex clustering and its variants, bridging these insights to good intuition established in the classical clustering literature. 

Not only does feature selection reduce dimension while retaining interpretability of the features (compared to a generic dimension reduction pre-processing step), but feature weights obtained by the algorithm improve clustering quality while serving as informative and interpretable quantities themselves. Our algorithm returns simple alternating updates that resemble classical clustering algorithms, while conferring several advantages over traditional approaches. In particular, we show that sacrificing convexity and exact coalescence of centroids enables algorithmic and theoretical benefits that we consider well worth the tradeoff. Fruitful avenues for future work include extending the framework to tensor settings \citep{sun2019dynamic}, and exploring other optimization frameworks such as semi-smooth Newton \citep{yuan2018efficient,sun2018convex}. % and stochastic descent algorithms \citep{panahi2017clustering} that may lead to further computational gains.
Related penalties involving $k$-NN affinities within a fusion penalty have recently been studied for non-parametric regression \citep{madrid2020adaptive}. Similar to their benefits in our setting, the $k$-NN terms enable ``manifold adaptivity", in addition to the fusion term which provides local adaptivity. It will be fruitful to explore the extent to which our contributions are useful in related contexts such as regression and trend filtering, as well as to potentially import their theoretical insights into the clustering setting.
\bibliographystyle{apalike}

\clearpage
\begin{center} {\LARGE \bf Appendix }
\end{center}
\appendix
\section{Proof of Theorem 1}
%\subsection{Proof of Theorem 1}
We now discuss the solution of the optimization problem $P_2$, defined in Section 2.2. Since $\bmu$ is kept fixed in this subproblem, we define the constant $D_l=\sum_{i=1}^n(x_{il}-\mu_{il})^2$ for brevity. Thus $f(\bmu,\bw)$ (equation (10)) can be we written as 
\begin{equation}\label{sap11}
    g(\bw)=\sum_{l=1}^p (w_l^2+\lambda w_l)D_l+ (\text{constant in } \bw). 
\end{equation}
Equation \eqref{sap11} is to be minimized subject to the constraints $\sum_{l=1}^pw_l=1$ and $w_l \ge 0$, for all $l=1,\dots,p$. We write the Lagrangian for this problem as follows:
\[\mathcal{L}= \sum_{l=1}^p (w_l^2+\lambda w_l)D_l - \alpha (\sum_{l=1}^pw_l-1) - \sum_{l=1}^p \xi_l w_l \]
From the Karush-Kuhn-Tucker (KKT) conditions of optimality, we have, 
\begin{equation}\label{sap2}
    \frac{\partial \mathcal{L}}{w_l}=0,\quad, \forall\, l=1,\dots,p.
\end{equation}
\begin{equation}\label{sap3}
    \sum_{l=1}^pw_l=1.
\end{equation}
\begin{equation}\label{sap4}
    w_l \ge 0, \quad \forall\, l=1,\dots,p.
\end{equation}
\begin{equation}\label{sap5}
    \xi_l \ge 0.
\end{equation}
\begin{equation}\label{sap6}
   \xi_l w_l =0,  \quad \forall\, l=1,\dots,p.
\end{equation}
Equation \eqref{sap2} implies that 
\begin{equation}\label{sap7}
    w_l = \frac{1}{2} \bigg(\frac{\alpha+\xi_l}{D_l}-\lambda\bigg)
\end{equation}
We now consider the following cases.
\paragraph{Case 1: $\frac{\alpha}{D_l} > \lambda$:}
Since $\xi_l \ge 0$, we conclude from equation \eqref{sap7} that $w_l > 0$. Thus, from equation \eqref{sap6}, we deduce that $\xi_l=0$. Plugging in this value in equation \eqref{sap7}, we get that $w_l=\frac{1}{2} \bigg(\frac{\alpha}{D_l}-\lambda\bigg)$.
\paragraph{Case 2: $\frac{\alpha}{D_l} \le \lambda$:} Suppose $w_l>0$: then equation \eqref{sap6} implies that $\xi_l=0$. Thus, from equation \eqref{sap7}, we would have $w_l = \frac{1}{2} \bigg(\frac{\alpha}{D_l}-\lambda\bigg) \le 0$, which is a contradiction. Thus, in this case, we must have $w_l=0$.

We now observe that since $\lambda,D_l \ge 0$, we may conclude that $\alpha > 0$. Otherwise, case 2 would always occur implying $w_l=0$, $\forall\, l=1,\dots,p$, which violates equation \eqref{sap3}. Thus, case 1 holds and the slack variables $\xi_l=0$. Together wtih equation \eqref{sap7}, we observe that $w_l=\frac{1}{2}S\big(\frac{\alpha}{D_l},\lambda\big)$, where recall the soft thresholding operator is given by 
\[S(x,y)=\begin{cases}
x-y & \text{ if } x \ge y\\
x+y & \text{ if } x \le -y\\
0 & \text{ Otherwise.} 
\end{cases} \]
 Moreover, together with constraint \eqref{sap3}, we observe that $\alpha$ must satisfy the equation: 
\[\frac{1}{2}S\bigg(\frac{\alpha}{D_l},\lambda\bigg)=1.\]
This concludes the proof.
\section{Proofs from Section 3}

\subsection{Proof of Theorem 2}

 \paragraph{Theorem 2}
	Suppose $\bx=\bu+\bepsilon$, where,  $\bepsilon \in \mathbb{R}^{np}$ is a vector of independent bounded random variables, with mean $0$ and $|\epsilon_i| \le M$, for all $i=1,\dots,np$. Suppose that $\hat{\bu}$ and $\widehat{\bW}$ are obtained form minimizing (14) (of main text), then if $\gamma' \ge \frac{(1+\lambda)M}{\sqrt{n^3p}}$
	\begin{align*}
	    \frac{1}{2np} \vvvert\hat{\bu}-\bu\vvvert_{\widehat{\bW}}^2\leq & M^2 (1+\lambda) \left[\frac{1}{\sqrt{n}} + c \sqrt{\frac{\log(1/\delta)}{np}} + c \frac{\log(1/\delta)}{np}\right]+\frac{\gamma' |\mathcal{E}|}{4}\\
	    &+ \gamma' \sum_{(i,j) \in \mathcal{E}}\|\bD_{\mathcal{C}(i,j)} \bu\|_2 +\gamma' \sum_{(i,j) \in \mathcal{E}}\|\bD_{\mathcal{C}(i,j)} \bu\|^2
	\end{align*}
%  	 $$\frac{1}{2np} \vvvert\hat{\bu}-\bu\vvvert_{\widehat{\bW}}^2\leq M^2 (1+\lambda) \left[\frac{1}{\sqrt{n}} + c \sqrt{\frac{\log(1/\delta)}{np}} + c \frac{\log(1/\delta)}{np}\right]+\frac{\gamma' |\mathcal{E}|}{4n^2} + \frac{\gamma'}{n} \sum_{(i,j) \in \mathcal{E}}\|\bD_{\mathcal{C}(i,j)} \bu\|_2 +\gamma' \sum_{(i,j) \in \mathcal{E}}\|\bD_{\mathcal{C}(i,j)} \bu\|^2 $$
 	holds with probability at least  $1-\delta$.
 \begin{proof}
 Let $\bD=\bA \bLambda \bV_{\bbeta}^\top$ be the singular value decomposition (SVD) of $\bD$, where $\bV_{\bbeta} \in \mathbb{R}^{np \times p(n-1)}$. We can construct $\bV_\alpha \in \mathbb{R}^{np \times p}$ such that $\bV=[\bV_\alpha,\bV_{\bbeta}]$ is an $np \times np$ orthonormal matrix. Thus, $\bV^\top \bV= \bV \bV^\top=I$ and $\bV_\alpha^\top \bV_{\bbeta}=0$.\par
 Next, denote ${\bbeta} = \bV_{\bbeta}^\top \bu $ and $\balpha=\bV_\alpha^\top \bu$, and let $\gamma'=\frac{\gamma}{np}$. Thus the optimization problem (14) (of main text) becomes 
 \begin{equation}
 \label{opt3}
     \min_{\balpha,\bbeta,W} \frac{1}{2np} (\bx-\bV_\alpha \balpha - \bV_{\bbeta} \bbeta)^\top \bW (\bx-\bV_\alpha \balpha - \bV_{\bbeta} \bbeta)+\gamma' \sum_{(i,j) \in \mathcal{E}} \|\bZ_{\mathcal{C}(i,j)} \bbeta\|_2^2
 \end{equation}
Let, $\hat{\balpha},\hat{\bbeta}, \widehat{\bW}$ be the minimizers of \eqref{opt3}. We first observe that $\hat{\bu}=\bV_\alpha \hat{\balpha} + \bV_{\bbeta} \hat{\bbeta}$. Let $\bZ=\bA \bLambda$. Let $\bZ^-$ be a pseudo-inverse of $\bZ$ such that $\bZ^- \bZ =\bI$. To simplify notation, we write $\vvvert\by\vvvert_{\bW}^2=\by^\top \bW \by$. By definition, we have
\begin{equation}
     \frac{1}{2np} \vvvert(\bx-\bV_\alpha \hat{\balpha} - \bV_{\bbeta} \hat{\bbeta})\vvvert_{\widehat{\bW}}^2+\gamma' \sum_{(i,j) \in \mathcal{E}} \|\bZ_{\mathcal{C}(i,j)} \hat{\bbeta}\|_2^2
     \leq \frac{1}{2np} \vvvert(\bx-\bV_\alpha \balpha - \bV_{\bbeta} \bbeta)\vvvert_{\widehat{\bW}}^2 +\gamma' \sum_{(i,j) \in \mathcal{E}} \|\bZ_{\mathcal{C}(i,j)} \bbeta\|_2^2
\end{equation}
Substituting $\bx$ by $\bu+\bepsilon$, we obtain
\begin{equation}
\label{s3}
  \frac{1}{2np}  \vvvert\bV_\alpha(\hat{\balpha}-\balpha)+ \bV_{\bbeta}(\hat{\bbeta}-\bbeta)\vvvert^2_{\widehat{\bW}} + \gamma' \sum_{(i,j) \in \mathcal{E}} \|\bZ_{\mathcal{C}(i,j)} \hat{\bbeta}\|_2^2 \leq \frac{1}{np} G(\hat{\balpha},\hat{\bbeta},\widehat{\bW})+ \gamma' \sum_{(i,j) \in \mathcal{E}} \|\bZ_{\mathcal{C}(i,j)} \bbeta\|_2^2,
\end{equation}
where $G(\hat{\balpha},\hat{\bbeta},\widehat{\bW})= \bepsilon^\top \widehat{\bW} [\bV_\alpha(\hat{\balpha}-\balpha)+\bV_{\bbeta}(\hat{\bbeta}-\bbeta)] $. Since $\hat{\balpha}$ is the minimizer of \eqref{opt3}, we can choose $\hat{\balpha}$ such that $\bx - \bV_{\bbeta} \hat{\bbeta} - \bV_\alpha \hat{\balpha}=\mathbf{0}$. That is, this $\hat{\balpha}$ satisfies
\begin{align*}
    \ha & = \bV_\alpha^\top (\bx - \bV_{\bbeta} \hb)\\
    & = \bV_\alpha^\top (\bu + \bepsilon - \bV_{\bbeta} \hb) \, = \, \balpha + \bV_\alpha^\top \bepsilon.
\end{align*}
Thus, we are now in a position to bound
\begingroup
\allowdisplaybreaks
\begin{align}
    \frac{1}{np} |G(\ha,\hb,\widehat{\bW})| & = \frac{1}{np} |\bepsilon^\top \widehat{\bW} [\bV_\alpha(\ha-\balpha)+\bV_{\bbeta}(\hb-\bbeta)] | \nonumber \\
    & = \frac{1}{np} |\bepsilon^\top \widehat{\bW} [\bV_\alpha \bV_\alpha^\top \bepsilon +\bV_{\bbeta}(\hb-\bbeta)] | \nonumber \\
    & \leq \frac{1}{np} \bepsilon^\top \widehat{\bW} \bV_\alpha \bV_\alpha^\top \bepsilon + \frac{1}{np} |\bepsilon^\top \widehat{\bW} \bV_{\bbeta}(\hb-\bbeta) | \nonumber \\
    & = \frac{1}{np} \bepsilon^\top \widehat{\bW} \bV_\alpha \bV_\alpha^\top \bepsilon + \frac{1}{np} |\bepsilon^\top \widehat{\bW}  \bV_{\bbeta} \bZ^- \bZ (\hb-\bbeta) | \label{l2} \\
    & = \frac{1}{np} \bepsilon^\top \widehat{\bW} \bV_\alpha \bV_\alpha^\top \bepsilon +
    \frac{1}{np} \bigg|\sum_{(i,j) \in \mathcal{E}} \bepsilon^\top \widehat{\bW} \bV_{\bbeta} \bZ^-_{\mathcal{C}(i,j)} \bZ_{\mathcal{C}(i,j)} (\hb-\bbeta) \bigg| \nonumber \\
    & \leq \frac{1}{np} \bepsilon^\top \widehat{\bW} \bV_\alpha \bV_\alpha^\top \bepsilon + \frac{1}{np} \sum_{(i,j) \in \mathcal{E}}  \bigg| \bepsilon^\top \widehat{\bW} \bV_{\bbeta} \bZ^-_{\mathcal{C}(i,j)} \bZ_{\mathcal{C}(i,j)} (\hb-\bbeta) \bigg| \nonumber \\
    &  \leq \frac{1}{np} \bepsilon^\top \widehat{\bW} \bV_\alpha \bV_\alpha^\top \bepsilon + \frac{1}{np} \sum_{(i,j) \in \mathcal{E}}  \| \bepsilon^\top \widehat{\bW} \bV_{\bbeta} \bZ^-_{\mathcal{C}(i,j)}\|_2 \|\bZ_{\mathcal{C}(i,j)} (\hb-\bbeta) \|_2 \label{l1} \\
    &  \leq \frac{1}{np} \bepsilon^\top \widehat{\bW} \bV_\alpha \bV_\alpha^\top \bepsilon + \frac{1}{np}   \max_{(i,j) \in \mathcal{E}}\| \bepsilon^\top \widehat{\bW} \bV_{\bbeta} \bZ^-_{\mathcal{C}(i,j)}\|_2 \sum_{(i,j) \in \mathcal{E}} \|\bZ_{\mathcal{C}(i,j)} (\hb-\bbeta) \|_2 \nonumber 
\end{align}
\endgroup
We note that equation \eqref{l2} follows from the definition of $\bZ^-$ as pesudo-inverse of $\bZ$,  and inequality \eqref{l1} follows from the Cauchy-Schwartz inequality. Next, we derive high-probability bounds for the terms  $ \frac{1}{np} \bepsilon^\top \widehat{\bW} \bV_\alpha \bV_\alpha^\top \bepsilon $ and $\max_{(i,j) \in \mathcal{E}}\| \bepsilon^\top \widehat{\bW} \bV_{\bbeta} \bZ^-_{\mathcal{C}(i,j)}\|_2$. %  that hold with a very high probability.
\paragraph{Bounds for $ \frac{1}{np} \bepsilon^\top \widehat{\bW} \bV_\alpha \bV_\alpha^\top \bepsilon $:}
Denoting the spectral norm of the matrix $\bTheta$ by  $\|\bTheta\|_{sp}$, note
$\|\widehat{\bW} \bV_\alpha \bV_\alpha^\top\|_{sp} \leq \|\widehat{\bW}\|_{sp} \| \bV_\alpha \bV_\alpha^\top\|_{sp} \leq (1 + \lambda)$. %Here $\|\bTheta\|_{sp}$ denotes the spectral norm of the matrix $\bTheta$. 
 Now let $\sigma^2$ denote the variance of an individual $\epsilon_i$. 
% \begingroup
% \allowdisplaybreaks
% \begin{align*}
% \|\widehat{\bW} \bV_\alpha \bV_\alpha^\top\|_F & = tr(\widehat{\bW} \bV_\alpha \bV_\alpha^\top \bV_\alpha \bV_\alpha^\top \widehat{\bW})\\
% & = tr(\widehat{\bW} \bV_\alpha \bV_\alpha^\top \widehat{\bW})\\
% & = tr( \widehat{\bW}^2 \bV_\alpha \bV_\alpha ^\top)\\
% & = tr( \widehat{\bW}^2 \sum_{i=1}^p \bv_i \bv_i^\top)\\
% & = \sum_{i=1}^p tr( \widehat{\bW}^2  \bv_i \bv_i^\top)\\
% & = \sum_{i=1}^p tr(\bv_i^\top \widehat{\bW}^2  \bv_i )\\
% & \leq \sum_{i=1}^p \Lambda_{max}(\widehat{\bW}^2)\\
% & \leq p(1+\lambda)^2.
% \end{align*}
% \endgroup
By Lemma 3.2 of  \citep{10.1214/20-EJP422}, with probability at least $1-e^{-t}$, we have 
\begingroup
\allowdisplaybreaks
\begin{align}
    &\sup_{\bW \in \mathcal{W}} \bepsilon^\top \bW \bV_\alpha \bV_\alpha^\top \bepsilon \nonumber \\
    & \le \mathbb{E} \left( \sup_{\bW \in \mathcal{W}} \bepsilon^\top \bW \bV_\alpha \bV_\alpha^\top \bepsilon \right) + c \left[ M \sqrt{t} \mathbb{E} \sup_{\bW \in \mathcal{W}} \| \bW \bV_\alpha \bV_\alpha^\top \bepsilon\|_{sp} + t M^2 \sup_{\bW \in \mathcal{W}} \|\bW\|\right] \nonumber\\
    & \le \mathbb{E} \left( \sup_{\bW \in \mathcal{W}} \text{tr}( \bW \bV_\alpha \bV_\alpha^\top \bepsilon \bepsilon^\top)\right) + c \left[ M \sqrt{t} \mathbb{E} \sup_{\bW \in \mathcal{W}} \| \bW \bV_\alpha \bV_\alpha^\top\|_{sp} \|\bepsilon\|_2 + t M^2 (1+\lambda) \right] \nonumber \\
    & \le \mathbb{E} \left(\sup_{\bW \in \mathcal{W}} \sqrt{\text{tr}(\bW^2)}\sqrt{\text{tr}\left((\bV_\alpha \bV_\alpha^\top \bepsilon \bepsilon^\top)^\top (\bV_\alpha \bV_\alpha^\top \bepsilon \bepsilon^\top)\right)} \right) + c \left[ M \sqrt{t} (1+\lambda) \mathbb{E}  \|\bepsilon\| + t M^2 (1+\lambda) \right] \label{j1}\\
    & = \sup_{\bW \in \mathcal{W}} \sqrt{\text{tr}(\bW^2)} \mathbb{E} \left(\sqrt{\text{tr}\left( \bepsilon \bepsilon^\top \bV_\alpha \bV_\alpha^\top  \bV_\alpha \bV_\alpha^\top \bepsilon \bepsilon^\top\right)} \right) + c \left[ M \sqrt{t} (1+\lambda) \mathbb{E}  \|\bepsilon\| + t M^2 (1+\lambda) \right] \nonumber \\
    & = \sup_{\bW \in \mathcal{W}} \sqrt{\text{tr}(\bW^2)} \mathbb{E} \left(\sqrt{\text{tr}\left( \bepsilon \bepsilon^\top \bV_\alpha \bV_\alpha^\top \bepsilon \bepsilon^\top\right)} \right) + c \left[ M \sqrt{t} (1+\lambda) \mathbb{E}  \|\bepsilon\| + t M^2 (1+\lambda) \right] \nonumber \\
    & = \sup_{\bW \in \mathcal{W}} \sqrt{\text{tr}(\bW^2)} \mathbb{E} \left(\sqrt{\text{tr}\left( \|\bepsilon\|^2_2  \bV_\alpha \bV_\alpha^\top \bepsilon \bepsilon^\top\right)} \right) + c \left[ M \sqrt{t} (1+\lambda) \mathbb{E}  \|\bepsilon\| + t M^2 (1+\lambda) \right] \nonumber \\
    & = \sup_{\bW \in \mathcal{W}} \sqrt{\text{tr}(\bW^2)} \mathbb{E} \left(\|\bepsilon\|_2 \sqrt{\text{tr}\left(   \bV_\alpha \bV_\alpha^\top \bepsilon \bepsilon^\top\right)} \right) + c \left[ M \sqrt{t} (1+\lambda) \mathbb{E}  \|\bepsilon\| + t M^2 (1+\lambda) \right] \nonumber \\
    & \le \sup_{\bW \in \mathcal{W}} \sqrt{np}M\sqrt{\text{tr}(\bW^2)} \mathbb{E} \left( \sqrt{\text{tr}\left(   \bV_\alpha \bV_\alpha^\top \bepsilon \bepsilon^\top\right)} \right) + c \left[ M \sqrt{t} (1+\lambda) \sqrt{np} M  + t M^2 (1+\lambda) \right] \label{j2}\\
    & \le  M \sqrt{np}\sup_{\bW \in \mathcal{W}} \sqrt{\text{tr}(\bW^2)}   \sqrt{\mathbb{E} \text{tr}\left(   \bV_\alpha \bV_\alpha^\top \bepsilon \bepsilon^\top\right)}  + c \left[ M^2 \sqrt{np} \sqrt{t} (1+\lambda)  + t M^2 (1+\lambda) \right] \label{j3}\\
    & = M \sqrt{np}\sup_{\bW \in \mathcal{W}} \sqrt{\text{tr}(\bW^2)}   \sqrt{ \text{tr}\left(   \bV_\alpha \bV_\alpha^\top \Sigma \right)}  + c \left[ M^2 \sqrt{np} \sqrt{t} (1+\lambda)  + t M^2 (1+\lambda) \right] \nonumber \\
    & = M \sqrt{np}\sup_{\bW \in \mathcal{W}} \sqrt{\text{tr}(\bW^2)}   \sqrt{\sigma^2 \text{tr}\left(   \bV_\alpha \bV_\alpha^\top  \right)}  + c \left[ M^2 \sqrt{np} \sqrt{t} (1+\lambda)  + t M^2 (1+\lambda) \right] \nonumber \\
    & = \sigma M \sqrt{np}\sup_{\bW \in \mathcal{W}} \sqrt{\text{tr}(\bW^2)}  + c \left[ M^2 \sqrt{np} \sqrt{t} (1+\lambda)  + t M^2 (1+\lambda) \right] \nonumber\\
    & \le M^2 \sqrt{np}\sup_{\bW \in \mathcal{W}} \sqrt{\text{tr}(\bW^2)}  + c \left[ M^2 \sqrt{np} \sqrt{t} (1+\lambda)  + t M^2 (1+\lambda) \right]\label{j4}\\
    & \le  M^2 (1+\lambda) \sqrt{np^2}   + c \left[ M^2 \sqrt{np} \sqrt{t} (1+\lambda)  + t M^2 (1+\lambda) \right]\label{j5}\\
    & = M^2 (1+\lambda) \left[p \sqrt{n} + c \sqrt{np} \sqrt{t} + ct\right]. \nonumber
\end{align}
\endgroup
We remark that inequality \eqref{j1} follows from Cauchy-Schwartz for the inner product $\langle \bA ,\bB \rangle = \bA^\top \bB$. The inequality \eqref{j2} follows from the fact that $\|\bepsilon\|_2$ is bounded above by $\sqrt{np}M$. Inequality \eqref{j3} is an application of Jensen's inequality, and \eqref{j4} follows from observing that $\sigma \le M$ (for instance, apply Problem 2.5 (b) of \cite{wainwright2019high} and observe that  $\epsilon_i$ are $M$-subgaussian). Finally, inequality \eqref{j5} follows from observing that
\[ \text{tr}(\bW^2) = p \sum_{l=1}^p (w_l^2 + \lambda w_l)^2 \le p (1+\lambda) \sum_{l=1}^p (w_l^2 + \lambda w_l) \le p (1+\lambda) \sum_{l=1}^p (w_l + \lambda ) = p(1+\lambda)^2.  \]
Thus, from the above analysis, we arrive at 
\[\mathbb{P}\left(\frac{1}{np}\sup_{\bW \in \mathcal{W}} \bepsilon^\top W \bV_\alpha \bV_\alpha^\top \bepsilon > M^2 (1+\lambda) \left[\frac{1}{\sqrt{n}} + c \frac{1}{\sqrt{np}} \sqrt{t} + \frac{ct}{np}\right] \right) \le e^{-t}\]
Taking $t=\log\left(\frac{1}{\delta}\right)$, we get,
\begin{equation}
    \mathbb{P}\left(\frac{1}{np}\sup_{\bW \in \mathcal{W}} \bepsilon^\top W \bV_\alpha \bV_\alpha^\top \bepsilon > M^2 (1+\lambda) \left[\frac{1}{\sqrt{n}} + c \sqrt{\frac{\log(1/\delta)}{np}} + c \frac{\log(1/\delta)}{np}\right] \right) \le \delta
\end{equation}
Thus, with probability at least $1-\delta$, 
\begin{equation}\label{s1}
\frac{1}{np} \bepsilon^\top \widehat{\bW} \bV_\alpha \bV_\alpha^\top \bepsilon \le \frac{1}{np}\sup_{\bW \in \mathcal{W}} \bepsilon^\top W \bV_\alpha \bV_\alpha^\top \bepsilon \le  M^2 (1+\lambda) \left[\frac{1}{\sqrt{n}} + c \sqrt{\frac{\log(1/\delta)}{np}} + c \frac{\log(1/\delta)}{np}\right]     
\end{equation}

\paragraph{Bounds for $\max_{(i,j) \in \mathcal{E}}\| \bepsilon^\top \widehat{\bW} \bV_{\bbeta} \bZ^-_{\mathcal{C}(i,j)}\|_2$}
Let $\be_j$ be the $j$-th coordinate vector of length $p {n \choose 2}$. Let $y_j=\be_j^\top (\bZ^-_{\mathcal{C}(i,j)})^\top  \bV_{\bbeta}^\top \widehat{\bW} \bepsilon $. Now note that $\Lambda_{max}(\widehat{\bW}) \leq (1+\lambda)$, $\Lambda_{max}(\bV_{\bbeta}) = 1$ and $\Lambda_{max}(\bZ^-)=\frac{1}{\sqrt{n}}$. Thus $y_j$ is a univariate, bounded random variable with $|y_j| \le \frac{(1+\lambda)M}{\sqrt{n}}$. Thus, $$ \max_{(i,j) \in \mathcal{E}}\| \bepsilon^\top \widehat{\bW} \bV_{\bbeta} \bZ^-\|_\infty = \max_{j}|y_j| \le \frac{(1+\lambda)M}{\sqrt{n}}.$$
Now since there are only $p$ indices in $\mathcal{C}(i,j)$, we have 
\begin{equation}\label{s2}
  \frac{1}{np}\max_{(i,j) \in \mathcal{E}}\| \bepsilon^\top \widehat{\bW} \bV_{\bbeta} \bZ^-_\mathcal{C}(i,j)\|_2 \leq \sqrt{p} \frac{1}{np}\max_{(i,j) \in \mathcal{E}}\| \bepsilon^\top \widehat{\bW} \bV_{\bbeta} \bZ^-_\mathcal{C}(i,j)\|_\infty\le 
\frac{(1+\lambda)M}{\sqrt{n^3p}}.  
\end{equation}
%We see that when $\gamma' > \frac{(1+\lambda)M}{\sqrt{np}}$, $\frac{\gamma'}{n} \ge  \frac{1}{np} \max_{(i,j) \in \mathcal{E}}\| \bepsilon^\top \widehat{\bW} \bV_{\bbeta} \bZ^-_\mathcal{C}(i,j)\|_2$
We see that  $\gamma^\prime \ge \frac{(1+\lambda)M}{\sqrt{n^3p}}$ implies that $\gamma^\prime \ge  \frac{1}{np} \max_{(i,j) \in \mathcal{E}}\| \bepsilon^\top \widehat{\bW} \bV_{\bbeta} \bZ^-_\mathcal{C}(i,j)\|_2.$

% \begin{equation}
% \label{s2}
% P\bigg(\frac{\gamma'}{n} < \frac{1}{np} \max_{(i,j) \in \mathcal{E}}\| \bepsilon^\top \widehat{\bW} \bV_{\bbeta} Z^-_\mathcal{C}(i,j)\|_2 \bigg) \leq \frac{2}{p {n \choose 2}}
% \end{equation} 
Combining equations (\ref{s1}) and (\ref{s2}), we see that
% \begin{equation}
% \label{s4}
% \frac{1}{np}|G(\ha,\hb,\widehat{\bW})| \leq M^2 (1+\lambda) \left[\frac{1}{\sqrt{n}} + c \sqrt{\frac{\log(1/\delta)}{np}} + c \frac{\log(1/\delta)}{np}\right]+ \frac{\gamma'}{n} \sum_{(i,j) \in \mathcal{E}}\|\bZ_{\mathcal{C}(i,j)}(\hb-\bbeta)\|_2 
% \end{equation}

\begin{equation}
\label{s4}
\frac{1}{np}|G(\ha,\hb,\widehat{\bW})| \leq M^2 (1+\lambda) \left[\frac{1}{\sqrt{n}} + c \sqrt{\frac{\log(1/\delta)}{np}} + c \frac{\log(1/\delta)}{np}\right]+ \gamma^\prime \sum_{(i,j) \in \mathcal{E}}\|\bZ_{\mathcal{C}(i,j)}(\hb-\bbeta)\|_2 
\end{equation}
holds with probability at least $1-\delta$. Thus, from equations (\ref{s3}) and (\ref{s4}), we deduce that with probability at least $1-\delta$,
% \begin{align*}
% & \frac{1}{2np}\|\bV_\alpha(\ha-\balpha)+\bV_{\bbeta} (\hb-\bbeta)\|^2_{\widehat{\bW}}+\gamma' \sum_{(i,j) \in \mathcal{E}}  \|\bZ_{\mathcal{C}(i,j)}\hb\|_2^2\\ 
% &\leq M^2 (1+\lambda) \left[\frac{1}{\sqrt{n}} + c \sqrt{\frac{\log(1/\delta)}{np}} + c \frac{\log(1/\delta)}{np}\right]+ \frac{\gamma'}{n} \sum_{(i,j) \in \mathcal{E}}\|\bZ_{\mathcal{C}(i,j)}(\hb-\bbeta)\|_2+\gamma' \sum_{(i,j) \in \mathcal{E}}\|\bZ_{\mathcal{C}(i,j)}\bbeta\|_2^2
% \end{align*}
 
\begin{align*}
& \frac{1}{2np}\vvvert\bV_\alpha(\ha-\balpha)+\bV_{\bbeta} (\hb-\bbeta)\vvvert^2_{\widehat{\bW}}+\gamma' \sum_{(i,j) \in \mathcal{E}}  \|\bZ_{\mathcal{C}(i,j)}\hb\|_2^2\\ 
&\leq M^2 (1+\lambda) \left[\frac{1}{\sqrt{n}} + c \sqrt{\frac{\log(1/\delta)}{np}} + c \frac{\log(1/\delta)}{np}\right]+ \gamma' \sum_{(i,j) \in \mathcal{E}}\|\bZ_{\mathcal{C}(i,j)}(\hb-\bbeta)\|_2+\gamma' \sum_{(i,j) \in \mathcal{E}}\|\bZ_{\mathcal{C}(i,j)}\bbeta\|_2^2
\end{align*} 
Finally, upon rearranging terms, we obtain that with probability at least $1-\delta$, 
% \begingroup
% \allowdisplaybreaks
% \begin{align}
% & \frac{1}{2np}\|\bV_\alpha(\ha-\balpha)+\bV_{\bbeta} (\hb-\bbeta)\|^2_{\widehat{\bW}} \nonumber\\
% & \leq M^2 (1+\lambda) \left[\frac{1}{\sqrt{n}} + c \sqrt{\frac{\log(1/\delta)}{np}} + c \frac{\log(1/\delta)}{np}\right]+ \gamma' \sum_{(i,j) \in \mathcal{E}}\bigg[\frac{\|\bZ_{\mathcal{C}(i,j)}\hb\|_2}{n}-\|\bZ_{\mathcal{C}(i,j)}\hb\|_2^2\bigg]\nonumber \\
% &+ \frac{\gamma'}{n} \sum_{(i,j) \in \mathcal{E}}\|\bZ_{\mathcal{C}(i,j)}\bbeta\|_2 +\gamma' \sum_{(i,j) \in \mathcal{E}}\|\bZ_{\mathcal{C}(i,j)}\bbeta\|_2^2 \nonumber \\
% & \leq M^2 (1+\lambda) \left[\frac{1}{\sqrt{n}} + c \sqrt{\frac{\log(1/\delta)}{np}} + c \frac{\log(1/\delta)}{np}\right]+ \gamma' |\mathcal{E}| \frac{1}{4n^2}+\frac{\gamma'}{n} \sum_{(i,j) \in \mathcal{E}}\|\bZ_{\mathcal{C}(i,j)}\bbeta\|_2 \nonumber \\
% &+\gamma' \sum_{(i,j) \in \mathcal{E}}\|\bZ_{\mathcal{C}(i,j)}\bbeta\|_2^2 \label{sap1} \\
% & = M^2 (1+\lambda) \left[\frac{1}{\sqrt{n}} + c \sqrt{\frac{\log(1/\delta)}{np}} + c \frac{\log(1/\delta)}{np}\right]+\frac{\gamma' |\mathcal{E}|}{4n^2} + \frac{\gamma'}{n} \sum_{(i,j) \in \mathcal{E}}\|\bD_{\mathcal{C}(i,j)} \bu\|_2 +\gamma' \sum_{(i,j) \in \mathcal{E}}\|\bD_{\mathcal{C}(i,j)} \bu\|^2 \nonumber
% \end{align}
% \endgroup

\begingroup
\allowdisplaybreaks
\begin{align}
& \frac{1}{2np}\vvvert\bV_\alpha(\ha-\balpha)+\bV_{\bbeta} (\hb-\bbeta)\vvvert^2_{\widehat{\bW}} \nonumber\\
& \leq M^2 (1+\lambda) \left[\frac{1}{\sqrt{n}} + c \sqrt{\frac{\log(1/\delta)}{np}} + c \frac{\log(1/\delta)}{np}\right]+ \gamma' \sum_{(i,j) \in \mathcal{E}}\bigg[\|\bZ_{\mathcal{C}(i,j)}\hb\|_2-\|\bZ_{\mathcal{C}(i,j)}\hb\|_2^2\bigg]\nonumber \\
&+ \gamma' \sum_{(i,j) \in \mathcal{E}}\|\bZ_{\mathcal{C}(i,j)}\bbeta\|_2 +\gamma' \sum_{(i,j) \in \mathcal{E}}\|\bZ_{\mathcal{C}(i,j)}\bbeta\|_2^2 \nonumber \\
& \leq M^2 (1+\lambda) \left[\frac{1}{\sqrt{n}} + c \sqrt{\frac{\log(1/\delta)}{np}} + c \frac{\log(1/\delta)}{np}\right]+ \frac{\gamma' |\mathcal{E}|}{4} + \gamma' \sum_{(i,j) \in \mathcal{E}}\|\bZ_{\mathcal{C}(i,j)}\bbeta\|_2 \nonumber \\
&+\gamma' \sum_{(i,j) \in \mathcal{E}}\|\bZ_{\mathcal{C}(i,j)}\bbeta\|_2^2 \label{sap1} \\
& = M^2 (1+\lambda) \left[\frac{1}{\sqrt{n}} + c \sqrt{\frac{\log(1/\delta)}{np}} + c \frac{\log(1/\delta)}{np}\right]+\frac{\gamma' |\mathcal{E}|}{4} + \gamma' \sum_{(i,j) \in \mathcal{E}}\|\bD_{\mathcal{C}(i,j)} \bu\|_2 +\gamma' \sum_{(i,j) \in \mathcal{E}}\|\bD_{\mathcal{C}(i,j)} \bu\|^2 \nonumber
\end{align}
\endgroup

%Equation \eqref{sap1} follows from the fact that the maximum value of $f(z) = \frac{z}{n} - z^2$ is $\frac{1}{4n^2}$.

Equation \eqref{sap1} follows from the fact that the maximum value of $f(z) = z - z^2$ is $\frac{1}{4}$.

%$$\frac{1}{2np}\|\bV_\alpha(\ha-\balpha)+\bV_{\bbeta} (\hb-\bbeta)\|^2_{\widehat{\bW}} \leq\sigma^2 (1+\lambda) \bigg[\frac{1}{n}+\sqrt{\frac{\log(np)}{n^2p}}\bigg]+ \frac{\gamma'} \sum_{(i,j) \in \mathcal{E}}[\frac{\|\bZ_{\mathcal{C}(i,j)}\hb\|_2}{n}-\bZ_{\mathcal{C}(i,j)}\hb\|_2^2]+ \frac{\gamma'}{2} \sum_{(i,j) \in \mathcal{E}}\|\bZ_{\mathcal{C}(i,j)}\bbeta\|_2$$
 \end{proof}

\subsection{Proof of Corollary 1}
\paragraph{Corollary 1}Suppose $\|\bD_{\mathcal{C}(i,j)}\bu\| \le C$, for all $1 \le i,j\le n$, for some constant $C$ and $\gamma^\prime = \mathcal{O}(n^{-3/2}p^{-1/2})$. If $n=o(p)$, then $\frac{1}{2np} \vvvert\hat{\bu}-\bu\vvvert_{\widehat{\bW}}^2 \xrightarrow{\mathbb{P}}0$, as, $n,p \to \infty$.  
\begin{proof}
For any fixed $\delta>0$, we know from Theorem 2 that with probability at least $1-\delta$,
\begingroup
\allowdisplaybreaks
\begin{align*}
    \frac{1}{2np} \vvvert\hat{\bu}-\bu\vvvert_{\widehat{\bW}}^2\leq & M^2 (1+\lambda) \left[\frac{1}{\sqrt{n}} + c \sqrt{\frac{\log(1/\delta)}{np}} + c \frac{\log(1/\delta)}{np}\right]+\frac{\gamma' |\mathcal{E}|}{4}\\
	    &+ \gamma' \sum_{(i,j) \in \mathcal{E}}\|\bD_{\mathcal{C}(i,j)} \bu\|_2 +\gamma' \sum_{(i,j) \in \mathcal{E}}\|\bD_{\mathcal{C}(i,j)} \bu\|^2\\
	    \le & M^2 (1+\lambda) \left[\frac{1}{\sqrt{n}} + c \sqrt{\frac{\log(1/\delta)}{np}} + c \frac{\log(1/\delta)}{np}\right]+\frac{\gamma' |\mathcal{E}|}{4}+  (C+C^2) \gamma'|\mathcal{E}|\\
	    \le & M^2 (1+\lambda) \left[\frac{1}{\sqrt{n}} + c \sqrt{\frac{\log(1/\delta)}{np}} + c \frac{\log(1/\delta)}{np}\right]+\frac{\gamma' n^2}{4}+  (C+C^2)n^2 \gamma'\\
	    \le & M^2 (1+\lambda) \left[\frac{1}{\sqrt{n}} + c \sqrt{\frac{\log(1/\delta)}{np}} + c \frac{\log(1/\delta)}{np}\right] + \mathcal{O}\left(\sqrt{\frac{n}{p}}\right)\\
	    \to & 0.
\end{align*}
\endgroup
Thus, for any fixed $\epsilon>0$, $ \mathbb{P}\left(\frac{1}{2np} \vvvert\hat{\bu}-\bu\vvvert_{\widehat{\bW}}^2 > \epsilon\right) \le \delta$ as $n$ and $p$ become large. Hence, $\frac{1}{2np} \vvvert\hat{\bu}-\bu\vvvert_{\widehat{\bW}}^2 \xrightarrow{\mathbb{P}} 0$.
\end{proof}

\subsection{Proof of Corollary 2}
\paragraph{Corollary 2}Suppose $\|\bD_{\mathcal{C}(i,j)}\bu\| \le C$, for all $1 \le i,j \le n$, for some constant $C$ and $\gamma^\prime = \mathcal{O}(n^{-3/2}p^{-1/2})$. If $|\mathcal{E}|=\mathcal{O}(n)$, then $\frac{1}{2np} \vvvert\hat{\bu}-\bu\vvvert_{\widehat{\bW}}^2 =\mathcal{O}_p\left(n^{-1/2}\right)$.    
\begin{proof}
For any fixed $\delta>0$, we know from Theorem 2 that with probability at least $1-\delta$,
\begingroup
\allowdisplaybreaks
\begin{align*}
    \frac{1}{2np} \vvvert\hat{\bu}-\bu\vvvert_{\widehat{\bW}}^2\leq & M^2 (1+\lambda) \left[\frac{1}{\sqrt{n}} + c \sqrt{\frac{\log(1/\delta)}{np}} + c \frac{\log(1/\delta)}{np}\right]+\frac{\gamma' |\mathcal{E}|}{4}\\
	    &+ \gamma' \sum_{(i,j) \in \mathcal{E}}\|\bD_{\mathcal{C}(i,j)} \bu\|_2 +\gamma' \sum_{(i,j) \in \mathcal{E}}\|\bD_{\mathcal{C}(i,j)} \bu\|^2\\
	    \le & M^2 (1+\lambda) \left[\frac{1}{\sqrt{n}} + c \sqrt{\frac{\log(1/\delta)}{np}} + c \frac{\log(1/\delta)}{np}\right]+\frac{\gamma' |\mathcal{E}|}{4}+  (C+C^2) \gamma'|\mathcal{E}|\\
	    \le & M^2 (1+\lambda) \left[\frac{1}{\sqrt{n}} + c \sqrt{\frac{\log(1/\delta)}{np}} + c \frac{\log(1/\delta)}{np}\right]+\gamma^\prime \mathcal{O}(n)\\
	    \le & M^2 (1+\lambda) \left[\frac{1}{\sqrt{n}} + c \sqrt{\frac{\log(1/\delta)}{np}} + c \frac{\log(1/\delta)}{np}\right] + \mathcal{O}\left(\sqrt{\frac{1}{np}}\right)\\
	    %\to & 0.
	    = & \mathcal{O}(n^{-1/2})
\end{align*}
\endgroup
Thus, $\sqrt{n} \frac{1}{2np} \vvvert\hat{\bu}-\bu\vvvert_{\widehat{\bW}}^2 = \mathcal{O}(1)$. This implies that there exists a constant $C^\prime$, such that $ \mathbb{P}\left(\sqrt{n} \frac{1}{2np} \vvvert\hat{\bu}-\bu\vvvert_{\widehat{\bW}}^2 \le C^\prime \right) \ge 1 - \delta$, for all $n \in \mathbb{N}$. Hence, $\sqrt{n} \frac{1}{2np} \vvvert\hat{\bu}-\bu\vvvert_{\widehat{\bW}}^2$ is tight.

% Thus, for any fixed $\epsilon>0$, $ \mathbb{P}\left(\frac{1}{2np} \vvvert\hat{\bu}-\bu\vvvert_{\widehat{\bW}}^2 > \epsilon\right) \le \delta$ as $n$ and $p$ become large. Hence, $\frac{1}{2np} \vvvert\hat{\bu}-\bu\vvvert_{\widehat{\bW}}^2 \xrightarrow{\mathbb{P}} 0$.
\end{proof}

\section{Proof of Theorem 3}

\begin{proof}
Let $g_t$ be the value of the objective function (9) (of the main text) at iteration $t$. From the update steps of Algorithm 1, it is clear that $g_{t+1} \le g_t$, for all $t \in \mathbb{N}$. Thus the sequence $\{g_t\}_{t=1}^\infty$ is forms a decreasing sequence, and moreover $g_t \ge 0, \, \forall t\in \mathbb{N}$. Thus $\{g_t\}_{t=1}^\infty$ converges by the monotone convergence theorem. Thus, for all $\epsilon>0$, there exists $T \in \mathbb{N}$, such that $g_T -g_{T+1} \le \epsilon$, so that an absolute or relative convergence criterion based on Eq. (9) is satisfied in finite iterations . %for all $t \ge T$. The 
Because the objective is biconvex, applying Theorem 5.1 of \cite{tseng2001convergence} immediately implies that the limit point is a coordinate-wise minimum of equation (9).
\end{proof}

\section{Additional simulation study, figures, and tables}

\subsection{Dendrograms, simulation study 4.2}
\begin{figure}[!h]
    \centering
            \begin{subfigure}[t]{0.48\textwidth}
    \centering
        \includegraphics[height=.8\textwidth,width=.8\textwidth]{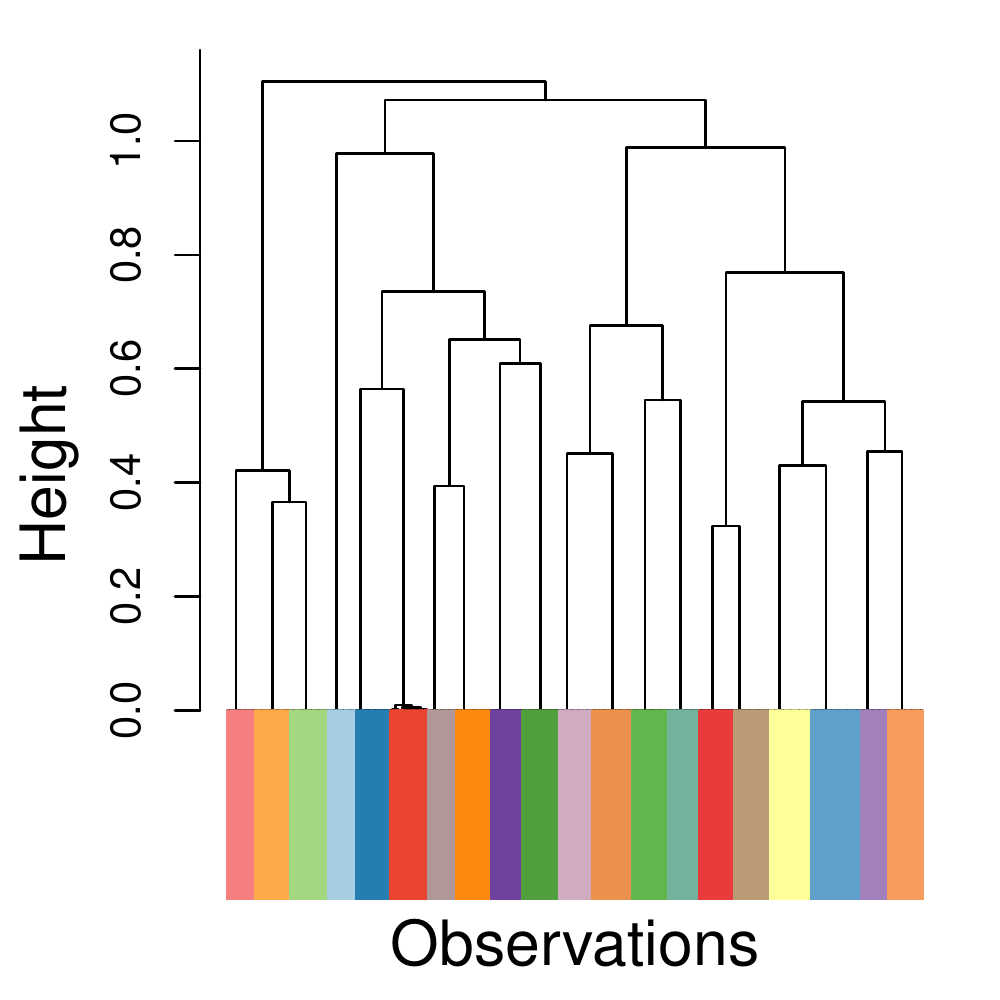}
        \caption{Biconvex}
    \end{subfigure}
    ~
                \begin{subfigure}[t]{0.48\textwidth}
    \centering
        \includegraphics[height=.8\textwidth,width=.8\textwidth]{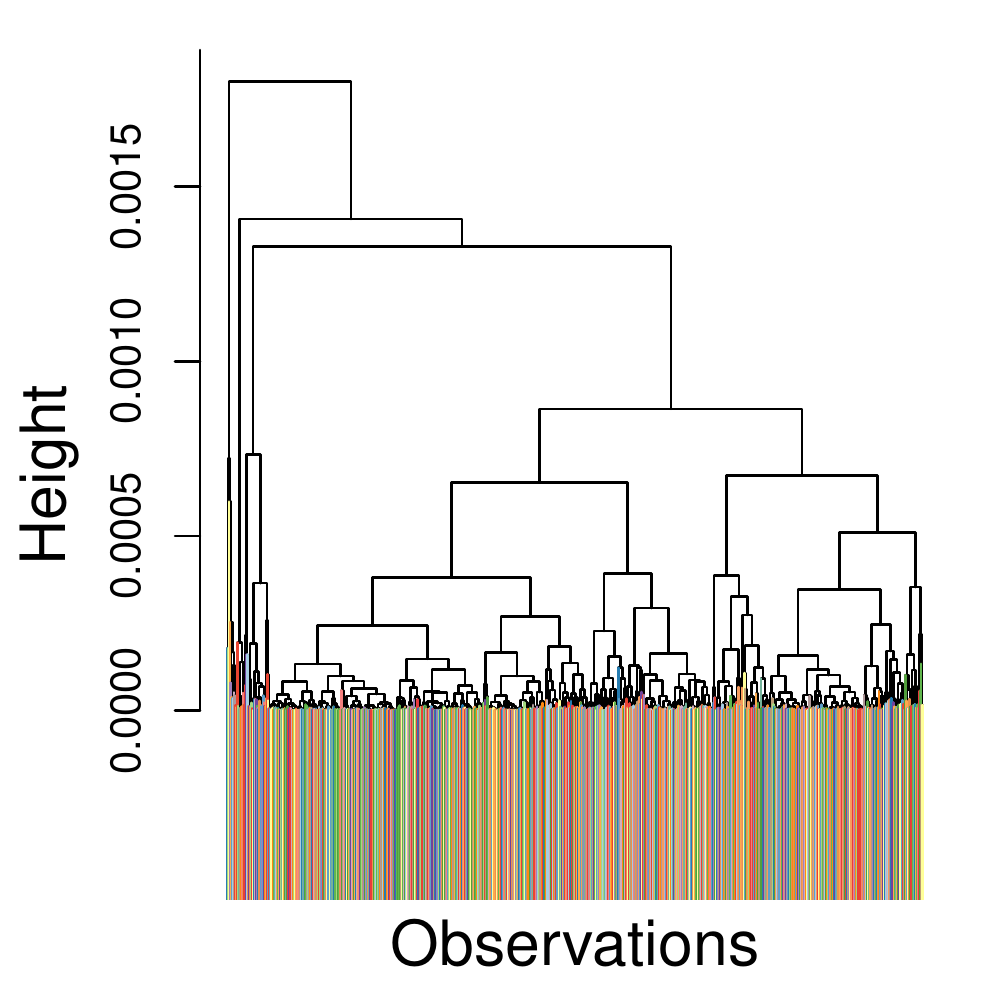}
        \caption{Sparse Hierarchical Clustering}
    \end{subfigure}
    \caption{Dendrograms, color coded with the ground truth, produced by biconvex clustering and sparse hierarchical clustering algorithms on the simulated data in ``Clustering accuracy" of Section 4.2 (of main text). Biconvex clustering perfectly recovers the ground truth using a dynamic tree cut, and it is clear that this can be achieved for a generous range of heights. In contrast sparse hierarchical clustering failed to recover the true clusters, and the dendrogram structure in panel (b)  provides insight on why this is the case. }
    \label{dendros}
\end{figure}

\subsection{Sensitivity to Random Restarts}
\label{dd}
\begin{figure}[!htbp]
    \centering
            \begin{subfigure}[t]{0.3\textwidth}
    \centering
        \includegraphics[height=\textwidth,width=\textwidth]{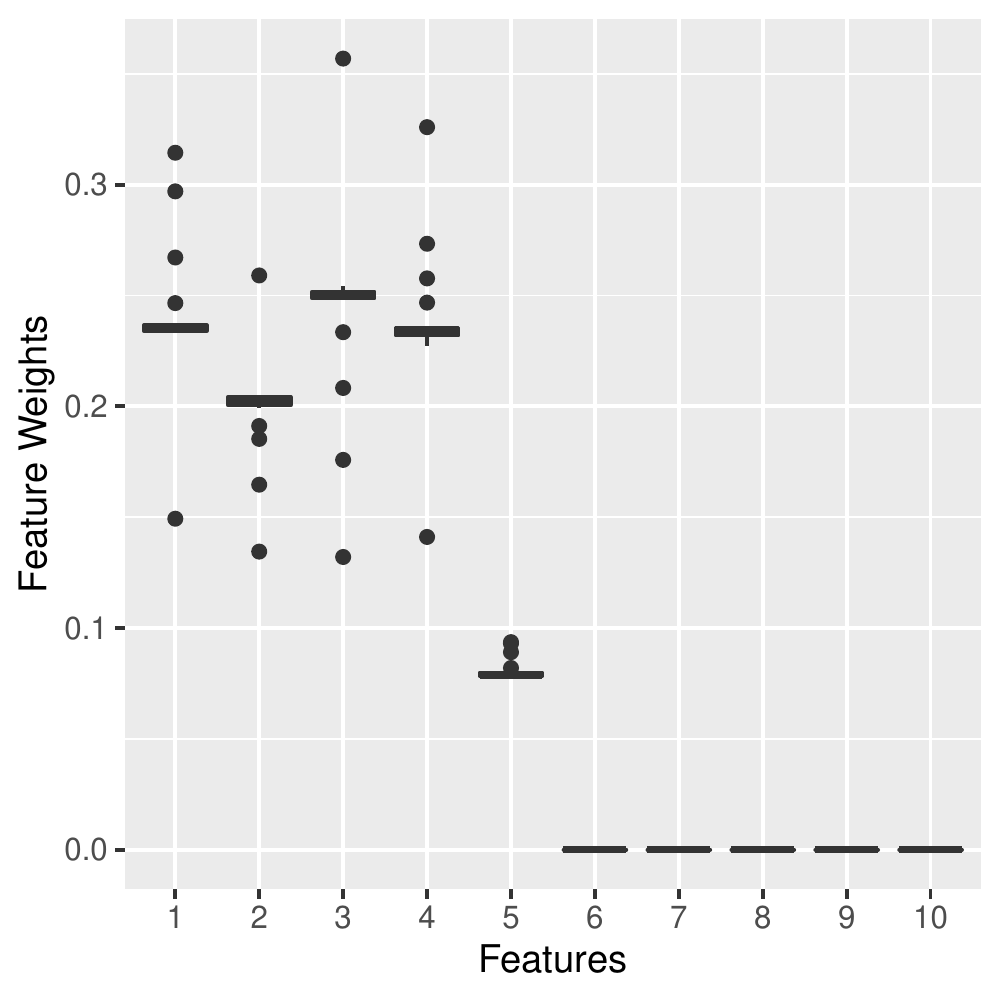}
        \caption{\# Irrelevant features = 5}
       % \label{ae_1}
    \end{subfigure}
    ~
    \begin{subfigure}[t]{0.3\textwidth}
    \centering
        \includegraphics[height=\textwidth,width=\textwidth]{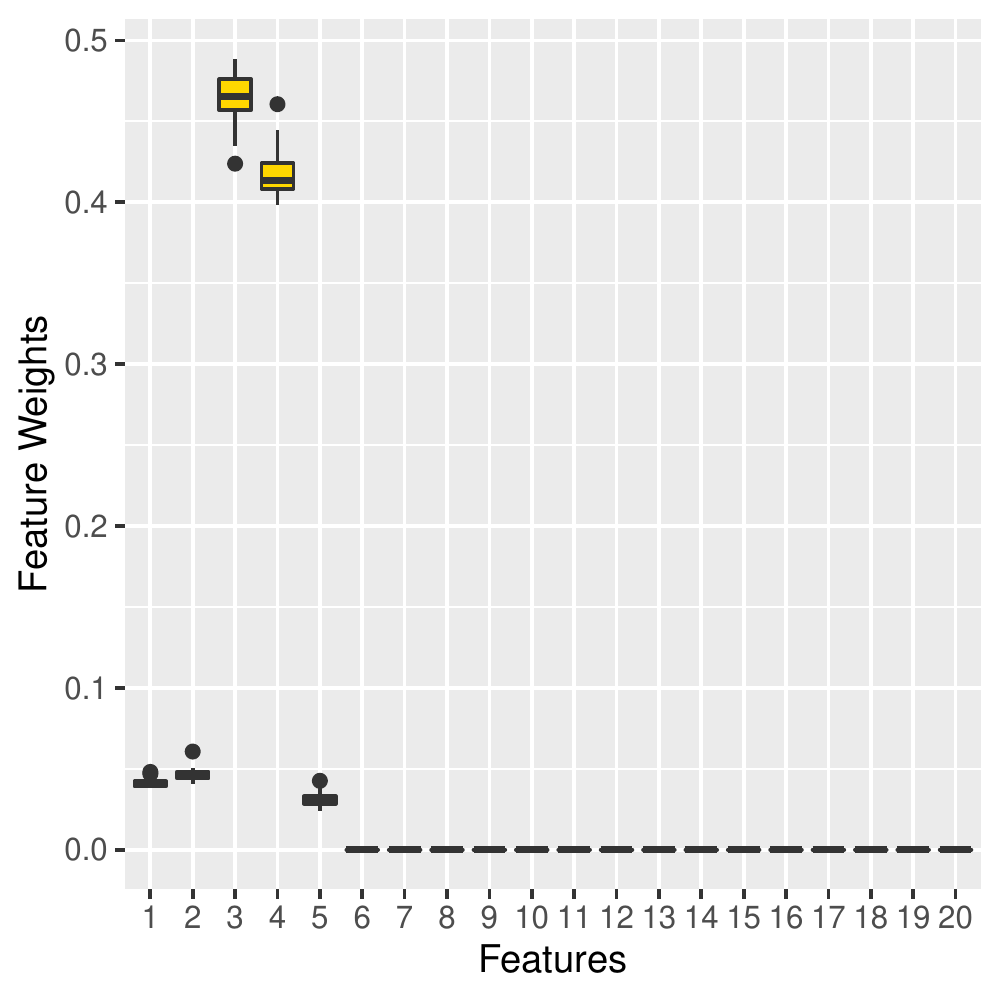}
        \caption{\#Irrelevant features = 15}
        %\label{ae_2}
    \end{subfigure}
    ~
    \begin{subfigure}[t]{0.3\textwidth}
    \centering
        \includegraphics[height=\textwidth,width=\textwidth]{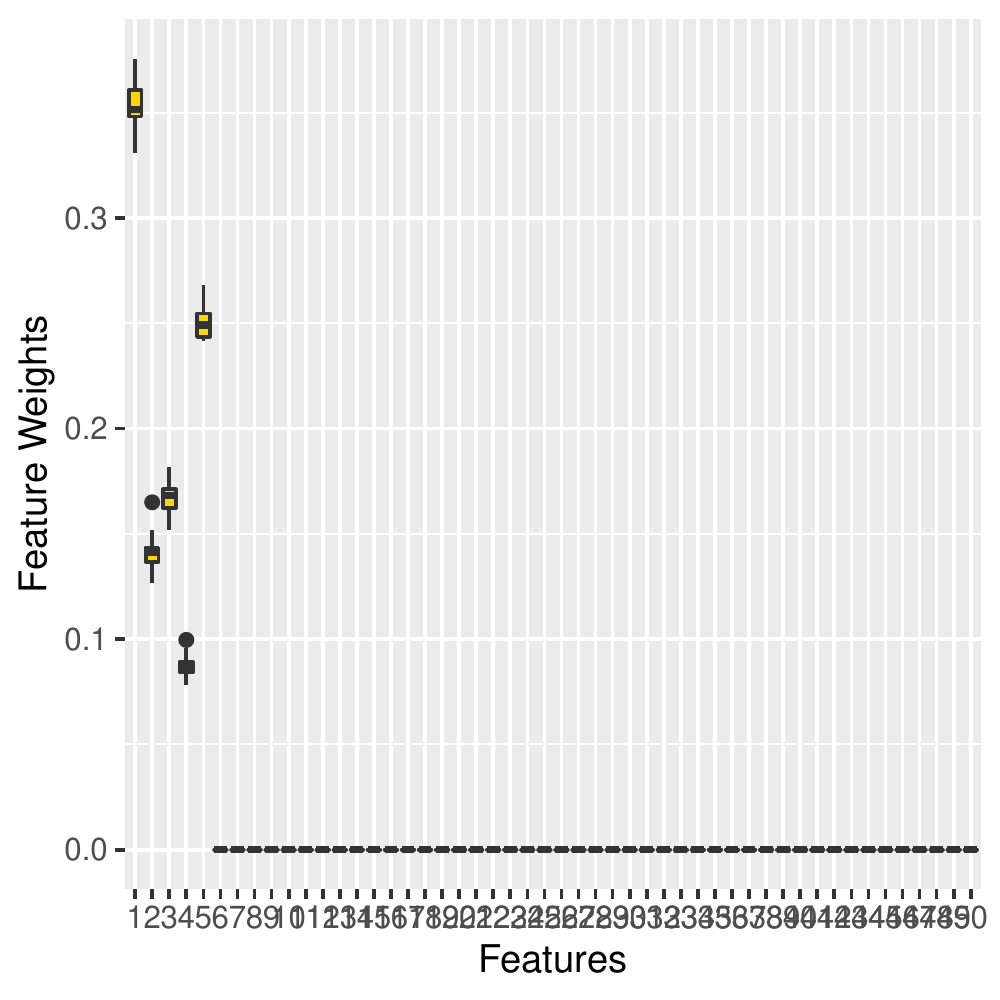}
        \caption{\#Irrelevant features = 85}
        %\label{ae_3}
    \end{subfigure}
                \caption{Box plot of top ranked features selected by BCC algorithm under random initialization of $\bw$ as the number of irrelevant features (d) varies (detailed in Section \ref{dd}). We see the performance of the algorithm appears to be fairly stable even in high dimensions.}
        \label{rw}
\end{figure}

In this section, we will study the effects of random initialization of the feature weights.  We generated three datasets each with the first five features revealing the cluster structure of the dataset. We then append $d$ many features to the dataset, each generated from a standard normal distribution. The datasets are generated using the procedures described in Section 4, ``Feature selection" (of main text). For each of the datasets, we start with a randomly chosen initialization of feature weights and iterate until convergence. We initialize each of the feature weights from $Unif(0,1)$, and repeat the experiment over $100$ trials. Boxplots of the resulting feature weight estimates are shown in Figure \ref{rw}. It can be easily observed from Figure \ref{rw} that the algorithm can consistently identifies the relevant variables, while giving no weight to the unimportant ones. It also consistently recovers the cluster structure in each of the experiments, demonstrating stability to initial guess of $\bw$.

\subsection{Tuning  $\gamma$ and $\lambda$}
We recommend setting the parameters $\gamma$ and $\lambda$ in similar fashion to \cite{wang2018sparse,chi2017convex}. The idea is to formulate a matrix completion problem, similar to that done in section 5.1 of \cite{chi2017convex}, and select the hyperparameter values that result in minimum prediction (test) error value on the hold-out dataset. Let $\mathcal{T}\subset \{1,\dots,n\} \times \{1,\dots,p\}$ denote indices of the held-out data. One may select a small fraction of elements as the validation set, i.e.  $|\mathcal{T}| \approx 0.1 \times np$ for about 10\% of the data. We then solve according to the objective evaluated over the remaining data:
\[\min_{\bU,\bw}\left\{\sum_{(i,l) \in \mathcal{T}^c}(w_l^2+\lambda w_l) (x_{il}-u_{il})^2 + +\gamma \sum_{i,j=1;i\neq j}^n\sum_{l=1}^p \phi_{ij} (\mu_{il}-\mu_{jl})^2\right\} \text{ subject to } \sum_{l=1}^p w_l = 1.\]
Now, we choose $(\lambda^\ast,\gamma^\ast)$ such that the measure of fit on the hold-out dataset $\sum_{(i,l) \in \mathcal{T}} (x_{il} - u_{il})^2$ is minimized over a grid of $(\lambda,\gamma)$ values.

\subsection{Clustering solution visualizations on movement data}

\begin{figure}[htbp]
    \centering
            \begin{subfigure}[t]{0.23\textwidth}
    \centering
        \includegraphics[height=\textwidth,width=\textwidth]{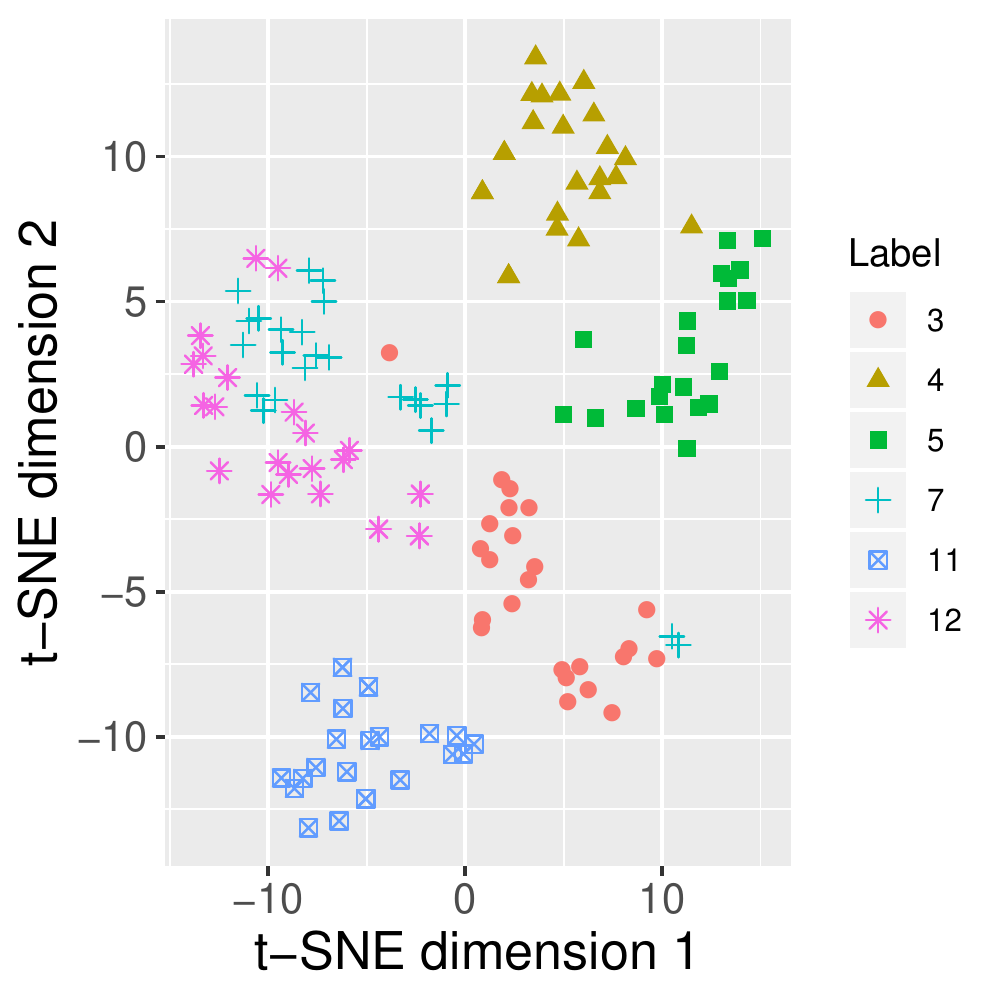}
        \caption{Ground Truth}
        \label{libras_gt}
    \end{subfigure}
    ~
    \begin{subfigure}[t]{0.23\textwidth}
    \centering
        \includegraphics[height=\textwidth,width=\textwidth]{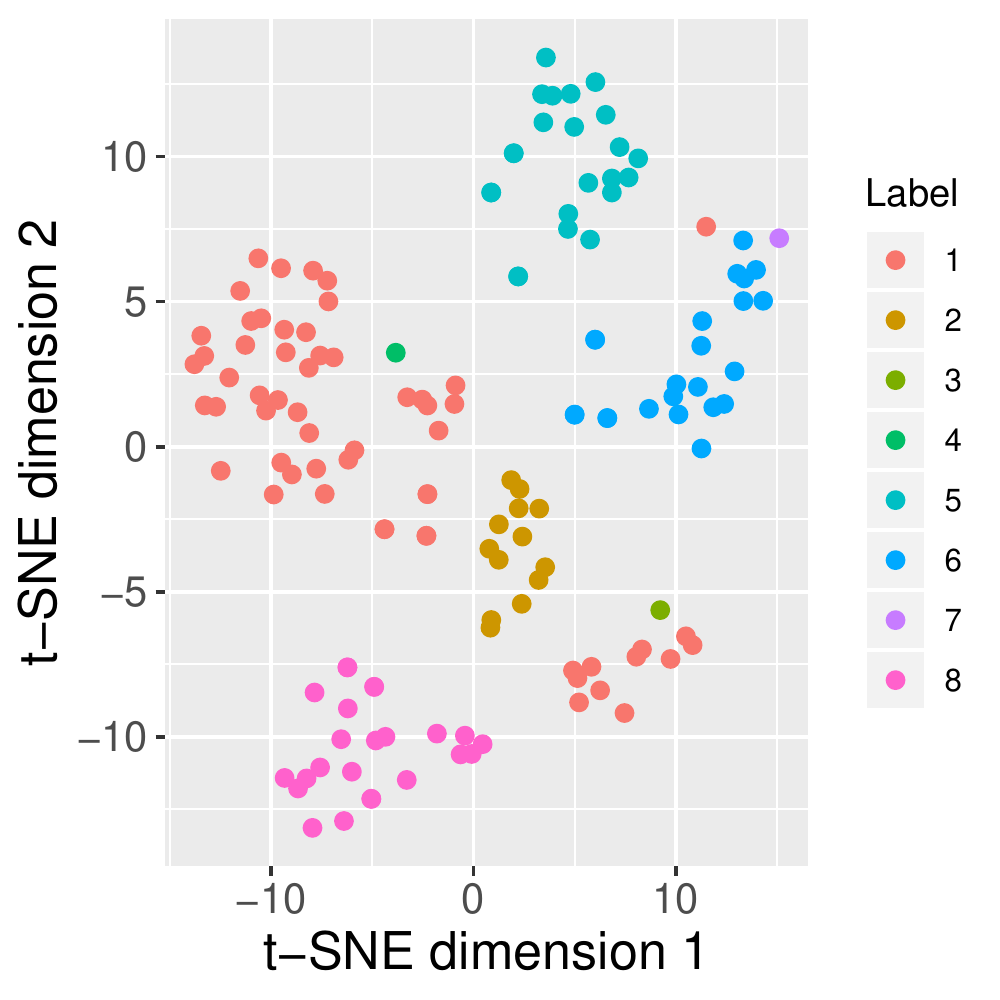}
        \caption{Convex Clustering}
        %\label{scvx_partition}
    \end{subfigure}
    ~
    \begin{subfigure}[t]{0.23\textwidth}
    \centering
        \includegraphics[height=\textwidth,width=\textwidth]{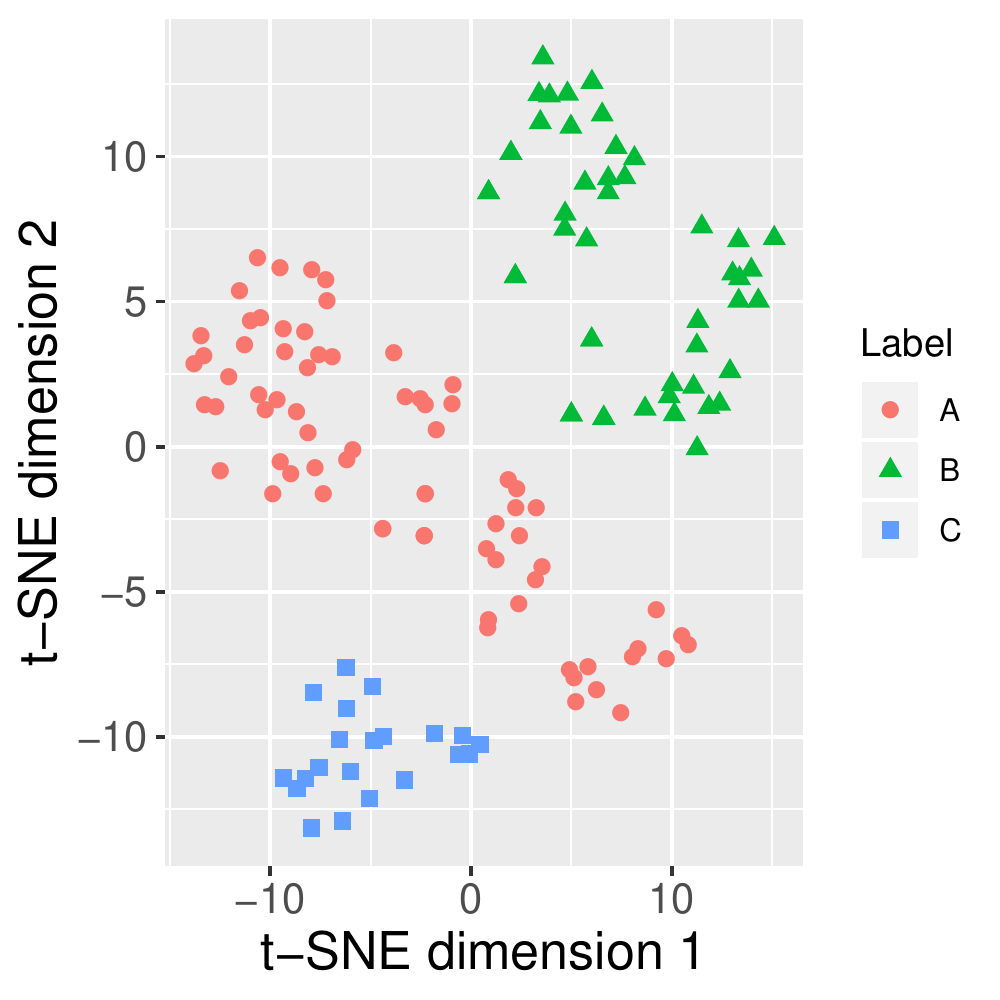}
        \caption{Sparse Convex}
       % \label{scvx_partition}
    \end{subfigure}
    ~
    \begin{subfigure}[t]{0.23\textwidth}
    \centering
        \includegraphics[height=\textwidth,width=\textwidth]{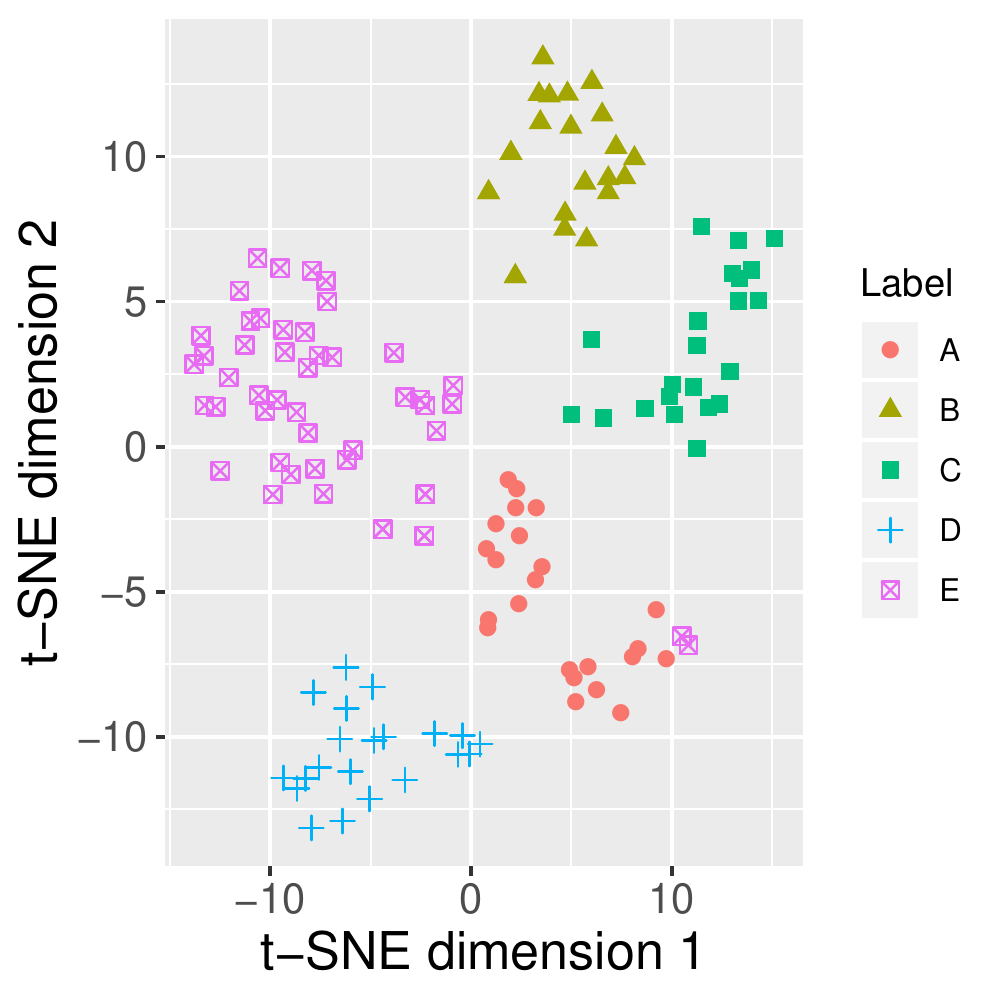}
        \caption{Biconvex}
        \label{libras_clw}
    \end{subfigure}
    ~
    \begin{subfigure}[t]{0.23\textwidth}
    \centering
        \includegraphics[height=\textwidth,width=\textwidth]{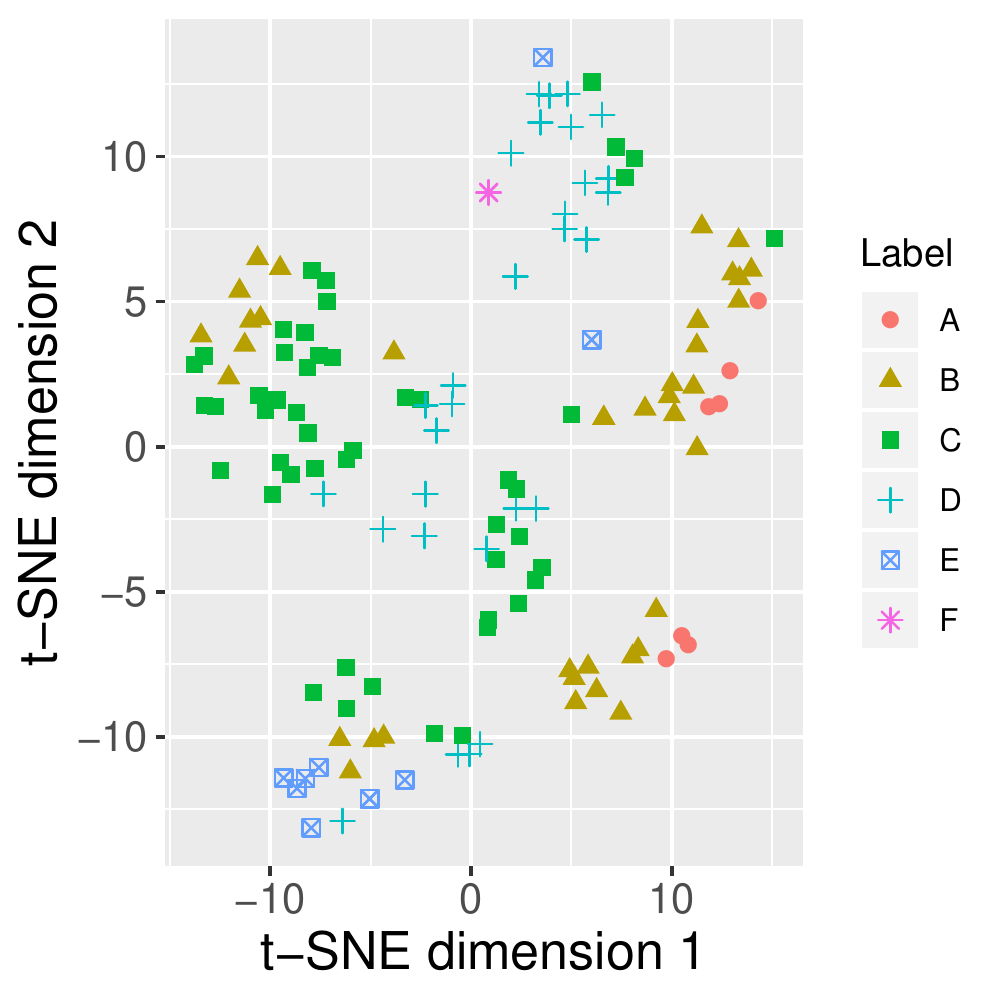}
        \caption{\footnotesize Sparse Hierarchical}
        %\label{scvx_partition}
    \end{subfigure}
    ~
    \begin{subfigure}[t]{0.23\textwidth}
    \centering
        \includegraphics[height=\textwidth,width=\textwidth]{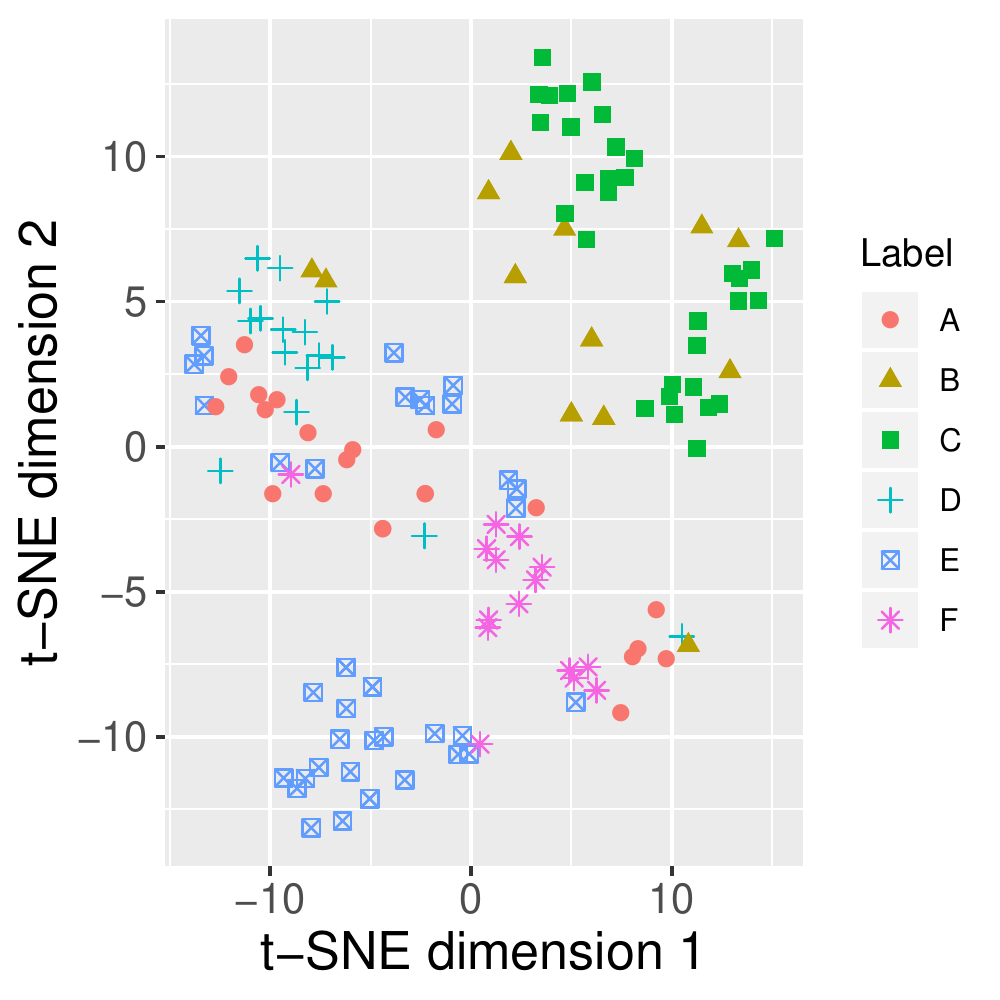}
        \caption{Sparse $k$-means}
       % \label{scvx_partition}
    \end{subfigure}
    ~
    \begin{subfigure}[t]{0.23\textwidth}
    \centering
        \includegraphics[height=\textwidth,width=\textwidth]{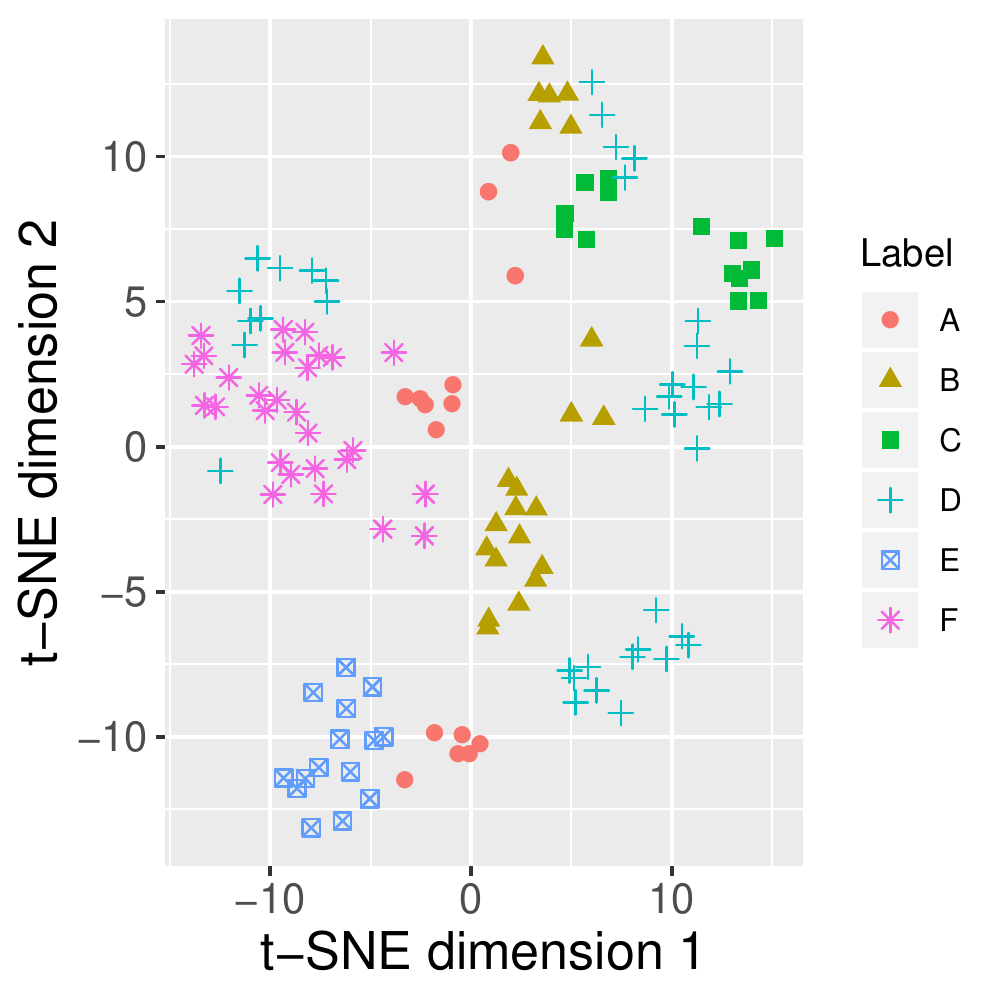}
        \caption{Average Linkage}
        \label{libras_hc}
    \end{subfigure}
                \caption{Comparison of solutions in t-SNE embedding, Libras  dataset.}
        \label{libras fig}
\end{figure}

\begin{figure}[!htbp]
    \centering
            \begin{subfigure}[t]{0.3\textwidth}
    \centering
        \includegraphics[height=\textwidth,width=\textwidth]{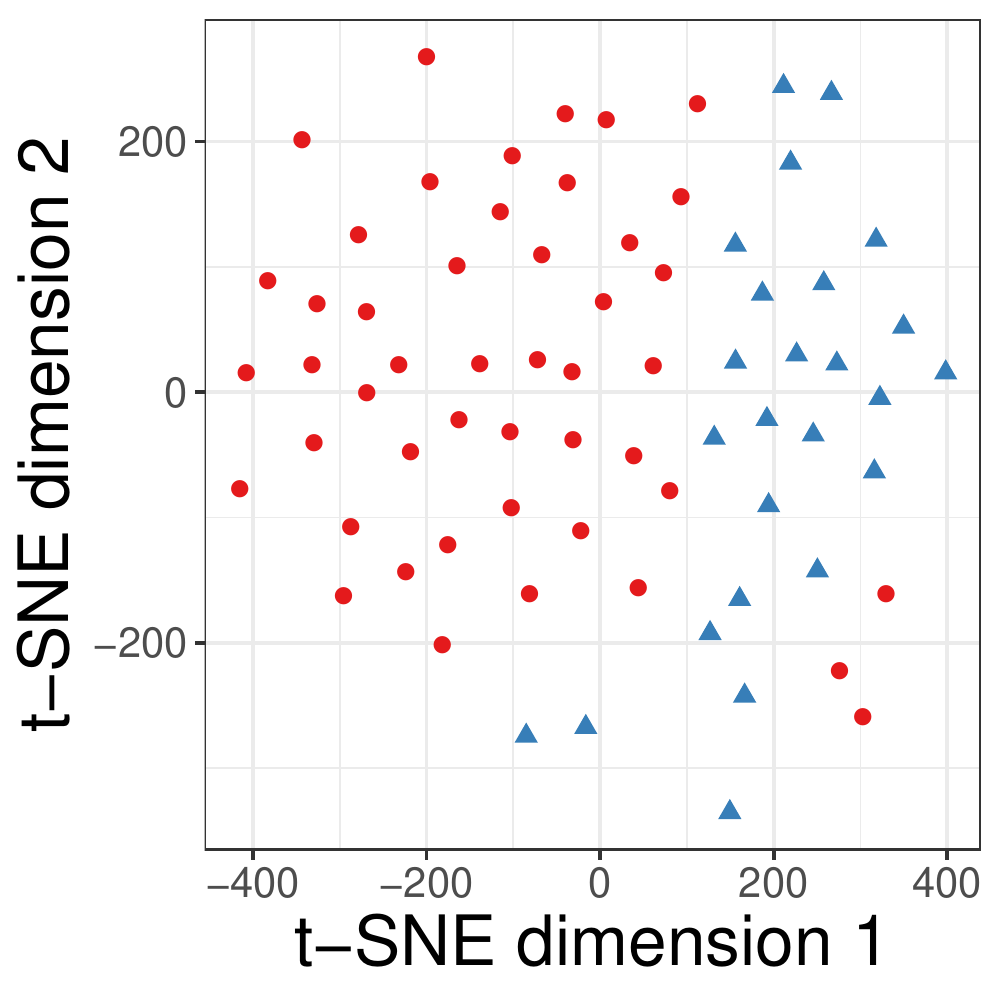}
        \caption{Ground Truth}
    \end{subfigure}
    ~
    \begin{subfigure}[t]{0.3\textwidth}
    \centering
        \includegraphics[height=\textwidth,width=\textwidth]{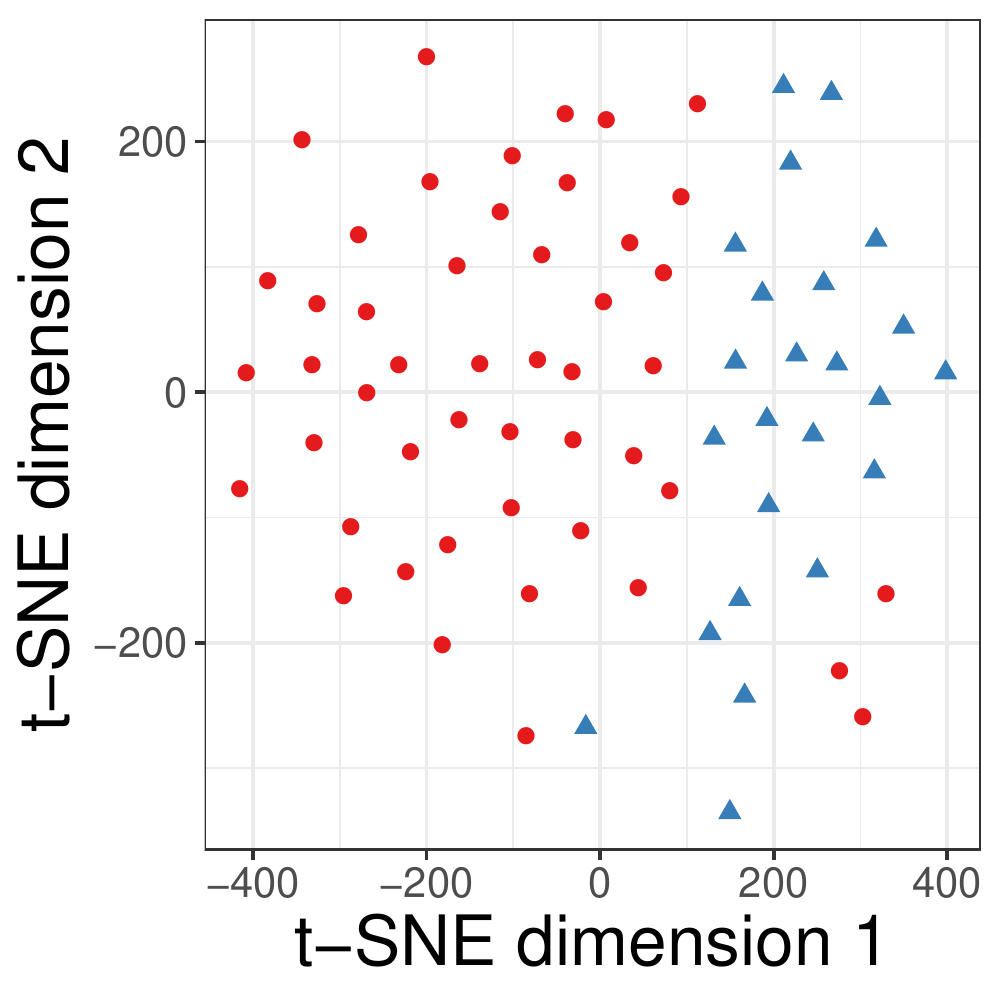}
        \caption{Biconvex}
    \end{subfigure}
     ~
    \begin{subfigure}[t]{0.3\textwidth}
    \centering
        \includegraphics[height=\textwidth,width=\textwidth]{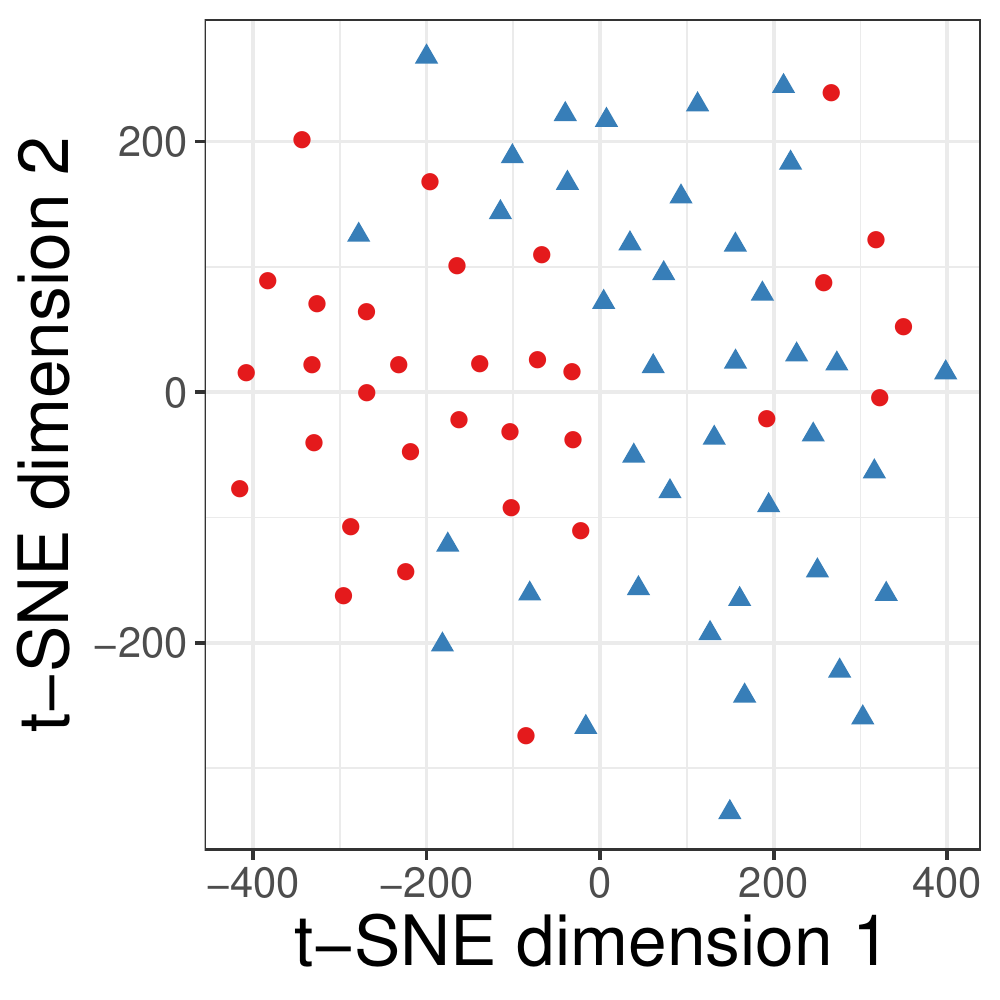}
        \caption{Sparse $k$-means}
      % \label{luk_gt}
    \end{subfigure}
                \caption{Comparison of solutions in t-SNE embedding, leukemia dataset. We see that the solution obtained under our method closely resembles the ground truth. }
        \label{fig:luk}
\end{figure}
\end{document}